\documentclass[11pt,a4paper,oneside]{article}
\pdfoutput=1
\usepackage{jheppub}
\usepackage{hyperref}
\usepackage{graphicx,xcolor}
\usepackage{sectsty,amssymb,amsmath}
\usepackage{enumerate}
\usepackage{booktabs}
\usepackage{multirow}
\usepackage{slashed}
\usepackage[utf8]{inputenc} 
\usepackage{subcaption}
\usepackage{siunitx}
\usepackage{here}
\usepackage{cleveref}
\usepackage{tikz}
\usepackage[normalem]{ulem} % provides strikethrough command \sout
\newcommand\redsout{\bgroup\markoverwith{\textcolor{red}{\rule[0.5ex]{2pt}{0.4pt}}}\ULon}
\allowdisplaybreaks % Allow align environment to be split over pages

\usepackage{chngcntr}
\counterwithin{figure}{section} 
\counterwithin{table}{section}
\counterwithin{equation}{section}

\def\({\left(}
\def\){\right)}
\def\be{\begin{equation}}
\def\ee{\end{equation}}
\def\bes{\begin{subequations}}
\def\ees{\end{subequations}}
\def\bea{\begin{eqnarray}}
\def\eea{\end{eqnarray}}
\def\bry{\begin{array}}
\def\ery{\end{array}}
\def\bit{\begin{itemize}}
\def\eit{\end{itemize}}
\def\ben{\begin{enumerate}}
\def\een{\end{enumerate}}

\def\tr{\textrm{Tr}}

\def\hc{\text{h.c.}}

\def\dst{\displaystyle}

\def\La{\mathcal{L}}
\def\f{\frac}
\def\eq{Eq.~\eqref}

\def\demub{\partial_{\mu}}

\def\demua{\partial^{\mu}}

\begin{document}

%=============================================================================
\title{A Simplified Model of Heavy Vector Singlets for the LHC and Future Colliders}

\author[a]{Michael J.~Baker,}    
\author[a]{Timothy Martonhelyi,}   
\author[a]{Andrea Thamm,}     
\author[b]{Riccardo Torre} 

\emailAdd{mjbaker@umass.edu}   
\emailAdd{tmartonhelyi@umass.edu}    
\emailAdd{athamm@umass.edu}
\emailAdd{riccardo.torre@ge.infn.it}

\affiliation[a]{Department of Physics, University of Massachusetts Amherst, MA 01003, USA}
\affiliation[b]{INFN, Sezione di Genova, Via Dodecaneso 33, I-16146 Genova, Italy}

\date{\today}

%=============================================================================

\abstract{
We study a simplified model of two colourless heavy vector resonances in the singlet representation of $SU(2)_{L}$, with zero and unit hypercharge. 
We discuss mixing with the Standard Model gauge bosons due to electroweak symmetry breaking, semi-analytic formulae for production at proton colliders, requirements to obey the narrow width approximation and selected low energy constraints.
We show current LHC constraints and sensitivity projections for the HL-LHC, HE-LHC, SPPC and FCC-hh on the charged and neutral heavy vectors. 
The utility of the simplified model Lagrangian is demonstrated by matching these results onto three explicit models: a weakly coupled abelian extension of the Standard Model gauge group, a weakly coupled non-abelian extension and a strongly coupled minimal composite Higgs model. 
All our results are presented in terms of physical resonance masses, using expressions which are accurate even at vector masses near the electroweak scale due to a parameter inversion we derive. We discuss the importance of this inversion and point out that its effect, and the effects of electroweak symmetry breaking, can remain important up to resonance masses of several TeV. 
Finally, we clarify the relation between this simplified model and the Heavy Vector Triplet (HVT) model, a simplified model for heavy $SU(2)_{L}$ triplets with zero hypercharge, and provide exact and approximate matching relations.
}

%-----------------------------------------------------------------------------
\maketitle
%-----------------------------------------------------------------------------

%=================================================
\section{Introduction}
\label{sec:introduction}
%=================================================

Heavy vector resonances appear in a wide variety of models beyond the Standard Model (SM). For example, they appear as gauge bosons of spontaneously broken abelian or non-abelian gauge symmetries that extend the SM gauge group \cite{
%Z'
Barger:1980ix,
Hewett:1989dr,
Cvetic:1995zs,
Rizzo:2006wq,
Agashe:2007hh,
Langacker:2008yv,
Salvioni:2010p2769,
Accomando:2013ve,
%W'
Agashe:2009bj,
Schmaltz:2010p2610,
Grojean:2011vu,
Langacker:1989xa,
Frank:2010p2250,
Accomando:2011up,
Accomando:2011gt,
Dobrescu:2021vak} and as heavy vector mesons of new strongly coupled sectors \cite{
Chanowitz:1993fc,
Langacker:2008yv,
Barbieri:2008cc,
Barbieri:2009p33,
Agashe:2009dg,
Agashe:2009ve,
Cata:2009iy,
Barbieri:2010mn,
CarcamoHernandez:2010wpm,
CarcamoHernandez:2010qxf,
Accomando:2011gt,
Falkowski:2011ua,
Contino:2011np,
Chanowitz:2011ew,
Bellazzini:2012tv,
Accomando:2012us,
Low:2015uha,
Accomando:2016mvz,
Liu:2018hum,
Liu:2019bua,
DeCurtis:2021fdm,
Greco:2014aza,
Liu:2023jta}. This makes them prime targets for collider searches.

While heavy vector resonances can have very different theoretical origins, their phenomenology at colliders depends on just a few properties.  Heavy vectors that couple to quarks or gauge bosons can be produced via Drell-Yan processes or vector boson fusion \cite{Baker:2022zxv}, respectively. After production, the heavy particles may decay into jets, heavy quarks, gauge bosons, Higgs bosons, and leptons.  This wide range of possible decay channels makes heavy vectors a good benchmark for collider searches.  Furthermore, heavy vectors are constrained in a wide variety of ways as the ATLAS and CMS collaborations have searched extensively for narrow resonances in these final states \cite{
%charged
ATLAS:2019fgd, ATLAS:2017eqx,ATLAS:2018uca,CMS:2021mux,CMS:2017zod,ATLAS:2018iui,ATLAS:2019nat,ATLAS:2016hal,CMS:2022pjv,CMS:2017fgc,CMS:2016rqm,ATLAS:2020fry,ATLAS:2017jag,CMS:2021klu,CMS:2018dff,ATLAS:2017otj,CMS:2021xor,CMS:2018sdh,CMS:2021itu,CMS:2018ygj,ATLAS:2018sxj,CMS:2021zxu,ATLAS:2017xel,ATLAS:2017ptz,
%neutral
ATLAS:2019erb,ATLAS:2017fih,CMS:2021ctt,CMS:2018ipm,CMS:2016cfx,ATLAS:2017eiz,CMS:2016xbv,CMS:2022zoc,ATLAS:2023taw,ATLAS:2020lks,CMS:2017ucf}.

Simplified models provide a powerful framework for studying the phenomenological features of a particle, 
regardless of its theoretical origin, provided it is the lightest new particle in the full theory (like the pion in QCD). The simplified model can be viewed as a bridge between explicit models and experimental searches. On one side, the simplified model can be analytically matched onto any explicit model, typically reducing the number of free parameters. On the other side, experimental results can be expressed in terms of a general set of simplified model parameters. The experimental limits can then easily be translated into any explicit model, which makes the experimental results extremely general and versatile. However, care has to be taken to remain within the realm of validity of the simplified model. This can be ensured by restricting attention to narrow resonances, which can be treated in the narrow width approximation.  Experimental searches then constrain strictly on-shell quantities, such as $\sigma \times \text{BR}$. Resonance searches focusing on a narrow resonance region, where finite widths effects are negligible, ensure model independence \cite{Accomando:2011up,Choudhury:2011cg,Accomando:2013ve, Pappadopulo:2014tg}. 

Simplified models for heavy vector resonances have been studied extensively in the literature \cite{
Sullivan:2002p2617,
Rizzo:2007bk,
Salvioni:2009mt,
Grojean:2011vu,
Alves:2011dz,
Bauer:2009p1295,
Accomando:2010p2249,
delAguila:2010mx,
Han:2010p2696,
Barbieri:2011p2759,
Chiang:2011kq,
DeSimone:2012ul,
deBlas:2012tc,
AguilarSaavedra:2013hg,
Buchkremer:2013uj,
Lizana:2013vz,
Pappadopulo:2014tg,
Chivukula:2017lyk,
Chivukula:2021foa,
Saez:2018off,
Belyaev:2018xpf,
Baker:2022zxv}.\footnote{See Ref.~\cite{Dawson:2024ozw} for the effects of a neutral heavy resonance on the SM effective theory.} Here, we add for the first time:~(1) LHC constraints on the simplified model parameter space for a model of heavy vector singlets;
(2) sensitivity projections for the HL-LHC, HE-LHC, SPPC and FCC-hh on the heavy vector singlets in the simplified model framework;
(3) the implementation of a parameter inversion that allows us to express experimental results in terms of physical parameters, while correctly taking into account the effects of electroweak symmetry breaking; and
(4) a detailed discussion of the relation between simplified models of $\text{SU(2)}_L$ triplets and singlets.

Reference \cite{Pappadopulo:2014tg} explores triplets under $\text{SU(2)}_L$ with zero hypercharge, the Heavy Vector Triplet (HVT) simplified model. Here we follow a similar approach and explore the simplified model for colourless singlets under $\text{SU(2)}_L$ with unit and zero hypercharge, the Heavy Vector Singlet (HVS) simplified model. Individually, these vectors amount to a charged and neutral vector but combined they can describe a triplet under an $\text{SU(2)}_R$ symmetry. 
In \cref{sec:simplified-model}, we first discuss the simplified model Lagrangian.  We then perform a parameter inversion which allows us to treat the physical heavy singlet masses as input parameters and automatically reproduce the $Z$ boson mass, the Fermi constant and the fine structure constant.  To do this we take account of electroweak symmetry breaking (EWSB), to find the mixing with the electroweak gauge bosons, and compute the new Fermi constant in this model. We also compare the exact treatment accounting for EWSB effects and parameter inversion with a truncated method which ignores the effects of EWSB. We discuss vector singlet production via Drell-Yan processes at the LHC and future colliders, and their decay.  We also discuss low energy constraints from electroweak precision tests (EWPTs).
In \cref{sec:data-bounds}, we collect current LHC limits and provide constraints on the simplified model parameter space of the charged and neutral vectors. Furthermore, we provide sensitivity projections for the HL-LHC, HE-LHC, SPPC and FCC-hh and discuss their reach on the production cross-section of the heavy vectors.
In \cref{sec:explicit-models}, we match three explicit models onto the simplified model parameter space: a $U$(1) extension of the SM gauge group, a non-abelian extension and a strongly coupled minimal composite Higgs model. We discuss current LHC limits and constraints from electroweak precision tests in these explicit models and provide sensitivity projections for future colliders.
Finally, in \cref{sec:hvt}, we explore the relation of the simplified model for heavy vector singlets with the HVT. The triplet Lagrangian constitutes a subset of the singlet Lagrangian and we conclude that certain singlet parameters can be constrained by experimental results provided in terms of the triplet simplified model parameters.

%=================================================
\section{A Simplified Model for Heavy Vector Singlets}
\label{sec:simplified-model}
%=================================================

%-------------------------------------------------
\subsection{The Simplified Model Lagrangian}
%-------------------------------------------------

In addition to the SM field content, we consider a colourless charged vector $\mathcal{V}_{\mu}^{\pm}$ and a colourless neutral vector $\mathcal{V}^{0}_{\mu}$ which are singlets under $SU(2)_{L}$ and have unit and zero hypercharge, respectively. Assuming $CP$ conservation and in analogy with Refs.~\cite{deBlas:2012tc,Grojean:2011vu,Pappadopulo:2014tg}, we describe the new vectors by the phenomenological Lagrangian
\be\label{sml}
\La_{\text{ph}}=\mathcal{L}_{\mathcal{V}^{+}} + \mathcal{L}_{\mathcal{V}^{0}} + \mathcal{L}_{\text{mix}} \,,
\ee
where the Lagrangian terms for the charged vector are
\be
\bry{lll} \label{sml-charged}
\mathcal{L}_{\mathcal{V}^{+}} 
& = & \dst - \f{1}{2} D_{[\mu}\mathcal{V}^{+}_{\nu]}D^{[\mu}\mathcal{V}^{-\,\nu]} + m_{\mathcal{V}^{+}}^{2} \mathcal{V}^{+}_{\mu}  \mathcal{V}^{-\mu} \vspace{2mm}\\
&& \dst - i \f{g_{V}}{\sqrt{2}}c_{H}^{+} \mathcal{V}^{+}_\mu H^\dag {\overset{{}_{\leftrightarrow}}{D}}^\mu \tilde{H} + \f{g_{V}}{\sqrt{2}} c_{q}^{+} \mathcal{V}^{+}_{\mu} J^{\mu}_{q}+ \hc \vspace{2mm}\\
&& \dst + 2g_{V}^{2} c_{VVHH}^{+} \mathcal{V}^{+}_\mu \mathcal{V}^{- \, \mu} H^\dag H + i g' c_{VVB}^{+} B_{\mu\,\nu} \mathcal{V}^{+\,\mu} \mathcal{V}^{-\,\nu}\,, 
\ery
\ee
the Lagrangian terms for the neutral vector are
\be
\bry{lll} \label{sml-neutral}
\mathcal{L}_{\mathcal{V}^{0}}
& = & \dst - \f{1}{4} \partial_{[\mu}\mathcal{V}^{0}_{\nu]}\partial^{[\mu}\mathcal{V}^{0\,\nu]} + \frac{m_{\mathcal{V}^{0}}^{2}}{2} \mathcal{V}^{0}_{\mu}  \mathcal{V}^{0\,\mu} \vspace{2mm}\\
&& \dst +i \f{g_{V}}{2}  c_H^{0} \mathcal{V}^{0}_\mu H^\dag {\overset{{}_{\leftrightarrow}}{D}}^\mu H + \sum_{\Psi=Q,L,U,D,E} \f{g_{V}}{2} c_{\Psi}^{0} \mathcal{V}^{0}_{\mu} J^{\mu}_{\Psi} \vspace{2mm}\\
&& \dst + g_{V}^{2} c_{VVHH}^{0} \mathcal{V}^{0}_\mu \mathcal{V}^{0\,\mu} H^\dag H\,, 
\ery
\ee
and the Lagrangian terms for the mixing of the two new vectors are
\be
\bry{lll} \label{sml-mix}
\mathcal{L}_{\text{mix}} 
& = & \dst (i g_{V} c_{VVV}^{+} D_{[\mu}\mathcal{V}^{-}_{\nu]} \mathcal{V}^{0\,\mu} \mathcal{V}^{+\,  \nu} + \hc) + i g_{V} c_{VVV}^{0} \partial_{[\mu}\mathcal{V}^{0}_{\nu]} \mathcal{V}^{+\,\mu} \mathcal{V}^{-\,\nu}\,,
\ery
\ee
where $\tilde{H} = i \sigma_2 H^*$, all couplings are assumed to be real and $g_V$ is a redundant but useful parameter (see below). The fermion currents are
\begin{equation}
    \label{eq:fcurrents}
    J^{\mu}_{q} = \sum_{i,j=1}^{3} V_{ij}^R \bar{U}^{i}\gamma^{\mu}D^{j} \, , \qquad J^{\mu}_{\Psi}=\sum_{i=1}^{3}\bar{\Psi}^{i}\gamma^{\mu}\Psi^{i} \,,
\end{equation}
where fermion fields are represented as four-component Dirac spinors and $i,j=1,2,3$ counts the generations and $V_{ij}^R$ is the right-handed mixing matrix (which we take to be the identity in our phenomenological work).
The charged singlet could in principle couple to right-handed neutrinos, $N^{i}$, via $(g_{V}/\sqrt{2}) c_{N}^{+} \mathcal{V}^{+}_{\mu} J^{\mu}_{N}$ where $J^{\mu}_{N}=\sum_{i,j=1}^{3}U_{ij}^R\bar{N}^{i}\gamma^{\mu}E^{j}$ and $U_{ij}^R$ is the corresponding lepton right-handed mixing matrix. However, we will not consider the presence of right-handed neutrinos here, since our aim is to build a phenomenological model to describe heavy resonance searches with direct decay into SM particles.  This model will apply to explicit models containing right-handed neutrinos as long as they are heavier than the charged heavy vector.

The three Lagrangians $\mathcal{L}_{\mathcal{V}^{+}}$, $\mathcal{L}_{\mathcal{V}^{0}}$ and $\mathcal{L}_{\text{mix}}$ describe the dynamics of the charged and neutral vectors in isolation and the interactions among the two. We can consider the case where the two vectors belong to a single triplet representation of the $SU(2)_{R}$ subgroup of the SM custodial group. In that case, the charged and neutral $c$ parameters will be related, as discussed in \cref{subsec:4.2,subsec:4.3}. Alternatively, each vector can be considered in isolation by taking the mass of the other to infinity. 

In analogy to Ref.~\cite{Pappadopulo:2014tg}, the phenomenological Lagrangian in \cref{sml} is based on the following assumptions:
\ben
\item the interactions of the new vectors conserve $CP$;
\item only operators with mass dimension lower than or equal to four are included;
\item quadrilinear interactions among the heavy vectors, which are irrelevant for the hadron collider phenomenology we consider, are not included;
\item the kinetic mixing between the neutral singlet and the SM hypercharge gauge boson $B_{\mu}$ is not included since it can be eliminated from the Lagrangian by a field redefinition (see Appendix \ref{AppA} for details);
\item motivated by composite models \cite{Panico:2015jxa}, every insertion of the heavy vector fields, of $H$ and of the fermionic fields in the simplified model Lagrangian (but not the Standard Model terms) is weighted by $g_{V}$, while insertions of $B_{\mu}$ are weighted by the SM gauge coupling, $g'$ (note that we then divide $\mathcal{L}_\text{ph}$ by $g_V^2$ to canonically normalise the kinetic terms of the heavy vector singlets);
\item $g_{V}$ has dimension $[\hbar]^{-1/2}$, while the $c$ parameters are dimensionless constants.
\een
For an extensive discussion of these assumptions and their motivations we refer the reader to Ref.~\cite{Pappadopulo:2014tg}. Notice that no gauge invariant kinetic mixing is allowed for the charged singlet $\mathcal{V}^{+}_\mu$. Finally, let us stress again that the parameter $g_{V}$ is redundant and could be absorbed in a redefinition of the $c$ coefficients. However, it represents a convenient normalisation to make the power counting manifest.

The first lines of $\mathcal{L}_{\mathcal{V}^{+}}$ and $\mathcal{L}_{\mathcal{V}^{0}}$ in \cref{sml-charged,sml-neutral},
\be
\bry{lll}
\mathcal{L}_{\mathcal{V}^{+}} 
& \supset & \dst - \f{1}{2} D_{[\mu}\mathcal{V}^{+}_{\nu]}D^{[\mu}\mathcal{V}^{-\,\nu]} + m_{\mathcal{V}^{+}}^{2} \mathcal{V}^{+}_{\mu}  \mathcal{V}^{-\mu} \,, 
\\
\mathcal{L}_{\mathcal{V}^{0}}
& \supset & \dst - \f{1}{4} \partial_{[\mu}\mathcal{V}^{0}_{\nu]}\partial^{[\mu}\mathcal{V}^{0\,\nu]} + \frac{m_{\mathcal{V}^{0}}^{2}}{2} \mathcal{V}^{0}_{\mu}  \mathcal{V}^{0\,\mu} \,, 
\ery
\ee
contain the kinetic and mass terms of the two vectors. The vector $\mathcal{V}^{0}$ is a full singlet under the SM gauge group and its kinetic term is written in terms of the ordinary derivative $\partial_\mu$, while $\mathcal{V}^{+}$ is charged under the hypercharge $U(1)_{Y}$ gauge group and its covariant derivative is given in terms of $D_{\mu}$,
\be
D_{[\mu}\mathcal{V}^{+}_{\nu]}=D_{\mu}\mathcal{V}^{+}_{\nu} -D_{\nu}\mathcal{V}^{+}_{\mu}\,,\qquad D_{\mu}\mathcal{V}^{+}_{\nu}=\partial_\mu \mathcal{V}^{+}_{\nu}- i g' B_\mu \mathcal{V}^{+}_{\nu}\,,
\ee
where $g'$ is the hypercharge gauge coupling.

The first terms in the second lines of $\mathcal{L}_{\mathcal{V}^{+}}$ and $\mathcal{L}_{\mathcal{V}^{0}}$ in \cref{sml-charged,sml-neutral},
\be
\bry{lll} 
\mathcal{L}_{\mathcal{V}^{+}} 
& \supset & \dst - i \f{g_{V}}{\sqrt{2}}c_{H}^{+} \mathcal{V}^{+}_\mu H^\dag {\overset{{}_{\leftrightarrow}}{D}}^\mu \tilde{H} + \hc\,, \vspace{2mm} 
\\
\mathcal{L}_{\mathcal{V}^{0}}
& \supset & \dst +i \f{g_{V}}{2}  c_H^{0} \mathcal{V}^{0}_\mu H^\dag {\overset{{}_{\leftrightarrow}}{D}}^\mu H\,, 
\ery
\ee
contain the interactions of the heavy vectors with the Higgs currents, 
\be
\bry{lll} \label{hcurrents}
i H^\dag {\overset{{}_{\leftrightarrow}}{D}}^\mu \tilde{H}
&=& i H^\dag (D^\mu \tilde{H})-i (D^\mu H)^\dag \tilde{H}\,,\\
i H^\dag {\overset{{}_{\leftrightarrow}}{D}}^\mu H
&=& 
i H^\dag (D^\mu H)-i (D^\mu H)^\dag H\,,
\ery
\ee
which transform in the appropriate complex conjugate representation of the hypercharge gauge group.  The interactions of the heavy vector singlets with these currents are proportional to the couplings $g_{V}c_{H}^{+}$ and $g_{V}c_{H}^{0}$ and describe interactions of the new vectors with the physical Higgs boson, the three unphysical Goldstone bosons and the transverse SM gauge bosons.  After electroweak symmetry breaking the couplings $g_{V}c_{H}^{0}$ and $g_{V}c_{H}^{+}$ control the interactions of the new vectors with the longitudinal polarisations of the SM gauge bosons and with the Higgs boson.  These interactions also lead to mixing between the heavy vector singlets and the SM $W$ and $Z$ bosons.

The second terms in the second lines of $\mathcal{L}_{\mathcal{V}^{+}}$ and $\mathcal{L}_{\mathcal{V}^{0}}$ in \cref{sml-charged,sml-neutral},
\be
\bry{lll} 
\mathcal{L}_{\mathcal{V}^{+}} 
& \supset & \dst  \f{g_{V}}{\sqrt{2}} c_{q}^{+} \mathcal{V}^{+}_{\mu} J^{\mu}_{q}+ \hc \,, \vspace{2mm}
\\
\mathcal{L}_{\mathcal{V}^{0}}
& \supset & \dst  \sum_{\Psi=Q,L,U,D,E} \f{g_{V}}{2} c_{\Psi}^{0} \mathcal{V}^{0}_{\mu} J^{\mu}_{\Psi} \,, 
\ery
\ee
contain the interactions with the fermion currents, given in \cref{eq:fcurrents}. 
An important difference between our HVS Lagrangian in \cref{sml-charged,sml-neutral} and the left-handed HVT discussed in Ref.~\cite{Pappadopulo:2014tg} is that while the HVT couplings to fermions are normalised by $g^{2}/g_{V}$, the HVS couplings are just weighted by $g_{V}$. This is because the left-handed triplet fermion current of Ref.~\cite{Pappadopulo:2014tg} only couples to the SM $SU(2)_{L}$ gauge bosons, implying that in most explicit models this current couples to the heavy vectors only through mixing with the $W$.  For heavy singlets, the fermion currents in \cref{eq:fcurrents} do couple to the heavy vectors. This means that we can expect non-mixing-suppressed HVS couplings to fermions in many explicit models.  While both the charged and the neutral vectors couple to quarks, and therefore have unsuppressed Drell-Yan production, they differ in their couplings to leptons. While the neutral vector couples directly to SM leptons, the charged vector does not. A coupling to a left-handed lepton and a neutrino can be induced through mixing with the SM $W$ boson after EWSB, but this is mixing suppressed. This implies that searches for new vector resonances decaying into two lepton final states can set a constraint on the neutral vector singlet but are irrelevant for the charged one. The fact that the charged vector singlet is difficult to constrain has been stressed in Ref.~\cite{Grojean:2011vu}, where a simplified model approach to this state has been proposed. Here we extend this approach to take into account possible correlations with the neutral singlet
and to unify the notation with our previous work of Ref.~\cite{Pappadopulo:2014tg}.  The coupling combinations $g_{V}c_{q}^{+}$ and $g_{V}c_{Q,U,D}^{0}$ control the production of the charged and the neutral vector, respectively, while the coupling combinations $g_{V}c_{q}^{+}$ and $g_{V}c_{\Psi}^{0}$ contribute to the neutral and charged vector decays. Of course, as in the case of vector triplets, one can go beyond the assumption of family-universal couplings to fermions and consider different couplings to the first two generations and the third one, e.g., $c_{Q}^{0}\to (c_{Q}^{0},c_{Q3}^{0})$, $c_{U}^{0}\to (c_{U}^{0},c_{U3}^{0})$ and $c_{D}^{0}\to (c_{D}^{0},c_{D3}^{0})$. The same can be done for the charged sector: $c_{q}^{+}\to (c_{q}^{+},c_{q 3}^{+})$.

The third lines of $\mathcal{L}_{\mathcal{V}^{+}}$ and $\mathcal{L}_{\mathcal{V}^{0}}$ in \cref{sml-charged,sml-neutral},
\be
\bry{lll} 
\mathcal{L}_{\mathcal{V}^{+}} 
& \supset & \dst + 2g_{V}^{2} c_{VVHH}^{+} \mathcal{V}^{+}_\mu \mathcal{V}^{- \, \mu} H^\dag H + i g' c_{VVB}^{+} B_{\mu\,\nu} \mathcal{V}^{+\,\mu} \mathcal{V}^{-\,\nu}\,, \vspace{2mm}
\\
\mathcal{L}_{\mathcal{V}^{0}}
& \supset & \dst + g_{V}^{2} c_{VVHH}^{0} \mathcal{V}^{0}_\mu \mathcal{V}^{0\,\mu} H^\dag H\,, 
\ery
\ee
contain interactions of the new vectors with the SM bosons. The operators proportional to $c_{VVHH}^{+}$ and $c_{VVHH}^{0}$ describe interactions of the heavy vectors with two Higgses. After EWSB these terms also give rise to interactions with a single Higgs boson and to an additional contribution to the vector masses proportional to $g_{V}^{2} c_{VVHH}^{0,+}\hat{v}^{2}$, where $\hat{v}^{2}=2\left<H^{\dag}H\right>$ is the vacuum expectation value of the Higgs doublet $H$, which in this model is in general different to the SM EWSB scale $v\equiv (\sqrt{2}G_{F})^{-1/2}\approx 246$\,GeV. The operator proportional to $c_{VVB}^{+}$ describes an interaction of the charged vector with the neutral SM $\gamma$ and $Z$ and, after EWSB, is relevant for the $WZ$ and $W\gamma$ decays of the charged heavy vector due to mixing effects. The parameter $c_{VVB}^{+}$ is also related to the gyromagnetic ratio $\mathrm{g}_{\mathcal{V}^{+}}$ of $\mathcal{V}^{+}$ by the relation $\mathrm{g} _{\mathcal{V}^{+}}=1-c_{VVB}^{+}$ \cite{Grojean:2011vu}.

Finally $\mathcal{L}_{\text{mix}} $ in \cref{sml-mix},
\be
\bry{lll}
\mathcal{L}_{\text{mix}} 
& = & \dst (i g_{V} c_{VVV}^{+} D_{[\mu}\mathcal{V}^{-}_{\nu]} \mathcal{V}^{0\,\mu} \mathcal{V}^{+\,  \nu} + \hc) + i g_{V} c_{VVV}^{0} \partial_{[\mu}\mathcal{V}^{0}_{\nu]} \mathcal{V}^{+\,\mu} \mathcal{V}^{-\,\nu}\,,
\ery
\ee
describes trilinear interactions among the new charged and neutral heavy vector fields. All the operators describing bilinear and trilinear interactions of the heavy vectors with SM fields and among each other are not directly relevant for single resonance production or for the decay of the heavy vectors into SM final states, but they do give subdominant contributions due to mixing with the SM gauge bosons after EWSB. Note that we do not consider cascade decays where one heavy vector decays into the other (and maybe a SM field).

%-------------------------------------------------
\subsection{Physical Masses and Input Parameters}
%-------------------------------------------------

The interactions between the heavy vector singlets and the Standard Model particles alter usual standard model relations (in particular the gauge boson masses and the Fermi constant), and so change the Standard Model Lagrangian parameters required to reproduce well-measured experimental quantities.  To take this into account, we invert the new relations for the gauge boson masses and the Fermi constant to write the relevant Lagrangian parameters in terms of these well-measured quantities.  In this subsection we first compute mixing angles and physical masses after electroweak symmetry breaking, before computing the new contributions to the Fermi constant.  We then describe the parameter inversion process.

%-------------------------------------------------
\subsubsection{Electroweak Symmetry Breaking}
\label{2.1}
%-------------------------------------------------

Upon EWSB only the electromagnetic $U(1)_Q$ remains unbroken, the photon is identified with the combination $A_{\mu}=\cos\theta_{W}B_{\mu}+\sin\theta_{W}W^{3}_{\mu}$, the electric charge assumes the form $e= g g'/\sqrt{g^{2}+g^{\prime\,2}}$ with $\tan\theta_{W}=g^{\prime}/g$, and the other SM vector bosons acquire masses.  In the HVS model, the new heavy vectors will receive extra contributions to their masses from interactions with the Higgs boson, and they will mix with the massive SM vector bosons. Working in the unitary gauge, we will write the massive vector fields after electroweak symmetry breaking but prior to this mixing as $\hat{W}^\pm$, $\hat{Z}$, $\hat{V}^\pm$ and $\hat{V}^0$, so the physical fields after mixing can be written $W^\pm$, $Z$, $V^\pm$ and $V^0$. We can then write two-by-two mass matrices for the charged vectors in the basis $(\hat{W}^+, \hat{V}^+)$,
\be
\displaystyle
\left(
\bry{cc}
{\hat{m}}^{2}_{W^+} & -c_H^{+} \zeta^{+} {\hat{m}}_{W^+} {\hat{m}}_{V^{+}}\\ -c_H^{+} \zeta^{+} {\hat{m}}_{W^+} {\hat{m}}_{V^{+}} &  {\hat{m}}_{V^{+}}^{2}
\ery
\right)\,,\;\;\;\;\;
{\textrm{where}}\;\;
\left\{
\bry{l}
\displaystyle
{\hat{m}}_{W^+} = \f{g \hat{v}}{2}\vspace{1mm}\\
\displaystyle
{\hat{m}}_{V^{+}}^{2}=m_{\mathcal{V}^{+}}^{2}+ g_{V}^{2} c_{VVHH}^{+} \hat{v}^{2}\vspace{1mm}\\
\displaystyle
\zeta^{+}=\frac{g_{V} \hat{v}}{{\hat{m}}_{V^{+}}}
\ery
\right.\,,
\ee
and for the neutral vectors in the basis $(\hat{Z}, \hat{V}^0)$,
\be
\displaystyle
\left(
\bry{cc}
{\hat{m}}^{2}_{Z} & - c_H^{0} \zeta^{0} {\hat{m}}_{Z} {\hat{m}}_{V^{0}}\\
-c_H^{0} \zeta^{0} {\hat{m}}_{Z} {\hat{m}}_{V^{0}} &  {\hat{m}}_{V^{0}}^{2}
\ery
\right)\,,\;\;\;\;\;
{\textrm{where}}\;\;
\left\{
\bry{l}
\displaystyle
{\hat{m}}_{Z} = \f{g \hat{v}}{2 \cos \theta_W}\vspace{1mm}\\
\displaystyle
{\hat{m}}_{V^{0}}^{2}=m_{\mathcal{V}^{0}}^{2}+ g_{V}^{2} c_{VVHH}^{0} \hat{v}^{2}\vspace{1mm}\\
\displaystyle
\zeta^{0}=\frac{g_{V} \hat{v}}{{2\hat{m}}_{V^{0}}}
\ery
\right. \,,
\ee
and where $\theta_W$ is the Weinberg angle.  These matrices can be diagonalised by two different rotations with mixing angles
\begin{equation}
\bry{lll}
\label{eq: mixing angle}
\dst \tan2\theta^+ =&\, -2\dfrac{c_H^+\zeta^+ \hat{m}_{W^+}\hat{m}_{V^+}}{\hat{m}_{V^+}^2-\hat{m}_{W^+}^2}\,,\\
\dst \tan2\theta^0 =&\, -2\dfrac{c_H^0\zeta^0 \hat{m}_Z\hat{m}_{V^0}}{\hat{m}_{V^0}^2-\hat{m}_Z^2}\,,
\ery
\end{equation}
and the physical masses are given by
\begin{equation}
\label{eq: physical masses}
\bry{lll}
    \dst m_{W^+}^{2} & = \dfrac{1}{2} \dst\left(\hat{m}_{V^+}^2 + \hat{m}_{W^+}^2 - 
    \sqrt{(\hat{m}_{V^+}^2 - \hat{m}_{W^+}^2)^2 + \frac{16 g_V^2 c_H^{+ \, 2} \hat{m}_{W^+}^4}{g^2}} \right) \,, \vspace{2mm} \\
    \dst m_{V^+}^{2} & = \dfrac{1}{2} \dst\left(\hat{m}_{V^+}^2 + \hat{m}_{W^+}^2 + 
    \sqrt{(\hat{m}_{V^+}^2 - \hat{m}_{W^+}^2)^2 + \frac{16 g_V^2 c_H^{+ \, 2} \hat{m}_{W^+}^4}{g^2}} \right) \,, \vspace{2mm} \\
    \dst m_Z^{2} & = \dfrac{1}{2} \dst\left(\hat{m}_{V^0}^2 + \hat{m}_{Z}^2 - 
    \sqrt{(\hat{m}_{V^0}^2 - \hat{m}_{Z}^2)^2 + \frac{4 g_V^2 c_H^{0 \, 2} \hat{m}_{Z}^2 \hat{m}_{W^+}^2 }{ g^2}} \right) \,, \vspace{2mm} \\
    \dst m_{V^0}^{2} & = \dfrac{1}{2} \dst\left(\hat{m}_{V^0}^2 + \hat{m}_{Z}^2 + 
    \sqrt{(\hat{m}_{V^0}^2 - \hat{m}_{Z}^2)^2 + \frac{4 g_V^2 c_H^{0 \, 2} \hat{m}_{Z}^2 \hat{m}_{W^+}^2 }{ g^2}} \right) \,.
\ery
\end{equation}
In the following sections we will make plots and express quantities in terms of these physical masses, to ease the comparison with experimental results (unless otherwise specified). 

It was pointed out in Ref.~\cite{Pappadopulo:2014tg} that for the physical $W$ and $Z$ masses to be reproduced without unnatural cancellations in the determinant of the mass matrices, it is necessary to assume that the new vectors are parametrically heavier than the EW scale, i.e., with masses in the hundreds of GeV or TeV region.\footnote{The exact lower limit on the new heavy vector mass is model-dependent, as when \cref{eq: physical masses} is matched onto a given explicit model, the coupling $g_V$ will determine the scale at which $m_W$ and $m_Z$ are reproducible. For some strongly coupled models, this can be at masses of several TeV. However, generally speaking, this mass is typically low enough to already be ruled out by experiment.} Assuming $\hat{m}_{V^+},\hat{m}_{V^0}\gg \hat{m}_{W^+},\hat{m}_Z$, we find the physical masses to approximately be
\begin{equation}
\label{eq: approx physical masses}
\bry{ll}
    \dst m_{W^+,Z}^{2} & = \hat{m}_{W^+,Z}^2 \left[ 1 + \mathcal{O}\( \dfrac{\hat{m}_{W^+}^2}{\hat{m}_{V^+,V^0}^2} \) \right]\,, \vspace{2mm} \\
    \dst m_{V^+,V^0}^{2} & = \hat{m}_{V^+,V^0}^2 \left[ 1 + \mathcal{O}\( \dfrac{\hat{m}_{W^+}^4}{\hat{m}_{V^+,V^0}^4} \) \right]  \,.
\ery
\end{equation}
We see that the physical masses of the SM gauge bosons are changed from their SM values at order $\hat{m}_{W^+}^2/\hat{m}_V^2$ while the heavy vectors are changed only at order $\hat{m}_{W^+}^4/\hat{m}_V^4$.

In this work we generally consider the charged and neutral vectors to be independent.  However, they could be the components of an $SU(2)_{R}$ triplet. In this case, the mass splitting between the charged and neutral vectors is proportional to $g'^2/g_{V}^2$, so the mass splitting is suppressed if $g'^2 \ll g_V^2$.  This is a remnant of the fact that the hypercharge gauge coupling is the only source of $SU(2)_{R}$ breaking. The coupling $g_V$ will be the same in the charged and neutral sector and some of the $c$ parameters will be related, as we will see in \cref{subsec:4.2,subsec:4.3}. 

%-------------------------------------------------
\subsubsection{The Fermi Constant}
%-------------------------------------------------

The heavy vector singlets also contribute to the Fermi constant. Following mass diagonalisation, the Lagrangian of \cref{sml} has couplings of the $W$ and the $Z$ bosons to the fermion currents defined in \cref{eq:fcurrents}, 
\be
\bry{lll}\label{eq: W/Z suppressed L}
 \mathcal{L}_{\mathcal{V}^{+}} &\supset -\dfrac{g_V c_q^+}{\sqrt{2}}\sin\theta^+ W_\mu^+ J_q^\mu, \vspace{2mm} \\
 \mathcal{L}_{\mathcal{V}^{0}} &\supset -\dfrac{g_Vc_\Psi^0}{2}\sin\theta^0 Z_\mu J_\Psi^\mu.
\ery
\ee
While in the SM the Fermi constant is equal for leptons and quarks, this is no longer the case after mixing with the vector singlets. Upon integrating out the heavy vector singlets, the independent HVS couplings to left- and right-handed SM fermions contribute to several four-fermion operators. Instead of one Fermi constant as in the SM, the HVS model has many (in general, different) Fermi constants between left- and right-handed currents of leptons and quarks \cite{Dawson:2024ozw}. 

The best experimental measurement of these four-fermion operators is obtained from measurements of the muon lifetime \cite{MuLan:2010shf,MuLan:2012sih} which precisely determines the four lepton Fermi constant, $G_F^{\ell\ell}$.  The corresponding four-fermion operator comes from integrating the $W$ boson out of the interaction between two left-handed SM lepton currents. In the HVS model, this Fermi constant is given by
\begin{equation} \label{eq:Gf}
    G_F^{\ell \ell} = \dfrac{1}{\sqrt{2} \hat v^2} \frac{1}{1 - \dfrac{c_H^{+ \,2} g_V^2 \hat v^2}{\hat{m}_{V^{+}}^{2}}} \,.
\end{equation}

%-------------------------------------------------
\subsubsection{Using the Physical Masses and the Fermi Constant as Input Parameters}
%-------------------------------------------------

Using \cref{eq: physical masses,eq:Gf} we take the physical heavy vector masses $m_{V^{0}}$, $m_{V^{+}}$, the $Z$ boson mass $m_Z$ (which is better experimentally determined than the $W$ mass), the Fermi constant, $G_F^{\ell \ell}$, and the fine structure constant as input parameters. We use these inputs to iteratively rewrite 
$\hat m_{V^{+}}$ in terms of $m_{V^{+}}$, 
$\hat v$ in terms of $m_Z$, 
$\hat m_{V^{0}}$ in terms of $m_{V^{0}}$ and 
the SM coupling $g$ in terms of the Fermi constant $G_F^{\ell \ell}$.
In this parameter inversion we work to a precision of order $\mathcal{O}(g^2 g_V^2\hat{v}^4/\hat{m}_{V^{0,+}}^4)$. We can then express all quantities, such as partial widths and cross-sections, in terms of these five physical input parameters. Constraints imposed by this procedure are detailed in \cref{Appendix:InputParameters}.

We provide a \texttt{FeynRules} \cite{Alloul:2013vc,Christensen:2008py} and corresponding \texttt{MadGraph} \cite{Alwall:2011fk} UFO model with the Simplified Model Lagrangian in \cref{sml} implemented in the mass eigenstate basis and in the unitary gauge. The model files are registered in the HEPMDB model database \cite{hepmdb} with the unique identifier hepmdb:0724.0349 and are available to download \href{https://hepmdb.soton.ac.uk/hepmdb:0724.0349}{here}.

%-------------------------------------------------
\subsection{Collider Production and Decay}\label{sec: production and decay}
%-------------------------------------------------

In this subsection we discuss single production and decay of a heavy vector singlet at the LHC and future proton-proton colliders in the narrow width approximation. 

%-------------------------------------------------
\subsubsection{Production Cross-Section}
\label{sec:production-cross-section}
%-------------------------------------------------

The main production mechanisms relevant for vector resonance searches are Drell-Yan (DY) and vector boson fusion (VBF).  In this work we focus on DY production.  VBF is subdominant to DY in large regions of the parameter space, so we do not consider it here.

In the narrow width approximation for $s$-channel processes, 2-to-2 cross-sections can be factorised into a production cross-section times a branching ratio. We can then express the production cross-section in terms of the partial widths $\Gamma_{V\rightarrow ij}$ of the decay process $V\rightarrow ij$~\cite{ParticleDataGroup:2024pth},
\begin{equation}
\label{eq:Production cross-section}
    \sigma(pp \rightarrow V) = \sum_{i,j\in p} \dfrac{\Gamma_{V\rightarrow ij}}{m_V}\dfrac{16\pi^2(2J+1)}{(2S_i+1)(2S_j+1)}\dfrac{C}{C_iC_j}\dfrac{dL_{ij}}{d\hat{s}}\Bigg|_{\hat{s}=m_V^2}.
\end{equation}
The sum is performed over the relevant colliding partons in the two protons, $i,j=\{q,\bar{q}\}$, and $dL_{ij}/d\hat{s}|_{\hat{s}=m_V^2}$ is the corresponding parton luminosity evaluated at the resonance mass. The factors $J$ and $S_{i,j}$ are the spins of the resonance and the initial states, respectively ($J=1$ and $S_{i,j} = 1/2$ for quarks), and $C$ and $C_{i,j}$ are their colour factors ($C=1$ and $C_{i,j} = 3$ for quarks). Note that the only factors in \cref{eq:Production cross-section} that depend on the simplified model parameters are the partial widths and the physical masses of the heavy vectors.  
As  we only consider DY processes in this work, for production we only need the di-quark partial widths.  

Before electroweak symmetry breaking the charged vector $\mathcal{V}^+$ only couples to $\bar{U}^i\gamma^\mu D^i$.  After electroweak symmetry breaking the charged vector $\hat{V}^+$ mixes with $\hat{W}^+$, so will pick up additional couplings which will be mixing angle suppressed.  There will likewise be a small deviation in the coupling to $\bar{U}^i\gamma^\mu D^i$.  While we keep these mixing effects in our numerical work, if we neglect them then the partial width for the charged vector to decay into light quarks $q$ and $\bar q'$ is approximately
\begin{align}
\label{eq:charged-di-quark-widths}
\Gamma_{V^+\rightarrow q \bar q'} &\simeq C_i|V_{qq'}^R|^2\( g_V c_q^+ \)^2 \dfrac{m_{V^+}}{48\pi}\,,
\end{align}
where $C_i = 3$ is the colour factor of a quark. In this discussion we will only consider the dominant production channels of the charged vector, coming from $u\bar d$ and $d\bar u$.

Similarly, the neutral vector $\hat{V}^0$ mixes with the $\hat{Z}$ boson, with a subsequent impact on the couplings on EWSB.  While we keep these mixing effects in our numerical work, if we neglect them then the partial widths of the neutral vector to light quarks are approximately
\begin{align}
\Gamma_{V^0\rightarrow u \bar u} &\simeq  C_i g_V^2 \( c_Q^{0\,2} + c_U^{0\,2} \) \dfrac{m_{V^0}}{96\pi}, \vspace{2mm} \\
\Gamma_{V^0\rightarrow d \bar d} &\simeq  C_i g_V^2 \( c_Q^{0\,2} + c_D^{0\,2} \) \dfrac{m_{V^0}}{96\pi}.
\end{align}

\begin{figure}[tp]
    \centering
    \includegraphics[]{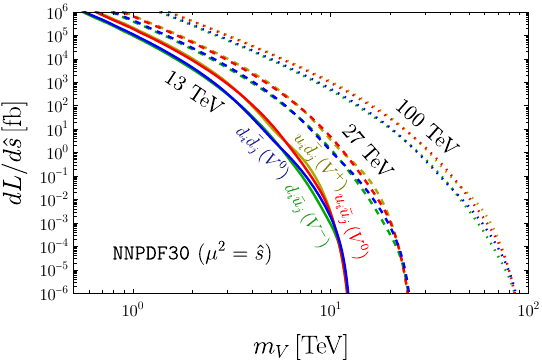}
    \caption{\small Values of the quark-antiquark parton luminosities for charged (yellow, green) and neutral (blue, red) initial states at the 13\,TeV LHC (solid), as well as for future 27\,TeV (dashed) and 100\,TeV (dotted) $pp$ colliders.}
    \label{fig:Parton luminosities}
\end{figure}

For the parton luminosity functions we used the \texttt{Mathematica} package \texttt{NNPDF} \cite{Hartland:2012ia} to generate universal fitted functions for the parton luminosities to leading order, with $\alpha_S = 0.118$ at the $Z$-mass scale. The differential quark-antiquark parton luminosities are shown in \cref{fig:Parton luminosities} as a function of the resonance mass $m_V$ (where $m_V$ stands for $m_{V^+}$ or $m_{V^0}$, as appropriate), for the 13\,TeV LHC (solid lines) as well as for future 27\,TeV (dashed) and 100\,TeV (dotted) colliders.  We see that the LHC luminosity drops steadily from around 2.5\,TeV to around 10\,TeV, where it starts to drop sharply.  At masses around 7.5\,TeV a 27\,TeV collider would have a luminosity almost three orders of magnitude larger than the LHC, and a 100\,TeV collider would have a luminosity almost five orders of magnitude larger.  Higher energy colliders also only suffer a sharp drop in luminosity at higher energies, when the resonance mass nears the beam centre of mass energy.  Note that the parton luminosity functions have an increasing uncertainty as the resonance mass approaches the centre of mass energy.  This does not lead to a large uncertainty in the LHC limits we set at $m_V \lesssim 5\,\text{TeV}$ but would become more important at masses that are a larger fraction of the centre of mass energy.

We do not investigate the role of VBF in this work, as it is heavily suppressed with respect to DY in most regions of the parameter space. This is mostly due to suppression from insertions of the fine structure constant in the vector boson PDFs.  While there are narrow regions of parameter space where DY is subdominant to VBF (like those discussed for the HVT in Refs.~\cite{Pappadopulo:2014tg,Baker:2022zxv}), we leave this to future work.

\begin{figure}[tp]
    \centering
    \includegraphics[width=0.49\textwidth]{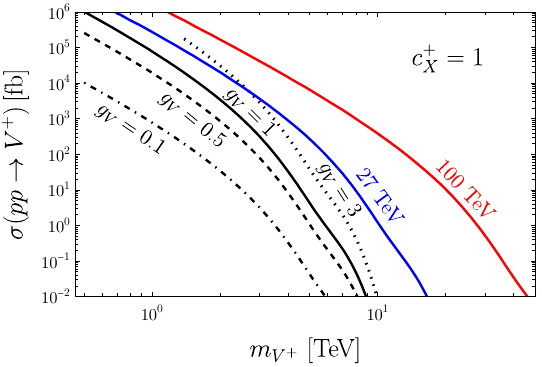}
    \includegraphics[width=0.49\textwidth]{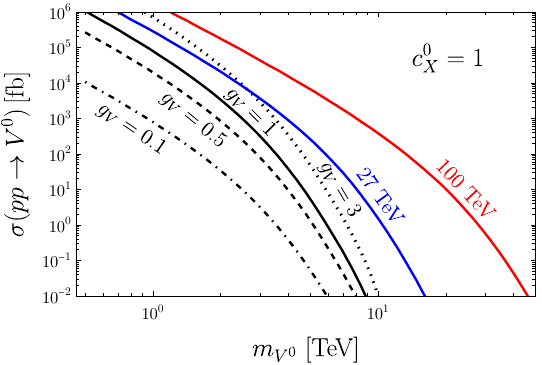}
    \caption{\small LHC production cross-sections for the heavy charged (left) and neutral (right) vectors as a function of the resonance mass, for a model with universal couplings $c_X^{+,0}=1$. The black lines correspond to different couplings $g_V$ for the values $g_V=0.1$ (dot-dashed), $0.5$ (dashed), $1$ (solid), and $3$ (dotted). The blue and red curves show the production cross-section for a 27\,TeV and 100\,TeV collider, respectively, for $g_V = 1$.}
    \label{fig:cross-sections}
\end{figure}

In \cref{fig:cross-sections} we show the LHC and future collider production cross-sections for the charged and neutral vector singlets, using \cref{eq:Production cross-section} for a benchmark case of universal couplings ($c_X^+=c_X^0=1$ for all $X$). We see that for both the charged and neutral vectors, the high-luminosity run of the LHC with $3$\,ab$^{-1}$ could produce tens of heavy vectors at 8\,TeV for $g_V \gtrsim 0.5$. At the same mass, this roughly extends to tens of thousands for a 27\,TeV collider such as the HE-LHC with $15$\,ab$^{-1}$, and tens of millions for a 100\,TeV collider such as the FCC-hh with $20$\,ab$^{-1}$. For this benchmark ($c_X = 1$), the production cross-section is always proportional to $g_V^2$, and the stronger this coupling, the larger the cross-section. In general, however, the relationship between $g_V$ and the cross-section will depend on the $c$ parameters in the given UV models.  We will see an example of a non-trivial relationship between $g_V$ and the cross-section in \cref{subsec:4.2,subsec:4.3}.

%-------------------------------------------------
\subsubsection{Narrow Width Approximation and Finite Width Effects}
%-------------------------------------------------

When using experimental data to set limits on the simplified model parameter space, the narrow width approximation (NWA) is very useful as it separates resonance production and decay.  However, this factorisation does not hold away from the peak in the invariant mass distribution and in an analysis using the NWA care should be taken not to include the tails.  Here we briefly review two well-known finite width effects \cite{Accomando:2011up,Choudhury:2011cg,Accomando:2013ve, Pappadopulo:2014tg} which can spoil the factorisation: the energy dependence of the pdfs and interference with the SM. 

Firstly, the factorisation of the total differential cross-section assumes that the parton luminosities are fairly constant within the peak region. Generally this means that the narrower the resonance, the better the agreement. Furthermore, it requires the resonance mass to be significantly below the kinematical production threshold of the collider, as the parton luminosities drop dramatically near this threshold, as can be seen in Fig.~\ref{fig:Parton luminosities}. In Ref.~\cite{Pappadopulo:2014tg} it was shown that this finite width effect remains small as long as $\Gamma/m_V \lesssim 0.1$ and an invariant mass interval no bigger than $[m_V - \Gamma, m_V + \Gamma]$ is considered in the experimental analyses.

Secondly, Feynman diagrams with virtual heavy vectors that share the same initial and final states as SM backgrounds will interfere at the amplitude level. While these interference effects can be sizeable, Ref.~\cite{Pappadopulo:2014tg} shows that by focusing on the peak region $[m_V-\Gamma, m_V+\Gamma]$ and treating the parton luminosities as constant within that window, the interference contribution to the signal is an odd function around the resonance peak and so cancels when integrated over a symmetric interval. In the peak region, the relative deviation between the total signal plus background (including interference effects) and the Breit-Wigner signal approximation plus background (excluding interference effects) is typically less than $10\%$.

\begin{figure}[tp]
    \centering
    \includegraphics[]{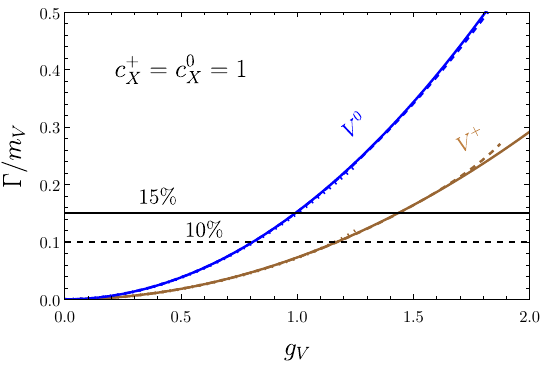}
    \caption{\small Total width over mass of the charged (brown) and neutral (blue) vectors as a function of $g_V$ for universal couplings and in the limit of large vector masses. The dotted and dashed lines show vector masses of $500\,$GeV and $750\,$GeV, respectively.
    }
    \label{fig:Total widths}
\end{figure}

In \cref{fig:Total widths} we show $\Gamma/m_V$ for the charged and neutral vectors for universal couplings as a function of $g_V$, in the limit of large vector masses. We see that for universal couplings the NWA requires $g_V\lesssim 1$. For heavy vector singlet masses near the electroweak scale, there is dependence on the masses due to mixing at the level of $\mathcal{O}(m_W^2/m_V^2)$. While \cref{fig:Total widths} shows that these mixing effects are fairly small for universal couplings, the mass dependence can be larger as we will see, e.g., in the strongly coupled model we consider in \cref{subsec:4.3} (Model E).

%-------------------------------------------------
\subsubsection{Partial Widths and Branching Ratios}
%-------------------------------------------------

We have given some partial widths for di-quark decays of the charged and neutral vectors in \cref{sec:production-cross-section}, since these enter the LHC production cross-sections.  In this section we present approximate expressions for the remaining partial widths, discuss the branching ratios and compare the approximate expressions to the full expressions.  

The charged vector decays into di-quarks and, as previously mentioned, acquires a (suppressed) coupling to leptons from mixing with the $\hat W$ boson.  This suppression is typically so large that di-lepton decays of charged heavy vector singlets are not relevant at colliders and can be neglected.  The charged vector also decays into di-boson final states, which we will discuss below.  The partial widths to $u \bar d$ and $c \bar s$ were given in \cref{eq:charged-di-quark-widths}, which we write here again for completeness, and they differ only by a right-handed mixing-matrix factor,
\begin{align}
\label{eq: Vc di-quark partial widths}
\Gamma_{V^+\rightarrow q \bar q'} &\simeq C_i|V_{qq'}^R|^2\( g_V c_q^+ \)^2 \dfrac{m_{V^+}}{48\pi}\,.
\end{align}
The partial width to $t \bar b$, keeping effects due to the top quark mass (but neglecting mixing effects), is approximately 
\begin{align}
\label{eq: Vc di-quark partial widths-tb}
    \Gamma_{V^{+}\to t \bar b} &\simeq\, C_i |V_{tb}^R|^2 \(g_{V} c_{q3}^{+}\)^{2}\frac{ m_{V^{+}}}{48 \pi }\left(1 - \f{3 m_{t}^{2}}{2 m_{V^{+}}^{2}} + \f{m_{t}^{6}}{2 m_{V^{+}}^{6}}\right)\,.
\end{align}
Even at 1\,TeV, effects from the top mass are negligible, and (for universal parameters) this agrees with \cref{eq:charged-di-quark-widths} to within a percent.

The neutral vector decays into di-quarks, di-leptons and di-bosons.  We will start with the fermion final states before discussing the di-boson final states.
After mixing with the $\hat{Z}$ boson, the neutral vector couplings to fermions consist of two contributions: the first comes directly from the Lagrangian, \cref{sml}, with an additional suppression factor of $\cos\theta^{0}$, while the second comes from the corresponding SM coupling of the $\hat{Z}$ boson suppressed by $\sin\theta^{0}$. For small $\theta^0$, these mixing effects are small.  If they are neglected, the partial widths can be written as
\begin{align}
\label{eq:di-lep di-top partial widths-qq}
\Gamma_{V^0\rightarrow u \bar u} &\simeq C_ig_V^2 \( c_Q^{0\,2} + c_U^{0\,2} \) \frac{m_{V^0}}{96\pi}, \qquad
\Gamma_{V^0\rightarrow d \bar d} \simeq C_ig_V^2 \( c_Q^{0\,2} + c_D^{0\,2} \) \frac{m_{V^0}}{96\pi}\,,\vspace{2mm}\\
\label{eq:di-lep di-top partial widths-ll}
\dst \Gamma_{V^{0}\to \ell \bar \ell}
&\simeq g_{V}^{2}\left(c_L^{0\,2}+c_{E}^{0\,2}\right)\frac{m_{V^{0}}}{96\pi}\,, \hspace{1.08cm}
\dst \Gamma_{V^{0}\to \nu \bar \nu}
\simeq \left(g_{V} c_L^{0}\right)^{2}\frac{m_{V^{0}}}{96\pi}\,,\vspace{2mm}\\
\label{eq:di-lep di-top partial widths-tt}
\dst \Gamma_{V^{0}\to t \bar t}
&\simeq  C_i g_{V}^{2} \frac{m_{V^{0}}}{96\pi }\(\left(c_{Q3}^{0\,2}+c_{U3}^{0\,2}\right)-\frac{m_{t}^{2}}{m_{V^{0}}^{2}} \left(c_{Q3}^{0\,2}-6 c_{Q3}^{0}c_{U3}^{0}+c_{U3}^{0\,2}\right)\)\sqrt{1-\f{4m_{t}^{2}}{m_{V^{0}}^{2}}}\,,
\end{align}
where we have kept mass corrections from the top quark. While it is useful to have these approximate expressions, in the numerical analysis in the following sections we retain the full mixing effects.

Note that while the widths of the charged vector only have contributions from right-handed fermions, the widths of the neutral vector receive contributions from both chiralities (we see that left- and right-handed $c$ parameters contribute to the $\ell \bar \ell$ and di-quark widths).  This means that unless polarised final states are studied, only combinations of the couplings enter measurements of the neutral vector.  The combinations $c_Q^{0\,2}+c_{U}^{0\,2}$ and $c_Q^{0\,2}+c_{D}^{0\,2}$ are relevant for production and for decay into di-quark final states, while  $c_L^{0}$ and $c_L^{0\,2}+c_{E}^{0\,2}$ are relevant for the decays into leptonic final states. For the neutral vector, a combination of different analyses would be required to disentangle the different parameters.

Comparing these fermionic widths of the HVS with those of the neutral and charged components of the HVT, we see that the widths of the charged vector into quarks are completely analogous.  The fermionic widths of the neutral vector singlet are larger for universal $c$ parameters, because the $SU(2)_{L}$ triplet only couples to left-handed fermions while the neutral singlet couples to both left and right-handed ones. 

We now turn to approximate expressions for the di-boson partial widths.  Since the heavy vectors couple mostly to longitudinal SM gauge bosons, we can use the Goldstone Boson Equivalence Theorem~\cite{Chanowitz:1985hj}.  In the approximate widths we also neglect the effect of gauge couplings $g$ and $g'$ (by taking $g=g'=0$) and include only the linear operators relevant for the decay.   While we derived the impact of electroweak symmetry breaking in the unitary gauge, to use the Goldstone Boson Equivalence Theorem we now work in the \textit{equivalent gauge} \cite{Wulzer:2013mza}.  We then write the SM Higgs doublet $H$ in terms of the physical Higgs boson $h$ and the Goldstone bosons $\pi_i$,
\begin{align}
\label{eq: Higgs Goldstone doublet}
    H = \begin{pmatrix}
    \displaystyle{\frac{\pi_2 + i\pi_1}{\sqrt{2}}} \\
    \displaystyle{\frac{\hat{v} + h -i\pi_3}{\sqrt{2}}}
    \end{pmatrix} \equiv \begin{pmatrix}
    i\pi^+ \\
    \displaystyle{\frac{\hat{v} + h -i\pi^0}{\sqrt{2}}}
    \end{pmatrix}.
\end{align}
Using the Goldstone Boson Equivalence Theorem, the longitudinal $W$ and $Z$ bosons are described by $\pi^\pm$ and $\pi^0$, respectively, in the high-energy limit. The Lagrangian in \cref{sml} can then be re-expressed in terms of these fields. In this limit a mixing of the form $V_{\mu}\demua \pi$ arises between the heavy vector singlets and the Standard Model Goldstone bosons, for both the charged and neutral singlets.  This mixing can be eliminated by a suitable shift of the vector fields
 \be
 \bry{lll}\label{eq: Field redefinitions}
     V_\mu^+ \rightarrow V_\mu^+ - \displaystyle{\frac{c_H^{+} \zeta^+}{m_{V^+}}}\partial_\mu\pi^+, \qquad V_\mu^{0} \rightarrow V_\mu^{0} - \displaystyle{\frac{c_H^{0} \zeta^0}{m_{V^{0}}}}\partial_\mu\pi^{0}\,,
 \ery
 \ee
followed by canonically normalising the Goldstone boson fields 
\be
\bry{lll}\label{eq: Rescaled Goldstones}
     \pi^+ \rightarrow \displaystyle{\frac{1}{\sqrt{1-{c_H^{+}}^2\zeta^{+\,2}}}}\pi^+, \qquad \pi^{0} \rightarrow \displaystyle{\frac{1}{\sqrt{1-{c_H^{0}}^2{\zeta^0}^2}}}\pi^{0}\,.
\ery
\ee
This leads to an $\mathcal{O}((\zeta^+)^2, (\zeta^0)^2)$ deviation in the partial widths, for the charged and neutral vectors, respectively, which we can neglect (recall that $\zeta^+ = g_V \hat v/\hat{m}_{V^+}$ and $\zeta^0 = g_V \hat v/2\hat{m}_{V^0}$).  For $c_X \lesssim 1$, $g_V \lesssim 1$ and $m_{W^+, Z} \ll m_{V^{+,0}}$, the partial decay widths into di-boson final states are then approximately given by
\be\label{dibosonswidths2}
\bry{lll}
\dst \Gamma_{V^+\to W^+_{L} Z_{L}} & \simeq 
\dfrac{(g_V{c_H^+})^2m_{V^+}}{48\pi}\,, \qquad
\dst \Gamma_{V^+\to W^+_Lh} & 
\simeq 
\dfrac{(g_V{c_H^+})^2m_{V^+}}{48\pi}, \vspace{1mm} \\
\dst \Gamma_{V^{0}\to W^+_LW^-_L} & \simeq
\dfrac{ (g_{V} c_{H}^0)^2 m_{V^{0}}}{192\pi}\,,\qquad
\dst \Gamma_{V^{0}\to Z_Lh} & \simeq
\dfrac{(g_{V} c_{H}^{0})^2 m_{V^{0}}}{192\pi }\, .
\ery
\ee
Compared to the di-boson widths of the HVT, the neutral singlet widths are identical while the charged singlet widths appear larger by a factor of four.  However, this factor comes from the definitions of the couplings in the Lagrangians as the charged component of the $SU(2)_L$ triplet contains an extra factor of $1/2$ from the $SU(2)_L$ generators.

\begin{figure}[tp]
    \centering
    \includegraphics[width=0.49\textwidth]{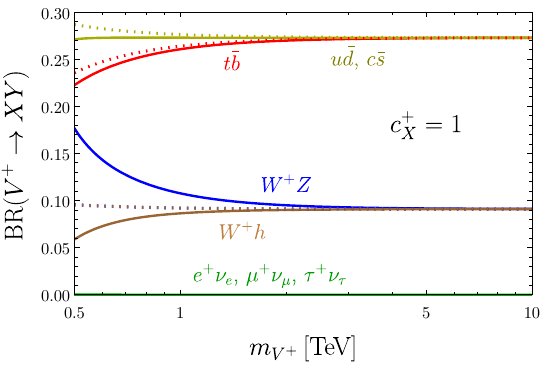}
    \includegraphics[width=0.49\textwidth]{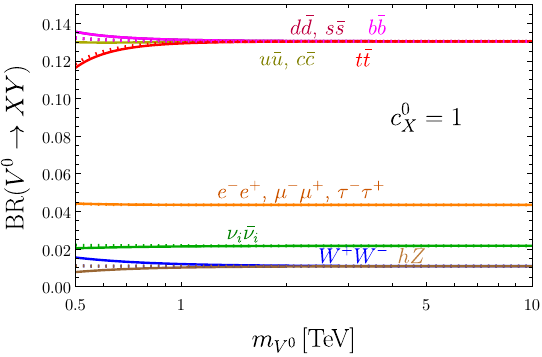}
    \caption{\small Branching ratios for the two body decays of the charged vector $V^{+}$ (left) and the neutral vector $V^0$ (right), for the case of universal couplings $c_X^{+}= c_X^0=1$. Here $g_V=1$, although in any case, the BRs depend only very weakly on $g_V$. The solid lines correspond to the full numerical expressions, while the dotted lines show the approximate expressions. In the left (right) panels, the neutral (charged) vector is decoupled.}
    \label{fig:BRs-universal}
\end{figure}

We have now given approximate partial widths for all decay channels of the heavy vector singlets.  We show the branching ratios of the charged and neutral heavy vector singlets as a function of their masses, \cref{fig:BRs-universal} (left) and (right), respectively, for universal couplings ($c_X^+ = c_X^0 = 1$ for all $X$).  The solid lines correspond to the full numerical expressions, which we obtain from our \texttt{FeynRules} implementation including the effects of electroweak symmetry breaking, the parameter inversion and which we use in the limits we set below, while the dotted lines correspond to the approximate expressions given above. Although universal couplings do not correspond to any of the UV models we consider in \cref{sec:explicit-models}, it is useful to discuss the branching ratios in the simplified model without referencing an explicit UV completion.  In \cref{sec:explicit-models} we show the branching ratios in two of the explicit UV models we explore. 

We see from the left panel of \cref{fig:BRs-universal} that for both the approximate and exact expressions, the charged vector dominantly decays into di-quarks, with di-boson decays being the only other relevant channel. The remaining channels into leptons are mixing-suppressed. Corrections from the top quark mass in \cref{eq: Vc di-quark partial widths-tb} lead to a suppression in the $t \bar b$ channel at masses below $1\,$TeV.  Comparing the exact and approximate branching ratios, we see that mixing effects are important when $m_{V^+} \lesssim 2\,\text{TeV}$, particularly for the di-boson channels.

In the right panel of \cref{fig:BRs-universal}, we show the branching ratios of $V^0$.  Again it dominantly decays into di-quarks, and top quark mass effects are important when $m_{V^0} \lesssim 1 \,\text{TeV}$.  The branching ratio into charged leptons is a factor of three smaller than the di-quark branching ratios due to a colour factor. Since the decay width into charged leptons receives a contribution from left-handed and right-handed fermion couplings, it is larger than the decay into (purely left-handed) neutrinos. The di-boson decays are even smaller.  We can see that mixing effects have a smaller but non-negligible impact when $m_{V^0} \lesssim 1\,\text{TeV}$.

In both panels of \cref{fig:BRs-universal} we show the branching ratios for $g_V = 1$.  We can see from our approximate expressions that all partial widths are proportional to $g_V^2$ when mixing effects are neglected, and this holds up to $\mathcal{O}(m_W^4/m_V^4)$.  The branching ratios are then independent of $g_V$ up to this order.  Although we saw in \cref{fig:Total widths} that the total widths are proportional to $g_V^2$ to high accuracy for masses below 1\,TeV, this is due to an accidental cancellation and the partial widths can have a different dependence on $g_V$ for these masses.  Since non-trivial $g_V$ dependence enters through mixing effects, the difference between the exact and approximate expressions can give us an idea of when the partial widths have a significant dependence on $g_V$.  For the charged vector, for example, the exact and approximate branching ratios agree within 5\% for $m_{V^+} \gtrsim 2\,\text{TeV}$ for $g_V=1$ and universal couplings, so we would expect dependence on $g_V$ for $m_{V^+} \lesssim 2\,\text{TeV}$. We will see later that this mass differs with other coupling choices. In the more realistic benchmark models we consider in \cref{sec:explicit-models}, the BRs for the weakly (strongly) coupled Model D (E) agree within 5\% for a neutral vector mass $m_{V^0} \gtrsim 1.4 \,(4) \, \text{TeV}$ at $g_V = 1\,(3)$. The effects of electroweak symmetry breaking can therefore be relevant up to TeV scale masses, and particularly so in strongly coupled models. It is thus important to consider these mixing effects in the context of LHC phenomenology.

Overall, the picture for LHC production and decay is similar to the HVT case.  The main production channel is Drell-Yan, with production cross-sections proportional to the combinations $g_{V}^{2}(c_Q^{0\,2}+c_{U}^{0\,2})$, $g_{V}^{2}(c_Q^{0\,2}+c_{D}^{0\,2})$, and $(g_{V} c_{q}^{+})^{2}$ for the $uu$, $dd$ and $u\bar{d}$ ($d\bar{u}$) partonic initial states, respectively. In the universal coupling scenario, the main decay channels are di-quark decays, while di-boson and di-lepton decays are reduced but not negligible.  The same limitations of the narrow width approximation are also present, and experimental analyses should take care to only consider the peak region (and not the tails).
There are, however, three main differences to the HVT model: first, the charged vector singlet has a very suppressed coupling to SM leptons compared to the charged component of the HVT; second, the vector singlets typically have enhanced couplings to fermions compared to the vector triplets;\footnote{The HVT coupling to fermions is typically generated through mixing with the $W$ boson and is therefore proportional to $g^2/g_V$. This is not the case for singlets where the coupling is proportional to $g_V$. This is reflected in our choice of normalisation in the simplified model Lagrangian.} and third, the neutral singlet couples to both left- and right-handed quarks and leptons, introducing more free parameters than in the triplet model.

%-------------------------------------------------
\subsection{Electroweak Precision Tests}
\label{EWPTs}
%-------------------------------------------------

While our main focus in the paper is on direct searches at collider experiments, constraints can also be placed on the vector singlets via indirect measurements, such as electroweak precision tests. In this section we describe the impact of heavy vector singlets on the electroweak $S$, $T$, $U$, $W$ and $Y$ parameters. 

In order to compute the $S$, $T$, $U$, $W$ and $Y$ parameters, we follow the approach of Ref.~\cite{Cacciapaglia:2006pk} (see also Ref.~\cite{Pappadopulo:2014tg}) where the heavy vectors are integrated out after certain field redefinitions.  
This technique is based on the assumptions that the strongest constraints come from (i) the oblique parameters and (ii) the couplings to leptons. For the neutral vector this approach is fully motivated~\cite{Cacciapaglia:2006pk}. For the charged singlet, which does not directly couple to SM leptons, the main constraints come from the $T$ parameter and $\Delta F=2$ processes~\cite{Grojean:2011vu, Langacker:1989xa}.  Taking $V^R$ to be the identity, where the $\Delta F=2$ constraints are the weakest, we find that both electroweak and $\Delta F=2$ constraints are far weaker than direct searches in all the cases we consider. One should however bear in mind that particular corners of the parameter space could require an ad hoc discussion of the constraints from precision measurements and flavour physics.

%---------------------------------------------------------
\begin{table}
\begin{center}
\begin{tabular}{cccccc}
$Y_Q$	&$Y_L$	&$Y_U$		&$Y_D$	&$Y_E$&$Y_H$ \\ \hline\hline
&&&&& \vspace{-4mm}\\ 
$\f{1}{6}$&$-\f{1}{2}$		&$\f{2}{3}$	&$-\f{1}{3}$		&$-1$ 		&$\f{1}{2}$ 	\vspace{1mm}	\\  \hline 
\hline
\end{tabular}
\end{center}
\caption{\small\label{Table:Hypercharges} Hypercharges of the SM fields where we use the convention $Q = T^3 + Y$, where $T^3 = \sigma^3/2$ is the $SU(2)$ charge.}
\end{table}
%---------------------------------------------------------

For the neutral vector the computation is analogous to the one described in Ref.~\cite{Cacciapaglia:2006pk} for non-universal $Z'$ models, with the gauge charges $Z_{i}$ substituted by the corresponding quantities in our notation, i.e., $g_{V} c_{i}^{0}/2$. 
In this case we integrate out the neutral vector after the field redefinition,\footnote{Note that the signs are different to those in Ref.~\cite{Cacciapaglia:2006pk}, due to the different convention for the hypercharge of the right-handed fermions.}
\be
B_{\mu} \to B_{\mu} - \f{g_{V}c_{E}^{0}}{2g' Y_{E}} V^{0}_{\mu}\,,\qquad W^{3}_{\mu}\to W^{3}_{\mu}-\f{g_{V}\(c_{E}^{0}Y_{L}-c_{L}^{0}Y_{E}\)}{g Y_{E}}  V^{0}_{\mu}\,,
\ee
where $Y_{\Psi}$ denotes the hypercharges of the SM $\Psi$ multiplets, given in Table \ref{Table:Hypercharges}.
The charged vector can be integrated out directly, without introducing any mixing with the $W$, since it does not directly couple to charged leptons.
We can treat the neutral and charged vectors independently and sum up their contributions to the precision observables since the interactions between them do not, after integrating out these fields, contribute to the leading order effective Lagrangian. 

We then get the following contributions to the oblique parameters at leading order in $m_W^2/m_{V^{(+,0)}}^2$~\cite{Cacciapaglia:2006pk,Grojean:2011vu}:
\begin{align}
    \label{eq:S}
     \hat{S} \equiv \frac{\alpha_\text{EM}(m_Z)}{4 \sin^2 \theta_W}S =&\, \frac{g_V^2 m_W^2}{2 g^2 g'^ 2m_{V^0}^2} (c_L^0 -c_E^0 - c_H^0 )(g'^2(2 c_L^0 - c_E^0) - g^2c_E^0) \,,
     \\
    \label{eq:T}
     \hat{T} \equiv \alpha_\text{EM}(m_Z) T =&\, \frac{g_V^2 m_W^2}{g^2 m_{V^0}^2}(c_L^0 - c_E^0 - c_H^0)^2 - \frac{g_V^2 (c_H^+)^2m_W^2}{g^2 m_{V^+}^2}\,,
     \\
    \label{eq:U}
     \hat{U} \equiv -\frac{\alpha_\text{EM}(m_Z)}{4 \sin^2 \theta_W} U =&\, \frac{g_V^2 m_W^2}{g^2 m_{V^0}^2}(c_L^0 - c_E^0 - c_H^0)(2 c_L^0 - c_E^0) \,,
     \\
     \label{eq:W}
     W =&\, \frac{g_V^2 m_W^2}{4 g^2 m_{V^0}^2}(2 c_L^0 - c_E^0)^2 \,,
     \\
     \label{eq:Y}
     Y =&\, \frac{g_V^2 m_W^2}{4 g'^2 m_{V^0}^2}(c_E^0)^2 \,,
\end{align}
where $\alpha_\text{EM}(m_Z)$ is the fine-structure constant at the scale of $m_Z$. We see that the oblique parameters are proportional to $g_V^2$, in contrast to the relatively weak $g_V$-dependence seen for the HVT. The EWPTs then lead to stronger constraints on the HVS model, especially when considering strongly coupled scenarios.

To set the constraints from EWPTs on the HVS parameter space, which will be shown in the next sections together with the constraints from collider searches, we use a three-dimensional $\chi^2$ fit for $S$, $T$ and $U$ using the experimental values given in Ref.~\cite{ParticleDataGroup:2024pth}, and, separately, a two-dimensional $\chi^2$ fit for $W$ and $Y$ using Refs.~\cite{CMS:2022krd,Strumia:2022qkt}.\footnote{Since Refs.~\cite{CMS:2022krd,Strumia:2022qkt} do not provide a correlation matrix between $W$ and $Y$ we assume zero correlation.}  We consider two separate fits because there are no current correlation matrices between all five of these parameters. Note that stronger constraints could be placed on the model by performing a complete global fit, or by using extended parameterisations of new physics effects (see, e.g., Ref.~\cite{Cacciapaglia:2006pk}). 

While the $S$, $T$ and $U$ parameters are best measured at lepton colliders, hadron colliders are best suited to determine $W$ and $Y$. Furthermore, the Drell-Yan processes relevant for $W$ and $Y$ grow with the center-of-mass energy of the collider, which makes future hadron colliders particularly sensitive~\cite{Farina:2016rws,Torre:2020aiz}. We will see this explicitly in our discussion in \cref{sec:explicit-models}.
Note that some limits from EWPTs at future colliders, in particular the HL-LHC, are less constraining than at the current LHC. This is unrealistic and is due to the projections being out-of-date.  As such, we see that there is a need for more up-to-date projections.

%=================================================
\section{Current and Future Limits from Collider Searches}
\label{sec:data-bounds}
%=================================================

\begin{table}
    \begin{center}
    {
    \begin{tabular}{l|c|c|c}
    \hline
    \hline
     Channel & Reference & Main background & Extrapolation \\
    \hline 
    $jj$ & \cite{ATLAS:2019fgd, ATLAS:2017eqx} & multi-jet & $\times$ \\
    $t \bar b$ & \cite{ATLAS:2018uca,CMS:2021mux,CMS:2017zod} & multi-jet & $\times$ \\
   $WZ \rightarrow 3 \ell \nu$ & \cite{ATLAS:2018iui} & DY $WZ$ & \checkmark \\
   $WZ \rightarrow jj$ & \cite{ATLAS:2019nat,ATLAS:2016hal,CMS:2022pjv,CMS:2017fgc,CMS:2016rqm} & multi-jet & $\times$ \\
    $WZ \rightarrow \ell\nu jj$ & \cite{ATLAS:2020fry,ATLAS:2017jag,CMS:2021klu,CMS:2018dff} & $W/Z$+jets & $\times$\\
   $WZ \rightarrow \ell\ell jj$ & \cite{ATLAS:2020fry,ATLAS:2017otj,CMS:2021xor,CMS:2018sdh} & $W/Z$+jets & $\times$\\
   $WZ \rightarrow \nu \nu jj$ & \cite{ATLAS:2017otj,CMS:2021itu,CMS:2018ygj} & $W/Z$+jets, $t\bar t$ in certain control regions & $\circ$ \\
   $W\gamma$ & \cite{ATLAS:2018sxj,CMS:2021zxu} & $\gamma$+jet, $\gamma+W$ & $\times$ \\
    $Wh \rightarrow \ell \nu  \bar b b$ & \cite{ATLAS:2017xel,CMS:2021klu} & $t \bar t$ & \checkmark\\
   $Wh \rightarrow jj \bar b b$ & \cite{ATLAS:2017ptz} & multi-jet & $\times$\\
    \hline 
    \hline
    \end{tabular}
    }
   \caption{\small Summary of ATLAS and CMS searches relevant for a charged heavy vector resonance, $V^+$. 
   Explanation of the symbols: \checkmark --
   extrapolation procedure works well, $\times$ -- extrapolation procedure is not appropriate, $\circ$ -- extrapolation procedure can be used with caution.
}
    \label{tab:charged searches}
    \end{center}
\end{table}

The ATLAS and CMS collaborations have performed a significant number of direct searches for heavy resonances decaying to various SM final states. \Cref{tab:charged searches,tab:neutral searches} provide a summary of the searches relevant for decaying charged and neutral vector bosons, respectively.\footnote{For a recent ATLAS combination of the channels $(W/Z)(W/Z) + (W/Z)H + \ell \ell + \ell \nu + \tau \nu$ see \cite{ATLAS:2022jsi}. Note that combinations such as this do not apply to the HVS in a straightforward way since they need to make assumptions about the branching ratios.} The majority of these analyses provide limits on the production cross-section times branching ratio, $\sigma\times$BR, as a function of the resonance mass.  In this section we use these searches to places limits on the simplified model parameter space, first for universal couplings ($c_X^{(0,+)} = 1$ for all $X$) and then letting the couplings vary (where we also include the indirect electroweak constraints).  We then make sensitivity projections for a range of possible future colliders (the HL-LHC, HE-LHC, the FCC and the SPPC).

\begin{table}
    \begin{center}
    {
    \begin{tabular}{l|c|c|c}
    \hline
    \hline
    Channel & Reference & Main background & Extrapolation \\
    \hline 
    $\ell \ell$ & \cite{ATLAS:2019erb,ATLAS:2017fih,CMS:2021ctt,CMS:2018ipm,CMS:2016cfx} & DY $\ell \bar \ell$ & \checkmark \\
    $\tau\tau$ & \cite{ATLAS:2017eiz,CMS:2016xbv} & di-jet, jet + $\tau$ & $\times$ \\
    $jj$ & \cite{ATLAS:2019fgd, ATLAS:2017eqx} & multi-jet & $\times$ \\
    $b\bar{b}$ & \cite{CMS:2022zoc} & multi-jet & $\times$ \\
    $t \bar{t}$ & \cite{ATLAS:2023taw,ATLAS:2020lks,CMS:2017ucf}  & $t \bar t$ & \checkmark \\
    $WW \rightarrow jj$ & \cite{ATLAS:2019nat,ATLAS:2016hal,CMS:2022pjv,CMS:2017fgc,CMS:2016rqm} & multi-jet & $\times$ \\
    $WW \rightarrow \ell\nu jj$ & \cite{ATLAS:2020fry,ATLAS:2017jag,CMS:2021klu,CMS:2018dff} & $W$+jets, $t \bar t$ (50\% in certain signal regions) & $\circ$ \\
    $Zh \rightarrow \ell \ell/\nu\nu \bar b b$ & \cite{ATLAS:2017xel} &  2-$\ell$: $Z+(b \bar b)$ (75\%)$, t \bar t$ (25\%) / 0-$\ell$: $t \bar t$& \checkmark\\
    $Zh \rightarrow jj \bar b b$ & \cite{ATLAS:2017ptz} & multi-jet & $\times$ \\
    \hline 
    \hline
    \end{tabular}
    }
    \caption{\small Summary of ATLAS and CMS searches relevant for a neutral heavy vector resonance, $V^0$.
    }
    \label{tab:neutral searches}
    \end{center}
\end{table}

%-------------------------------------------------
\subsection{Current LHC Limits with Universal Couplings}\label{subsec:universal-couplings-exp}
%-------------------------------------------------

We first consider the simple case where the vector singlets couple universally to the SM particles. That is, we set $c_X^+=c_X^0=1$ for all $X$ and use the most constraining searches to place limits on the heavy vector masses.  We first discuss the charged vector and then the neutral.  We do not consider channels where both the charged and neutral vectors are present, such as di-jet, so they can be studied independently.  We do not include di-jet limits because they are generally less sensitive than di-lepton and di-boson searches.

\begin{figure}[tp]
\begin{center}
    \includegraphics{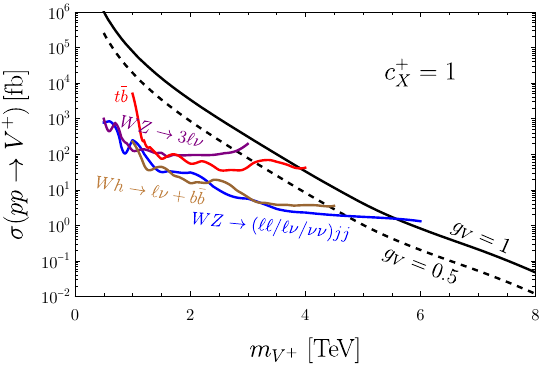}
    \caption{\small Experimental limits on the DY production cross-section of a charged heavy vector singlet $V^+$, for all $c$ couplings set to one. The blue and purple curves correspond to the ATLAS searches of a semi-leptonic di-boson final state \cite{ATLAS:2020fry} and fully-leptonic di-boson final state \cite{ATLAS:2018iui}, respectively. The red curve shows the CMS search corresponding to a $t \bar b$ final state \cite{CMS:2021mux}, and brown shows the CMS resonance search examining the $Wh\rightarrow \ell\nu + b\bar{b}$ channel \cite{CMS:2021klu}. The black curves show the theoretical DY production cross-section for $g_V=0.5$ (dashed) and $1$ (solid). The neutral vector is decoupled, with $m_{V^0}=100\,$TeV.}
    \label{Fig:Exp limits charged}
\end{center}
\end{figure}

In \cref{Fig:Exp limits charged} we take the most stringent limits on $\sigma\times$BR for a charged vector resonance and convert them to limits on the production cross-section of a charged vector singlet, $\sigma$, assuming universal couplings. The black dashed and solid curves correspond to our simplified model production cross-section with $g_V=0.5$ and $g_V=1$, respectively. A weakly coupled resonance with $g_V=0.5$ is excluded for masses below around $4.8$\,TeV, while for $g_V = 1$ the limit is $5.5$\,TeV.
The strongest bounds are provided by searches for di-boson decays into semi-leptonic final states (blue and brown) \cite{ATLAS:2020fry,CMS:2021klu}. As expected from the branching ratios, we see that decays into a $W$ boson and a single Higgs (brown) have comparable sensitivity to the combined $WZ$ channel (blue).  The combined fully-leptonic di-boson final state search (purple) \cite{ATLAS:2018iui} provides a weaker bound of around 3\,TeV despite lower backgrounds than in the semi-leptonic channel. This can be explained by the higher integrated luminosity of $139 \text{\,fb}^{-1}$ in the semi-leptonic search compared to $36.1 \text{\,fb}^{-1}$ in the fully leptonic. 
Although searches to $t \bar b$ (red) \cite{CMS:2021mux} are less constraining than di-boson channels, they can still set bounds up to $4\,$TeV for $g_V=1$.

\begin{figure}[tp]
\begin{center}
    \includegraphics{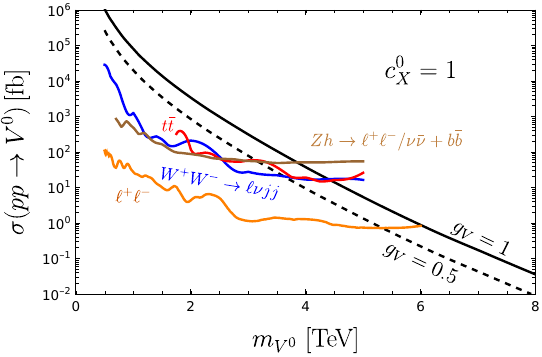}
    \caption{\small Experimental limits on the DY production cross-section of a neutral heavy vector singlet $V^0$, for all $c$ couplings set to one. The ATLAS searches shown correspond to the di-lepton final state \cite{ATLAS:2019erb} (orange), the semi-leptonic di-boson final state \cite{ATLAS:2020fry} (blue), the $t\bar{t}$ final state \cite{ATLAS:2020lks} (red), and the $Zh \rightarrow \ell^+\ell^-/\nu \bar \nu + b\bar{b}$ final state \cite{ATLAS:2017xel} (brown). The black curves show the DY production cross-section for $g_V=0.5$ (dashed) and $1$ (solid). The charged vector is decoupled, with $m_{V^+} = 100\,$TeV.}
    \label{Fig:Exp limits neutral}
\end{center}
\end{figure}

In \cref{Fig:Exp limits neutral} we show the most constraining searches for the neutral vector singlet, presented as limits on the production cross-section of $V^0$ with universal couplings. Here, the strongest limits are given by the di-lepton search (orange) \cite{ATLAS:2019erb}, which for a weakly coupled vector with $g_V=0.5$ can exclude resonance masses up to 5.2\,TeV, while for $g_V=1$ the limit is 6\,TeV. The di-boson semi-leptonic searches (blue and brown) \cite{ATLAS:2020fry,ATLAS:2017xel} provide weaker bounds, due to the small di-boson branching ratios of the neutral vector (see \cref{fig:BRs-universal}, right). 
The $t \bar t$ search (red) \cite{ATLAS:2020lks} is comparable to the di-boson searches, excluding masses up to 3.5--4.5\,TeV.

%-------------------------------------------------
\subsection{Current LHC Limits with Non-Universal Couplings}
%-------------------------------------------------

In this section we relax the universal coupling assumption from the previous section and interpret the current experimental searches as limits on our simplified model parameter space. When the couplings are free to vary, the charged and neutral vector singlets must be treated differently to each other. The LHC phenomenology of the charged vector depends only on two parameter combinations, $g_Vc_H^+$ and $g_Vc_q^+$, for a fixed mass $m_{V^+}$. We can then present exclusion contours in the $(g_Vc_H^+,g_Vc_q^+)$ plane, in analogy with the HVT.  For most explicit models, the $c$ parameters will be fixed, there is only one free coupling, $g_V$, and we can present exclusion limits in the $(m_{V^+},g_V)$ plane. The LHC phenomenology of the neutral vector is rather different, due to its independent couplings to left- and right-handed fermions. For a fixed mass $m_{V^0}$, there are six free parameters ($g_Vc_H^0$ and five different $g_Vc_\Psi^0$ couplings), so limits cannot be presented on a single two-dimensional plot. However, only certain parameter combinations enter into measurements of $\sigma\times$BR.  Using this information we can present exclusion contours in a somewhat reduced effective parameter space, as shown below.

%-------------------------------------------------
\subsubsection{Charged Vector Singlet}
%-------------------------------------------------

Although the charged HVS and the HVT both only have two relevant couplings, the charged singlet and the charged component of the triplet are phenomenologically very different.  While the triplet dominantly couples to two left-handed SM fermions, the singlet must couple to right-handed SM fermions. The charged heavy vector singlet dominantly couples to two right-handed quarks and only couples to leptons through mixing with the $W$ boson, which leads to a very suppressed branching ratio, $\text{BR}(V^{+}\to \ell^{+}\nu)\lesssim 10^{-2}\text{--}10^{-3}$ depending on the values of the $c$ parameters. This channel gives one of the strongest bounds on the vector triplet, but for charged singlets, unless there is a strong 
suppression of the coupling to right-handed quarks (which reduces the production cross-section), the leading decay channels are di-jets, $t \bar b$, and di-bosons. 

\begin{figure}[t!]
\begin{center}
    \includegraphics[width=\textwidth]{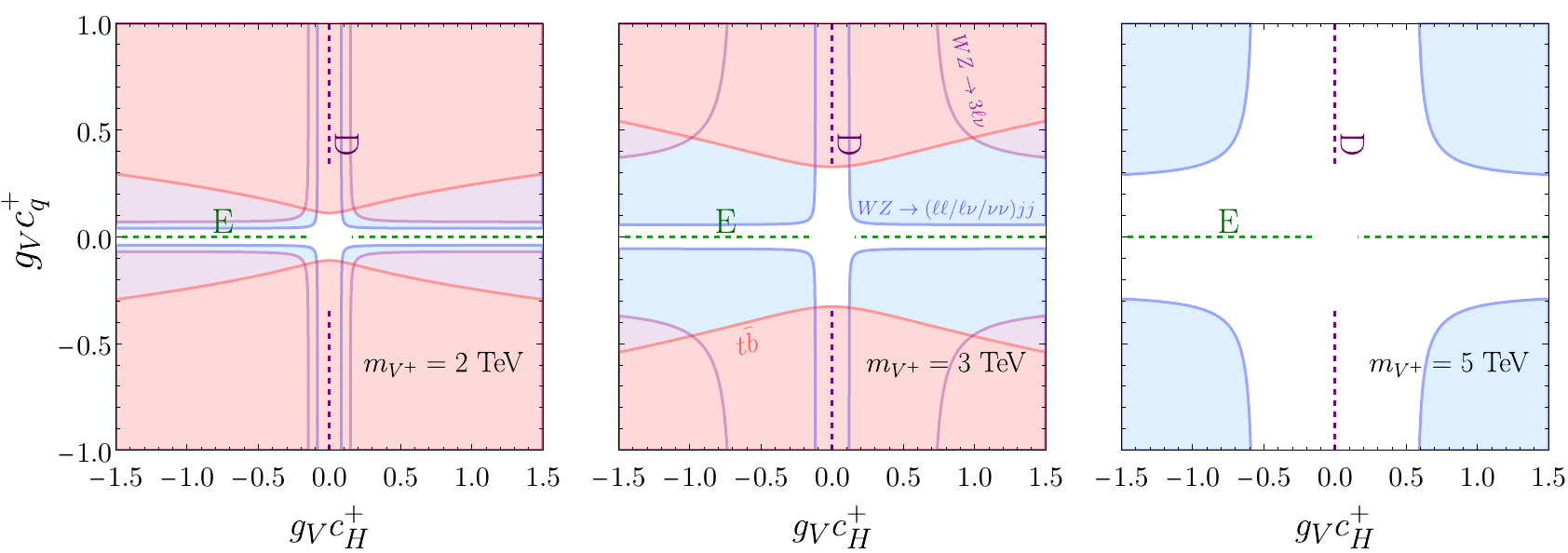}
    \caption{\small Current LHC constraints on a charged heavy vector singlet $V^+$ in the $(g_V c_{H}^{+}, g_V c_{q}^{+})$ plane for $m_{V^+} = $ $2, 3$ and $5$\,TeV (left, centre and right). The vector $V^0$ is decoupled, with $m_{V^0}$ set to 100\,TeV. The light blue regions show the exclusions for the ATLAS search of a combined semi-leptonic di-boson final state \cite{ATLAS:2020fry}, while the purple regions indicate the fully-leptonic di-boson CMS search of Ref.~\cite{ATLAS:2018iui}. Light red depicts limits from the CMS search for a charged vector decaying to a $t \bar b$ final state \cite{CMS:2021mux}. The dashed purple and green lines correspond to the \cref{sec:explicit-models} benchmark models D and E respectively, where $c_H^+$ and $c_q^+$ are constant and $g_V$ can vary.}
    \label{Fig:BoundsCharged1}
\end{center}
\end{figure}
In \cref{Fig:BoundsCharged1}, we show current exclusion bounds in the plane $(g_Vc_H^+,g_Vc_q^+)$ for three different resonance masses, $m_{V^+}=2, \ 3, \ \text{and} \ 5$\,TeV (left, centre and right). Both fully- and semi-leptonic di-boson final states \cite{ATLAS:2020fry,ATLAS:2018iui}, in purple and blue, respectively, rule out significant parts of the parameter space at $m_{V^+} = 2\,\text{TeV}$, and the sensitivity drops as the mass increases, with the fully-leptonic di-boson channel losing sensitivity around 3\,TeV and the semi-leptonic channel losing sensitivity around 5\,TeV.  At all masses, a narrow strip around $c_q^+\sim 0$ cannot be constrained by any DY search, as the production cross-section vanishes in this limit. While we do not consider it here, this region could potentially be probed with production via vector boson fusion (cf.~Ref.~\cite{Baker:2022zxv}). Small values of $c_H^+$ also cannot be probed by di-boson searches, since the branching ratio of the charged vector into SM bosons becomes negligible. In this case, di-quark final states such as $t \bar b$ \cite{CMS:2021mux} (in red) complement the di-boson searches, excluding some parameter space where $c_H^+ \sim 0$. For the electroweak constraints, the charged singlet only contributes to the $\hat{T}$ parameter and the constraints are too weak to appear in these figures. 

In \cref{Fig:BoundsCharged1} we also show the regions of parameter space corresponding to the explicit models D and E discussed in \cref{sec:explicit-models}. In these models the parameters $c_H^+$ and $c_q^+$ are fixed, so their corresponding lines in the figure represent permitted values of the coupling $g_V$. Model D is entirely excluded by $t \bar b$ searches up to 3\,TeV. Future di-quark searches are necessary to constrain this model at higher masses since di-boson searches are insensitive when $c_H^+=0$, as is the case in Model D. The charged component of Model E does not couple to quarks, so we cannot exclude it with DY production. A discussion of resonance production via vector boson fusion requires a dedicated analysis similar to \cite{Baker:2022zxv} which is beyond the scope of this paper.

%-------------------------------------------------
\subsubsection{Neutral Vector Singlet}
\label{sec: Ch3 neutral vector}
%-------------------------------------------------

Since the neutral singlet has six independent parameters, $g_Vc_H^0$ and $g_Vc_\Psi^0$ for \sloppy\mbox{$\Psi\in\{Q,L,U,D,E\}$}, it is challenging to present limits in this space in a meaningful way. However, we see from \cref{eq:di-lep di-top partial widths-qq,eq:di-lep di-top partial widths-ll,eq:di-lep di-top partial widths-tt} that the left- and right-handed fermion couplings can not easily be independently probed at the LHC.  If the heavy vector mass is large enough that these approximations hold, we can reduce the size of this parameter space to five independent parameters by defining the effective couplings,
\be \label{eq: effective couplings}
\bry{lll}
    (c_u^\text{eff})^2 &= (c_Q^0)^2 + (c_U^0)^2,\, \qquad (c_e^\text{eff})^2 &= (c_L^0)^2 + (c_E^0)^2\,,\\
    (c_d^\text{eff})^2 &= (c_Q^0)^2 + (c_D^0)^2\,, \qquad c_n^\text{eff} &= c_L^0\,.
\ery
\ee
Even though up- and down-type light jets cannot be distinguished, we have not combined the up- and down-type couplings since DY production is sensitive to the up and down quark content of the proton.
Decays to the $\nu \bar \nu$ final state only depend on $c_L^0$, and we define $c_n^\text{eff}=c_L^0$ to remain consistent in notation.  Current LHC constraints are only sensitive to $c_n^\text{eff}$ through the total width (which impacts the branching ratios).  While we retain this parameter for accuracy and to investigate the interplay with the electroweak constraints, the relevant parameter space for LHC physics is now just four-dimensional if this effect is neglected, which can be presented in a set of two-dimensional plots.

We can now show slices depicting the parameter dependence in the full five-dimensional parameter space by plotting contours on an array of two-dimensional plots.  To do this we first choose one effective parameter, say $c_u^\text{eff}$, and show each two-dimensional plot in the $(g_Vc_H^0,g_Vc_u^\text{eff})$ plane.  We then define the following ratios,
\begin{equation}
\label{eq: lambda parameters}
    \lambda_d = \dfrac{c_d^\text{eff}}{c_u^\text{eff}}, \qquad \lambda_e = \dfrac{c_e^\text{eff}}{c_u^\text{eff}}, \qquad \lambda_n = \dfrac{c_n^\text{eff}}{c_u^\text{eff}}.
\end{equation}
With these ratios, an array of plots corresponding to different fixed $\lambda$ values will show the dependence of the experimental constraints on all five effective couplings. Note that, due to the absence of right-handed neutrinos in the SM, $\lambda_e\geq \lambda_n$.

%-------------------------------------------------
\begin{figure}
    \centering
    \includegraphics[width=\textwidth]{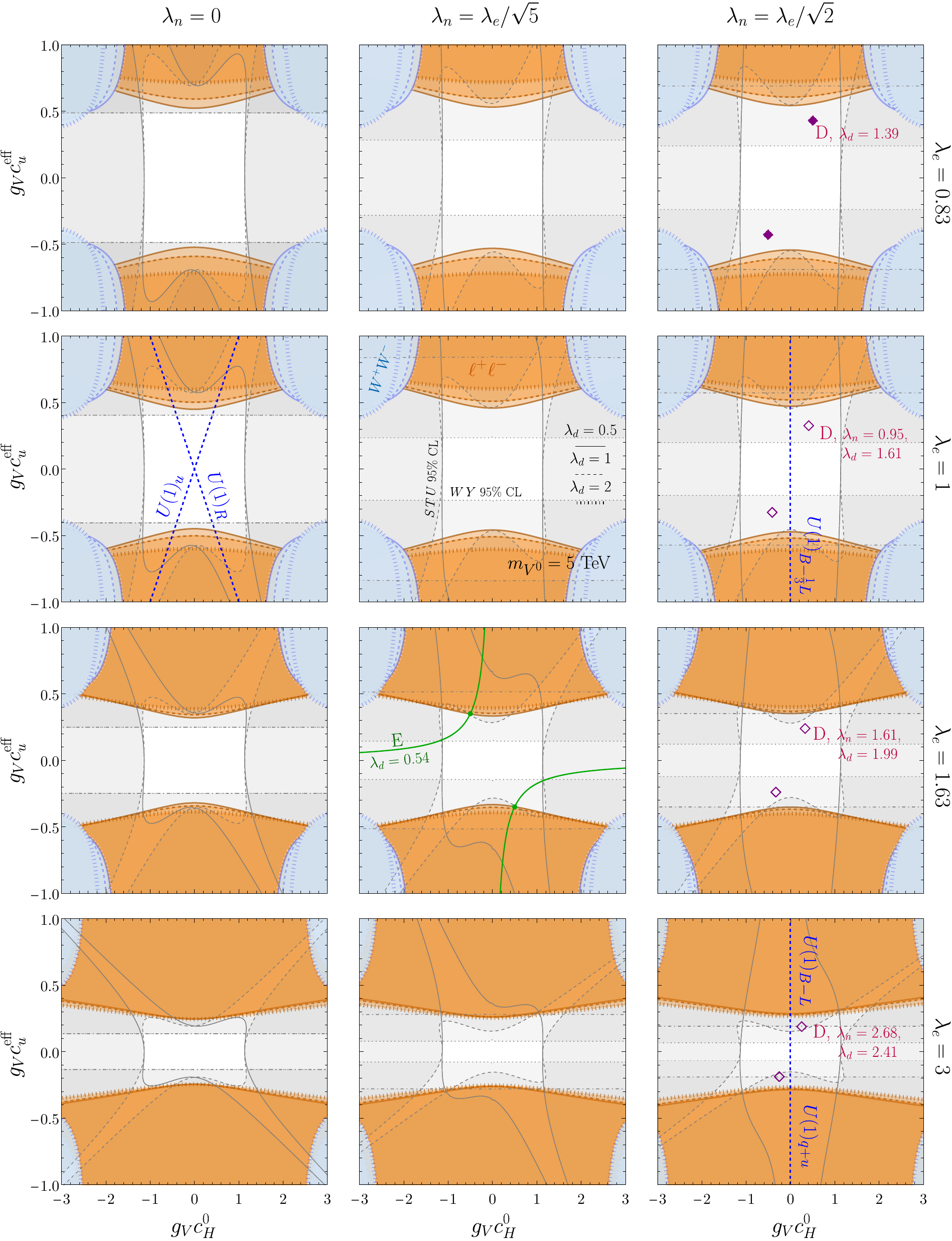} 
    \caption{\small Current LHC limits for a neutral heavy vector singlet $V^0$ at $m_{V^0}=5$\,TeV. See text for further details.
    }
    \label{fig:BoundsNeutral4x3}
\end{figure}
%-------------------------------------------------

We show this array of two-dimensional plots in \cref{fig:BoundsNeutral4x3}. The various two-dimensional plots in the array correspond to different lepton coupling strengths. From top to bottom, we show an increasing coupling to charged leptons $\lambda_e$.  In the rows, we fix $\lambda_n/\lambda_e$, which increases from left to right.  The ratio $\lambda_n/\lambda_e$ parameterises the relative size of $c_E^0$ and $c_L^0$, with the leftmost column giving $c_L^0=0$  and the rightmost column giving $c_E^0=c_L^0$.  In each two-dimensional plot, we show three sets of contours, one set for $\lambda_d = 0.5$, one for $\lambda_d = 1$ and one for $\lambda_d = 2$.

In all two-dimensional plots we show the leading di-lepton~\cite{ATLAS:2019erb} (orange) and di-boson~\cite{ATLAS:2020fry} (blue) constraints, assuming $m_{V^0} = 5\,\text{TeV}$.  The di-lepton constraints are the most stringent at $|g_V c_H^0| \lesssim 1$, while the di-boson constraints are more constraining when $g_Vc_u^\text{eff}$ and $g_Vc_H^0$ are quite large and the lepton couplings are small.  The regions disfavoured by the indirect electroweak constraints at $95\%$ CL are shown in grey for $S$, $T$ and $U$ (solid, dashed) and $W$ and $Y$ (dot-dashed, dotted). Expressing \cref{eq:S,eq:T,eq:U,eq:W,eq:Y} in terms of \cref{eq: effective couplings,eq: lambda parameters} requires that $(c_E^0)^2 = (c_u^\text{eff})^2(\lambda_e^2 - \lambda_n^2)$, such that we cannot determine the relative sign of $c_E^0$. The solid and dot-dashed lines show the EWPTs for a positive relative sign, while the dashed and dotted lines show the negative relative sign.  In all the slices we show, and depending on this relative sign, the direct searches can only probe regions that are disfavoured by the indirect searches.  However, the electroweak constraints depend on the full particle content of the model so the limits obtained by direct searches remain the most reliable.

As $\lambda_e$ increases, the di-lepton constraints get stronger (as is expected, since a larger coupling leads to a larger branching ratio into leptons).  Decreasing $\lambda_n$ does not have a strong impact on the direct constraints, since it only enters the total width of $V^0$, but does change the EWPT disfavoured regions, due to the dependence seen in \cref{eq:S,eq:T,eq:U,eq:W,eq:Y}.  We see that if electroweak information is not taken into account, it can be a good approximation to neglect $\lambda_n$ and only consider a four-dimensional parameter space (which can be presented in a single row of plots).  For $\lambda_e = 1.63$ and 3, the limits are fairly insensitive to changes in $\lambda_d$ (even though the production cross-section becomes larger for larger $\lambda_d$, this happens to be compensated by a reduction in the di-lepton branching ratio).  Note that $\lambda_d$ does not impact the EWPT contours.
 
Some of the UV models we consider in \cref{sec:explicit-models} can be represented as points or lines on the particular slices of parameter space we have chosen. In the top-right panel Model D (a model based on the extended $SU(3)_C\times SU(2)_L \times SU(2)_R \times U(1)_X$ gauge group) is shown as two points (the given combination of $\lambda_e$ and $\lambda_n$ fixes $g_V = \pm 0.61$ in this model). In the other right hand panels we have also shown Model D even though $\lambda_n \neq \lambda_e/\sqrt{2}$, since the constraints are only weakly dependent on $\lambda_n$.  The choice $\lambda_e = 1$ leads to $g_V = 0.53$, $\lambda_e = 1.63$ leads to $g_V = 0.48$ and $\lambda_e = 3$ leads to $g_V = 0.43$.  For several models based on an extra $U(1)_X$ gauge group, $g_V$ remains a free parameter so these models are depicted as lines, with $g_V=0$ appearing at the origin. Note that $U(1)_u$ corresponds to $U(1)_{q+xu}$ for $x\gg1$.  For Model E (a minimal composite Higgs model), $g_V$ is again unconstrained but $\lambda_d = 0.54$ is fixed.  The model is excluded by di-lepton searches at 5\,TeV for $g_V\lesssim 0.61$ (we will see that a small $g_V$ leads to large couplings of the neutral heavy vector singlet to fermions in this model).

%-------------------------------------------------
\subsection{Projections to Future Colliders with Universal Couplings}
\label{sec:future-projections}
%-------------------------------------------------

For the case of universal couplings, we now extrapolate the current limits discussed above to predict the future sensitivity of the forthcoming high-luminosity LHC (HL-LHC), and the proposed designs for a 27\,TeV high-energy LHC (HE-LHC) \cite{FCC:2018bvk,CidVidal:2018eel}, a 100\,TeV Future Circular Collider (FCC-hh) \cite{FCC:2018vvp} and a 100\,TeV Super Proton-Proton Collider (SPPC) \cite{CEPC-SPPCStudyGroup:2015csa}. We follow the method discussed in Ref.~\cite{Thamm:2015zwa}, which uses rescaling of the parton luminosities to provide simple extrapolations of cross-section limits to future hadron-hadron colliders. The specific energies and luminosities that we assume for these future colliders are given in \cref{tab:future-colliders}.

%-------------------------------------------------
\begin{table} \centering
\begin{tabular}{ c | c c c c}
  	 	& Collisions 	& $ \sqrt{s}$ [TeV]	& $L$ [ab$^{-1}$]  & References \\ \hline		
  HL-LHC 	& $pp$ 		& $14$ 			& $3$	    & \cite{ZurbanoFernandez:2020cco} \\
  HE-LHC 	& $pp$ 		& $27$ 			& $15$	    & \cite{FCC:2018bvk} \\
  SPPC  	& $pp$ 		& $100$ 		& $3$	& \cite{CEPC-SPPCStudyGroup:2015csa}\\
  FCC-hh 	& $pp$ 		& $100$ 		& $20$  & \cite{FCC:2018vvp}\\ 
\end{tabular} \caption{\small Benchmark centre-of-mass energies and integrated luminosities for various future collider proposals used in this work. }
\vspace{-.2cm}
\label{tab:future-colliders}
\end{table}
%-------------------------------------------------

The main idea that underlies this procedure is that, for a small window of partonic centre-of-mass energy centred around the resonance mass, the upper limit on the number of signal events depends exclusively on the number of background events in that window. By studying the scaling of the background with energy and luminosity, we can use current LHC bounds to obtain projected exclusions at future colliders. 

Equating the number of background events,
\begin{equation}\label{backgrounds}
    B(s_0,L_0,m_0) = B(s,L,m),
\end{equation}
where $\sqrt{s_0}$ and $L_0$ are the LHC centre-of-mass energy and integrated luminosity, and $\sqrt{s}$ and $L$ correspond to that of the new collider, we can determine the `equivalent mass' $m$, which describes the resonance mass where the number of
background events at the future collider is equal to the number of background events at
$m_0$ in the LHC analysis. As described in \cref{Appendix:extrapolation}, this can be written as
\begin{equation}
\label{backgrounds sum c_ij}
    \sum_{\{i,j\}} c_{ij} \dfrac{d\mathcal{L}_{ij}}{d\hat{s}}(m;\sqrt{s}) = \dfrac{L_0}{L}\sum_{\{i,j\}} c_{ij} \dfrac{d\mathcal{L}_{ij}}{d\hat{s}}(m_0;\sqrt{s_0})\,,
\end{equation}
where $c_{ij}$ are constants which depend on the dominant background processes.  If there is only one set of partons that produces the dominant background, the sum and the $c_{ij}$ coefficients cancel on both sides of the equation. If there is more than one set of partons, the $c_{ij}$ coefficients have to be computed and summed over. 

Since the number of background events will be the same for a heavy vector with mass $m$ at a future collider as for a heavy vector with mass $m_0$ at the LHC, we can estimate the expected limit on $\sigma \times \text{BR}$ at the future collider using
\begin{equation}
    \label{extrapolated cross-section}
    \text{limit}[\sigma \times \text{BR}](m;s,L) = 
    \min_{L' \leq L}
    \dfrac{L_0}{\sqrt{L L'}}
    \text{limit}[\sigma \times \text{BR}]_0(m_0;s_0,L_0)
    \,.
\end{equation}
For details see \cref{Appendix:extrapolation}.  This holds as long as the background and signal acceptances and efficiencies of the two experiments are similar, which we will assume.

This extrapolation procedure relies on the assumption that the exclusion limit is exclusively driven by the background. In principle, \cref{backgrounds sum c_ij} could include dominant and subdominant backgrounds. However, the extrapolation procedure would then become rather involved while still relying on the assumption that the background composition at the low and high-energy colliders are the same. At this point, a detailed signal and background analysis at the future collider will be more accurate. For this reason, we will restrict our attention to those signal channels where the dominant background is produced by just one or two partonic interactions.

\Cref{tab:charged searches,tab:neutral searches} list the dominant backgrounds for all search channels. For the charged heavy vector singlet, we apply the extrapolation procedure to the fully leptonically decaying $WZ$ search \cite{ATLAS:2018iui}. The dominant background to this process is the SM DY production of $WZ$ which requires the parton luminosity for $u \bar d$. Furthermore, we extrapolate the $Wh \to \ell \nu b \bar b$ search where $t \bar t$ is the dominant background, which is overwhelmingly produced via gluon fusion \cite{CMS:2021klu}. The extrapolation procedure could be applied to $WZ \to \nu \nu jj$, but while the dominant background originates from a SM $W$ or $Z$-boson plus jets, which \cref{backgrounds sum c_ij} does not capture easily, the $t \bar t$ background dominates in certain control regions. However, since the extrapolation procedure works reliably for the leptonically decaying di-boson channel, we do not include this semi-leptonic channel in our extrapolations. The backgrounds of the other signal channels are dominated by QCD processes or SM gauge boson plus jet production which is not easily captured by \cref{backgrounds sum c_ij}.

%-------------------------------------------------
\begin{figure}
    \centering
    \includegraphics[width=0.49\textwidth]{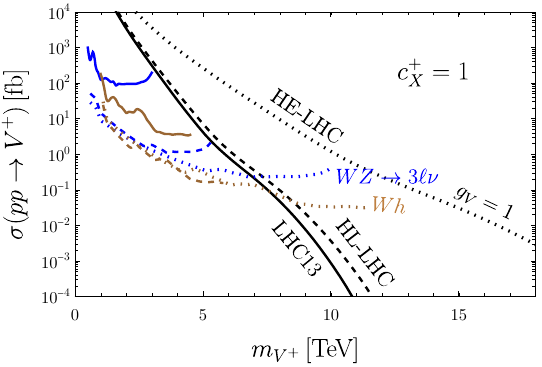}
    \includegraphics[width=0.49\textwidth]{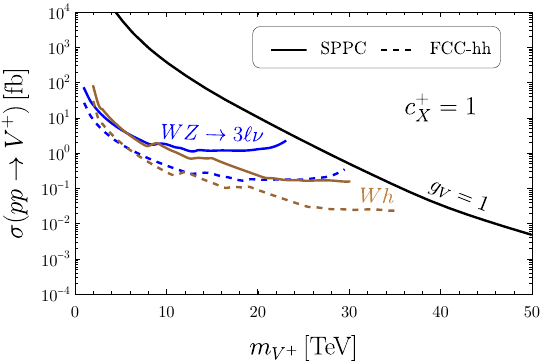}
    \caption{\small Limits on the production cross-section for a charged vector singlet $V^+$ with universal couplings at the LHC (left solid), and projected limits for the 14\,TeV HL-LHC with $3\text{\,ab}^{-1}$ (left dashed), 27\,TeV HE-LHC with $15\text{\,ab}^{-1}$ (left dotted), 100\,TeV SPPC with $3\text{\,ab}^{-1}$ (right solid) and FCC-hh with $20\text{\,ab}^{-1}$ (right dashed). The production cross-sections are shown in black, while blue corresponds to the fully-leptonic di-boson final state \cite{ATLAS:2018iui} at $36.1\text{\,fb}^{-1}$, and brown is the $Wh$ search of Ref.~\cite{CMS:2021klu} at $137\text{\,fb}^{-1}$.}
    \label{projected cross-sections_char}
\end{figure}
%-------------------------------------------------

\Cref{projected cross-sections_char} shows the production cross-section of a charged heavy vector $V^+$ with universal couplings at the LHC, the HL-LHC and HE-LHC (left), and at 100\,TeV colliders (right), along with the existing limits from the searches in \cref{tab:charged searches} and their projected sensitivities. As in \cref{subsec:universal-couplings-exp}, all $c_X^+$ couplings and $g_V$ are set to one. We see that the $Wh$ search of Ref.~\cite{CMS:2021klu} (brown) sets the current strongest limit at around 4--5\,TeV.  The HL-LHC would be expected to reach above 6\,TeV while the HE-LHC would be able to reach above 11--12\,TeV. Looking further ahead, the SPPC with 3\,ab$^{-1}$ would reach over 30\,TeV while the FCC-hh with 20\,ab$^{-1}$ would reach over 35\,TeV. Note that the exclusion limits are expected to continue at a constant production cross-section to higher masses since this is the no-background regime. Making this assumption, the mass reaches become 8\,TeV for the HL-LHC, 15\,TeV for the HE-LHC, 34\,TeV for the SPPC and 42\,TeV for the FCC-hh.

As shown in \cref{tab:neutral searches}, for neutral heavy resonances the extrapolation procedure can be applied to the di-lepton \cite{ATLAS:2019erb}, $t \bar t$ \cite{ATLAS:2020lks} and $Zh \to \nu \nu \bar b b$ \cite{ATLAS:2017xel} searches where the dominant backgrounds are DY $\ell^+ \ell^-$, $t \bar t$ and $t \bar t$, respectively. The DY $\ell^+ \ell^-$ background can be produced via $u \bar u$ and $d \bar d$ initial states. We compute the tree-level background processes analytically and obtain $c_{d \bar d} /c_{u \bar u} = 0.51$. The $t \bar t$ background is dominantly produced in gluon-gluon fusion and we use the $gg$ parton luminosities to estimate it. We also apply the extrapolation procedure to $W W \to \ell \nu jj$ \cite{ATLAS:2020fry}. Here the $W$ plus jet background competes with $t \bar t$ production. While we can not compute the former, we use the latter for our extrapolation. For any other signal channel, the multi-jet background dominates which can not be captured by our extrapolation procedure.

%-------------------------------------------------
\begin{figure}
    \centering
    \includegraphics[width=0.49\textwidth]{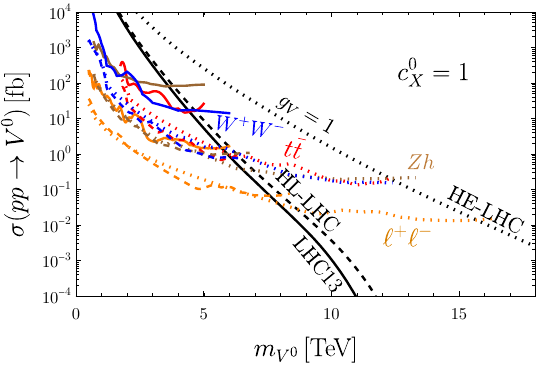}
    \includegraphics[width=0.49\textwidth]{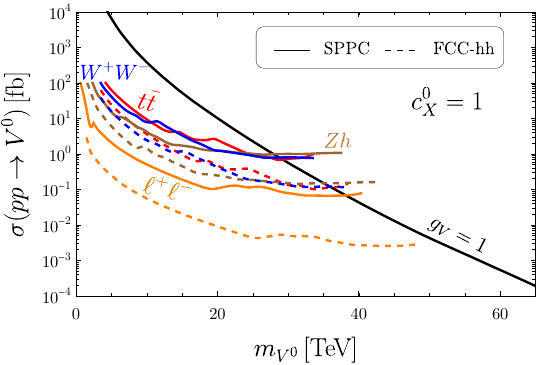}
    \caption{\small Limits on the production cross-section for a neutral vector singlet $V^0$ with universal couplings at the LHC (left solid), and projected limits for the 14\,TeV HL-LHC with $3\text{\,ab}^{-1}$ (left dashed), 27\,TeV HE-LHC with $15\text{\,ab}^{-1}$ (left dotted), 100\,TeV SPPC with $3\text{\,ab}^{-1}$ (right solid) and FCC-hh with $20\text{\,ab}^{-1}$ (right dashed). The production cross-sections are shown in black, while orange shows the di-lepton final state search of Ref.~\cite{ATLAS:2019erb} at $139\text{\,fb}^{-1}$, blue the semi-leptonic di-boson final state search of Ref.~\cite{ATLAS:2020fry} at $139\text{\,fb}^{-1}$, brown the di-boson search of Ref.~\cite{ATLAS:2017xel} at $36.1\text{\,fb}^{-1}$, and red the $t \bar t$ search of Ref.~\cite{ATLAS:2020lks} at $139\text{\,fb}^{-1}$.}
    \label{projected cross-sections_neut}
\end{figure}
%-------------------------------------------------

\Cref{projected cross-sections_neut} shows the production cross-sections and existing and projected limits for a neutral vector $V^0$ with universal couplings. In this case the di-lepton search of Ref.~\cite{ATLAS:2019erb} sets the strongest limits. The LHC has ruled out masses below 6\,TeV, the HL-LHC could reach 8\,TeV while the HE-LHC could reach 16\,TeV. The 100\,TeV SPPC with $3\text{\,ab}^{-1}$ could almost reach 40\,TeV, and a luminosity upgrade to $20\text{\,ab}^{-1}$, such as at the FCC, would push this to over 48\,TeV (or 51\,TeV when assuming a constant limit at higher masses). 

We see that the limits on $V^+$ and $V^0$ are comparable, even though the best searches in each case differ (di-boson for $V^+$ and di-lepton for $V^0$). The di-boson search for $V^+$ is stronger than that for $V^0$ because the branching ratio for $V^+$ into di-bosons is larger by a factor of around 10.

Detailed simulations were performed in Refs.~\cite{Helsens:2019bfw,FCC:2018bvk} to derive expected exclusion limits on a neutral resonance at future collider energies of 27\,TeV with $15\,$ab$^{-1}$ of integrated luminosity and at 100\,TeV with $30\,$ab$^{-1}$. In order to compare these limits to our extrapolations we note that the di-lepton exclusion limits in Refs.~\cite{Helsens:2019bfw,FCC:2018bvk} apply to the 2-to-2 cross-section $\sigma(pp \to \ell^+ \ell^-)$ and not to $\sigma(pp \to V^0) \times \text{BR}(V^0 \to \ell^+ \ell^-)$ in the narrow width approximation. The sizable tail at low masses in the invariant mass distributions in Refs.~\cite{Helsens:2019bfw,FCC:2018bvk} originates from the off-shell production of a heavy resonance with a non-negligible width. Since this low mass tail is not captured by the Breit-Wigner formula, factorization of the 2-to-2 cross-section is not justified (see, e.g., \cite{Pappadopulo:2014tg}). The sizeable tail leads to characteristic exclusion limits which lose sensitivity at large masses. Our extrapolation is based on \cite{ATLAS:2019erb} which uses generic signal shapes constructed from non-relativistic Breit–Wigner functions and presents exclusion limits on $\sigma(pp \to V^0) \times \text{BR}(V^0 \to \ell^+ \ell^-)$. This explains some inherent differences between Refs.~\cite{Helsens:2019bfw,FCC:2018bvk} and our di-lepton extrapolations. Quantitatively, our extrapolation of the limit on $\sigma\times \text{BR}$ for the di-lepton search at 27\,TeV agrees with the exclusions in \cite{FCC:2018bvk} up to a factor of 2 up to $m_{V^0} = 10\,$TeV but only up to a factor of 14 at $m_{V^0} = 14\,$TeV. Our 100\,TeV extrapolation in the di-lepton channel agrees well up to $m_{V^0} = 35\,$TeV but is stronger by a factor of 8 at $m_{V^0} = 50\,$TeV.  

Ref.~\cite{Helsens:2019bfw} also presents projected limits on the fully hadronic di-boson final state at 100\,TeV. We, however, extrapolate the semi-leptonic di-boson exclusion limit which we find to be a factor of 10 stronger. Note that the exclusion in Ref.~\cite{Helsens:2019bfw} was obtained for a spin-2 Randall-Sundrum graviton. Note further, that this descrepancy may be partially due to the fact that the LHC exclusion limit in the semi-leptonic final state \cite{ATLAS:2020fry} is stronger by a factor of 5-10 than the LHC limit in the fully hadronic channel \cite{ATLAS:2019nat}. 

Finally, our $t\bar t$ extrapolation agrees with that in Ref.~\cite{Helsens:2019bfw} up to a factor of 2 up to $m_{V^0} = 25\,$TeV and up to a factor of 8 at $m_{V^0} = 35\,$TeV.

%=================================================
\section{Matching Explicit Models onto the Simplified Lagrangian}
\label{sec:explicit-models}
%=================================================

To demonstrate how experimental limits given in terms of the simplified model parameters can easily be reinterpreted in explicit models, in this section we match three families of models onto the simplified model parameter space.  In Ref.~\cite{Pappadopulo:2014tg}, a similar matching was performed for the HVT. The example models were called A and B. To avoid confusion, here we name the models C, D and E.

%-------------------------------------------------
\subsection{Model C: New $U(1)_X$ Gauge Symmetry}
\label{subsec:4.1}
%-------------------------------------------------

If the gauge symmetry of the Standard Model is extended by a $U(1)_X$ symmetry, $G = SU(3)_C \times SU(2)_L \times U(1)_Y \times U(1)_X$, and the $U(1)_X$ is broken by, e.g., a Higgs mechanism, then the associated massive gauge boson can be described by $\mathcal{V}^0$ in our simplified Lagrangian, \cref{sml-neutral}.  After electroweak symmetry breaking and mass diagonalisation, $\mathcal{V}^0$ becomes $V^0$.  In this context $V^0$ is often referred to as $Z'$.  

%-------------------------------------------------
\begin{table}
\begin{center}
{$\begin{array}{cccc}
\hline
\hline
\text{Field} & U(1)_{B-xL} & U(1)_R & U(1)_{q+xu} \\ 
\hline
Q_L = (u_L, d_L)^T & 1/3 & 0 & 1/3 \\
u_R & 1/3 & -1/3 & x/3 \\
d_R & 1/3 & 1/3 & (2-x)/3 \\
L_L = (\nu_L, e_L)^T & -x & 0 & -1 \\
e_R & -x & 1/3 & -(2+x)/3 \\
H & 0 & -1/3 & (x-1)/3 \\
\hline
\hline
\end{array}$}
\\[0.1cm]
\caption{
\small Several $U(1)_X$ extensions of the SM and the $U(1)_X$ charges of the SM fermions and Higgs \cite{ParticleDataGroup:2024pth}.  
The right-handed group $U(1)_R$ corresponds to $U(1)_{d-xu}$, where the left-handed fermions are uncharged, with $x=1$.
}
\label{Table:u(1)-charges}
\end{center}
\end{table}
%-------------------------------------------------

While a wide range of possible extensions are considered in the literature~\cite{ParticleDataGroup:2024pth}, we here focus on a set of generation-independent extensions which require only the usual SM Higgs boson (some extensions require further scalars to generate the SM fermion masses).  The $U(1)_X$ models we consider and the SM gauge charges are shown in \cref{Table:u(1)-charges}, where $x$ can be any rational number. When the $U(1)_X$ charges are fixed, the model then has two free parameters: the gauge coupling $g_X$ and the $V^0$ mass $m_{V^0}$. 
 While in most cases anomaly cancellation requires additional fermions, we assume that these are heavy enough to not impact the HVS collider phenomenology.

%-------------------------------------------------
\begin{table}
\renewcommand{\arraystretch}{1.1}
\setlength{\tabcolsep}{0.23cm}
\begin{center}
\begin{tabular}{cccccc}
\hline
\hline
\rule{0pt}{\normalbaselineskip}
 & \multicolumn{3}{c}{{\bf Model C}} & {\bf Model D} & {\bf Model E}\\
 & $U(1)_{B-xL}$ & $U(1)_R$ & $U(1)_{q+xu}$ & $SU(2)_R\times U(1)_X$ & $SO(5)/SO(4)$ \\ 
 \hline
 $g_V$ & $\pm g_X$ & $\pm g_X$ & $\pm g_X$ & $\pm g_R$ & $\pm g_\rho$ \\
\hline 
\rule{0pt}{\normalbaselineskip}
$m_{\mathcal{V}^0}$ & $m_{\mathcal{V}^0}$ & $m_{\mathcal{V}^0}$ & $m_{\mathcal{V}^0}$ & $\dfrac{|g_V| v_R}{2 k_V}$ & $\dfrac{m_\rho}{k_V}$ \\[4mm]
$c_Q^0$ & $\dfrac{2}{3(1+x)}$ & $0$ & $\dfrac{2}{3(1+x)}$ & $- 2 Y_Q \dfrac{g'^2}{g_V^2 k_V}$ & $2 Y_Q \dfrac{g'^2}{g_V^2 k_V}$\\[4mm]
$c_U^0$ & $\dfrac{2}{3(1+x)}$ & $-\dfrac{1}{3}$ & $\dfrac{2x}{3(1+x)}$ &$ \dfrac{1}{k_V}  - 2 Y_U \dfrac{g'^2}{g_V^2 k_V}$ & $2 Y_U \dfrac{g'^2}{g_V^2 k_V}$ \\[4mm]
$c_D^0$ & $\dfrac{2}{3(1+x)}$ & $\dfrac{1}{3}$ & $\dfrac{2(2-x)}{3(1+x)}$ & $- \dfrac{1}{k_V}  - 2 Y_D \dfrac{g'^2}{g_V^2 k_V}$ & $2 Y_D \dfrac{g'^2}{g_V^2 k_V}$ \\[4mm]
$c_L^0$ & $-\dfrac{2x}{1+x}$ & $0$ & $-\dfrac{2}{1+x}$ &  $-2 Y_L \dfrac{g'^2}{g_V^2 k_V}$ & $2 Y_L \dfrac{g'^2}{g_V^2 k_V}$ \\[4mm]
$c_E^0$ & $-\dfrac{2x}{1+x}$ & $\dfrac{1}{3}$ & $-\dfrac{2(2+x)}{3(1+x)}$ &$ - \dfrac{1}{k_V}  - 2 Y_E \dfrac{g'^2}{g_V^2 k_V}$ & $2 Y_E \dfrac{g'^2}{g_V^2 k_V}$\\[4mm]
$c_H^0$ & $0$ & $-\dfrac{1}{3}$ & $\dfrac{2(x-1)}{3(1+x)}$ & $k_V$ & $- \dfrac{1}{k_V} \left(a_\rho^2 - \dfrac{g'^2}{g_V^2} \right)$ \\[4mm]
$c_{VVHH}^0$ & $0$ & $\dfrac{1}{36}$ & $\dfrac{1(x-1)^2}{9(1+x)^2}$ & $\dfrac{k_V^2}{4}$ & $- \dfrac{g'^2}{2 g_V^2 k_V^2} \left(a_\rho^2 - \dfrac{g'^2}{2 g_V^2} \right)$\\[4mm]
\hline
\rule{0pt}{\normalbaselineskip}
$m_{\mathcal{V}^+}$ & $\infty$ & $\infty$ & $\infty$ & $\dfrac{|g_V| v_R}{2}$ & $m_\rho$\\[4mm]
$c_q^+$ & - & - & - & $1$ & $0$ \\[4mm]
$c_H^+$ & - & - & - & $0$ & $-\dfrac{a_\rho^2}{2}$ \\[4mm]
$c_{VVHH}^+$ & - & - & - & $\dfrac{1}{4}$ & $0$ \\[4mm]
$c_{VVB}^+$ & - & - & - & $1$ & $1$ \\[4mm]
\hline
\rule{0pt}{\normalbaselineskip}
$c_{VVV}^0$ & - & - & - & $k_V$ & $- k_V$\\[4mm]
$c_{VVV}^+$ & - & - & - & $k_V$ & $- k_V$\\[4mm]
\hline
\hline
\end{tabular}
\\[0.1cm]
\caption{
\small Matching between the simplified model parameters and the explicit Models C, D and E discussed in sections \ref{subsec:4.1}, \ref{subsec:4.2} and \ref{subsec:4.3}. Note that $k_{V}=\sqrt{1-g^{\prime\,2}/g_{V}^{2}}$ and $a_{\rho}=m_\rho/g_V f$. Only the parameters which are relevant for the collider phenomenology are shown. The SM hypercharges $Y_{\Psi}$ are given in Table \ref{Table:Hypercharges} 
}
\label{Table:matching}
\end{center}
\end{table}
%-------------------------------------------------

We can now match these models onto \cref{sml-neutral}.  For a field of $U(1)_X$ charge $X$, the covariant derivative is
\begin{align}
    D_\mu^G = D_\mu^\text{SM} - i \dfrac{g_X}{1+x}X V^0_\mu \,.
\end{align}
Identifying $g_V = g_X$, the matching conditions for fermions and scalars are given by $c_\Psi^0 = 2 X_\Psi/(1+x)$, $c_H^0 = 2 X_H/(1+x)$ and $c_{VVHH}^0 = X_H^2/(1+x)^2$ where $X_i$ are the gauge charges taken from \cref{Table:u(1)-charges}. We have rescaled $g_X X$ by the factor $\dfrac{1}{1+x}$ so that the couplings remain perturbative when $x\gg1$ (e.g., to define the models $U(1)_L$ and $U(1)_u$).  Since these models do not contain a $V^+$, this must be decoupled in the simplified Lagrangian \cref{sml} by taking $m_{V^{+}}\to\infty$.  The matching relations are shown in \cref{Table:matching}, along with those for Models D and E which we discuss in \cref{subsec:4.2,subsec:4.3}.

We can now use the results from \cref{sec:data-bounds} to easily find the current LHC limits on these models. \Cref{fig:BoundsNeutral4x3} shows the current limits on $V^0$ at $m_{V^0} = 5$\,TeV with lines for various explicit models determined by the matching relations (note that we chose values of $\lambda_e$ and $\lambda_n$ which match many of the explicit models we consider).  Since we have fixed the mass, the models have one free parameter, $g_V$.  We can see from \cref{fig:BoundsNeutral4x3} that for all the $U(1)_X$ models we show, the main constraint at $m_{V^0} = 5$\,TeV comes from di-lepton searches.  The limits for the $U(1)_{B-L}$ model are $g_V c_u^\text{eff} = \sqrt{2}/3 g_V < 0.28$, so $g_V < 0.59$; $g_V < 0.73$ for the $U(1)_{B-\frac{1}{3}L}$ model; $g_V < 0.77$ for the $U(1)_R$ model; $g_V < 0.59$ for the $U(1)_{q+u}$ model; and $g_V < 0.77$ for the $U(1)_u$ model.  We see that constraints on the simplified model can simply provide exclusion contours for a wide variety of explicit models.  If, e.g., a future di-lepton search presented their limits as a sequence of panels for different values of $\lambda_e$ (the di-lepton searches are only sensitive to $\lambda_n$ through the total width) and with contours for different masses, accurate bounds could be determined for a wide variety of explicit models.

%-------------------------------------------------
\begin{figure}
    \centering
    \includegraphics[width=\textwidth]{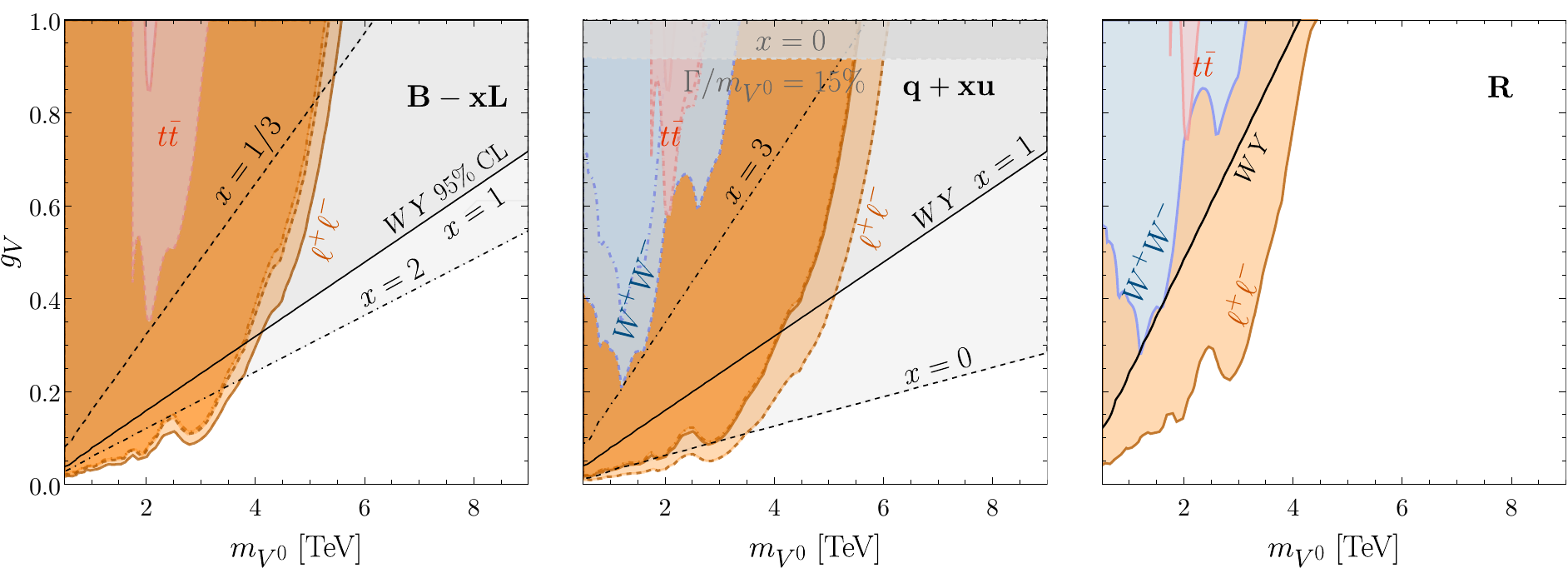}
    \caption{\small Exclusion limits from di-lepton \cite{ATLAS:2019erb} (orange), di-boson \cite{ATLAS:2020fry} (blue) and $t \bar t$ \cite{ATLAS:2020lks} (red) searches in models with gauge group extensions $U(1)_{B-xL}$ (left), $U(1)_{q+xu}$ (centre) and $U(1)_{R}$ (right) for sample values of $x$. The black contours show EWPT constraints on $W$ and $Y$ at $95\%$ CL. The light grey horizontal region in the centre panel indicates where, for the $U(1)_q$ model, $\Gamma/m_{V^0} \geq 15\%$.}
    \label{fig: U(1) mVgV}
\end{figure}
%-------------------------------------------------

To emphasise the accuracy and utility of this approach, we now use the results of the searches from \cref{sec:data-bounds} to obtain exclusion regions in the $(m_{V^0},g_V)$ plane for some explicit models in the usual way.  To do this we use the \texttt{NNPDF} parton distribution functions to construct the parton luminosities, then compute the HVS production cross-sections and branching ratios. We take care that all expressions include the effects of electroweak symmetry breaking and perform the parameter inversion.  We then digitise the experimental limits (or download the limits from \texttt{hepdata}).  We can then finally compare $\sigma \times \text{BR}$ to the experimental limits in the model's parameter space.  We first discuss these results, before comparing them to the simplified model approach.

\Cref{fig: U(1) mVgV} shows these exclusion regions for the gauge group extensions $U(1)_{B-xL}$ (left), $U(1)_{q+xu}$ (centre), and $U(1)_{R}$ (right) for some benchmark values of $x$ indicated by solid, dashed, and dot-dashed lines. As in the case of universal couplings, we see that current di-lepton searches (orange) lead to the tightest constraints, probing masses up to $\sim \,$4--5\,TeV for $g_V \sim g' \sim 0.36$. Di-boson (blue) and $t \bar t$ (red) searches can reach up to roughly $2$\,TeV for the same coupling strength, except for the $B-xL$ model, where the di-boson search does not provide any constraints because the HVS coupling to di-bosons, $c_H^0$, is zero. The EWPTs are shown as black lines, corresponding to our two-dimensional $\chi^2$ fit on $W$ and $Y$. They are straight since the oblique parameters given in \cref{EWPTs} are proportional to $g_V^2/m_{V^0}^2$. There are no constraints from $S$, $T$ and $U$ since $c_L^0 - c_E^0 - c_H^0 =0$ for these models \cite{Strumia:2022qkt}. Except for $U(1)_R$, indirect constraints on $W\, Y$ via EWPTs are more constraining than the direct collider searches at masses above 3.5--5.5\,TeV.  However, EWPTs may be sensitive to other particles present in the full model so when taking a simplified model approach the direct constraints are the most robust. We also indicate $\Gamma/m_{V^0} = 15\%$ by a grey region.  For $g_V < 1$, $V^0$ is only this wide for the $U(1)_q$ model, where values of $g_V > 0.92$ lead to $\Gamma/m_{V^0} > 15\%$. 

Comparing \cref{fig:BoundsNeutral4x3} and \cref{fig: U(1) mVgV} at $m_{V^0} = 5\,\text{TeV}$ the limits on $g_V$ for the $U(1)_{B-L}$,  $U(1)_{B-\frac{1}{3}L}$, $U(1)_{q+u}$ and $U(1)_R$ models agree, as they should.  However, the plots shown in \cref{fig: U(1) mVgV} each only provide bounds on a small range of models, while those shown in \cref{fig:BoundsNeutral4x3} show bounds on a wide range of models, including Models D and E which we discuss in \cref{subsec:4.2,subsec:4.3}.  Furthermore, for a new $Z'$ model, it is much easier to get a sense of the constraints from \cref{fig:BoundsNeutral4x3} than from \cref{fig: U(1) mVgV}.

%-------------------------------------------------
    \begin{figure}
    \centering
    \includegraphics[width=\textwidth,height=6.5cm, keepaspectratio]{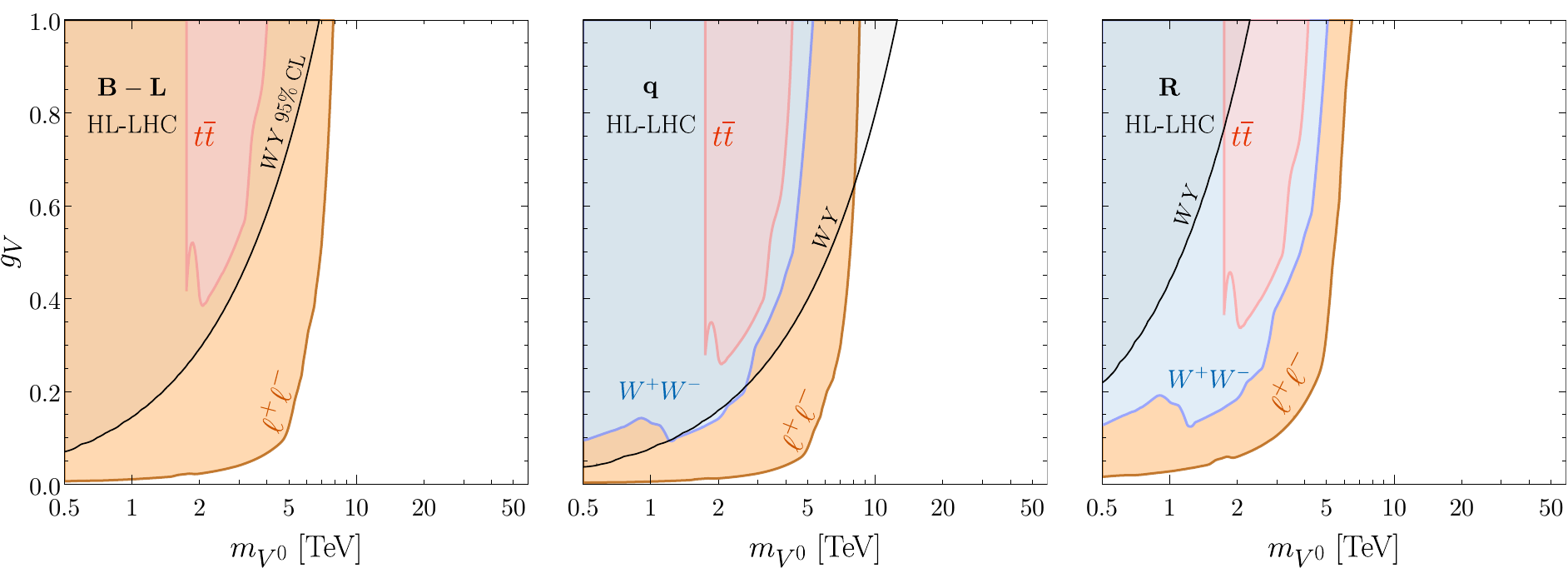}
    \includegraphics[width=\textwidth,height=6.5cm, keepaspectratio]{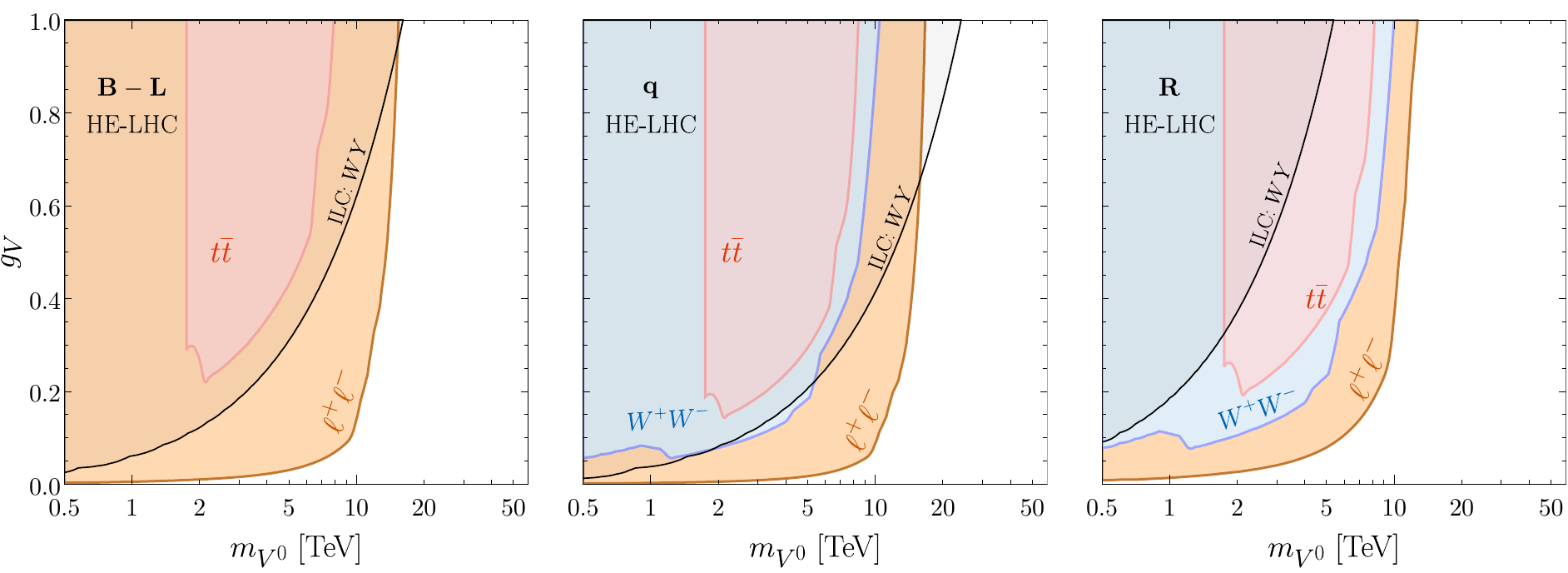}
    \includegraphics[width=\textwidth,height=6.5cm, keepaspectratio]{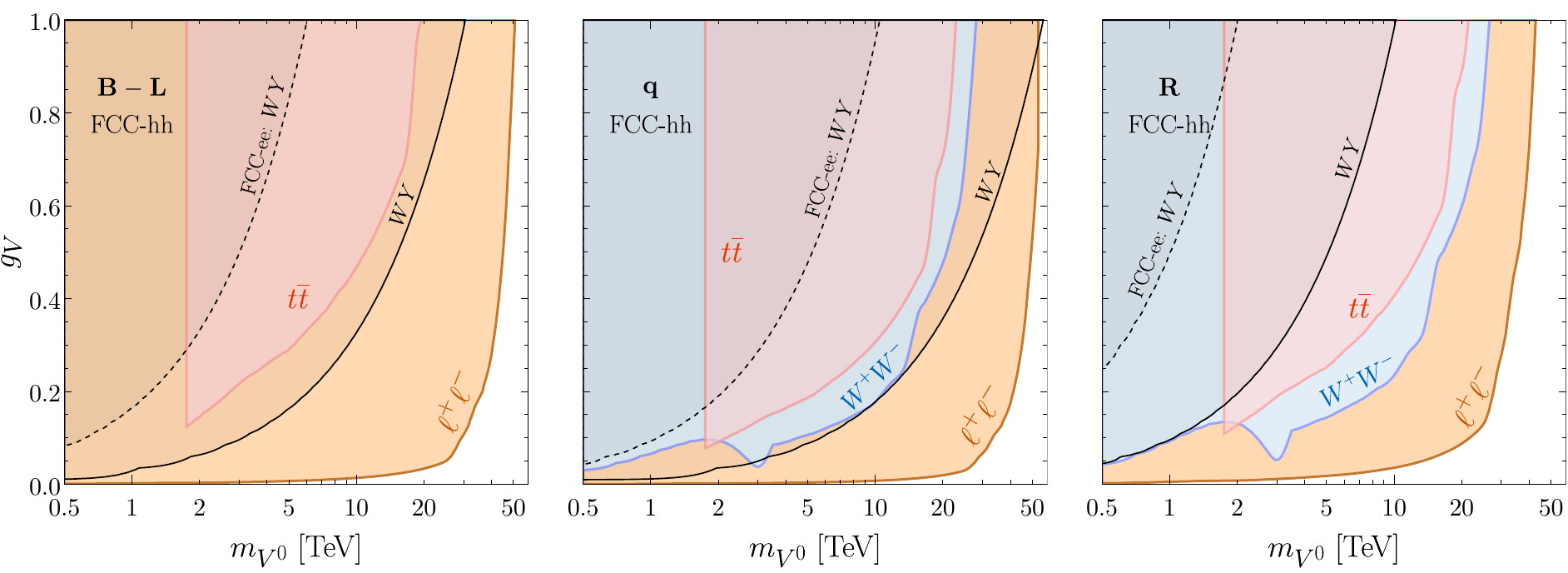}
    \caption{\small Extrapolations of Model C to the HL-LHC (top row), HE-LHC (centre row) and FCC-hh (bottom row) for selected benchmark models $U(1)_{B-L}$ (left column), $U(1)_q$ (centre column) and $U(1)_R$ (right column). The coloured regions show extrapolations of the corresponding searches in \cref{fig: U(1) mVgV}. The black contours show projected EWPT constraints on $W$ and $Y$ at $95\%$ CL for colliders which may produce results on a similar timescale to the HL-LHC, HE-LHC and FCC-hh~\cite{deBlas:2016ojx,Farina:2016rws}.}
    \label{fig: U(1) projections}
\end{figure}
%-------------------------------------------------

Finally, we take the projections to future colliders from the previous chapter and extrapolate to limits in this explicit model parameter space. Figure~\ref{fig: U(1) projections} shows extrapolations for the HL-LHC, HE-LHC, and FCC-hh in the $(m_{V^0},g_V)$ plane for three particular models. In addition we show projected sensitivities to $W\,Y$ which may be produced on a similar timescale~\cite{deBlas:2016ojx,Farina:2016rws}. For the HL-LHC (top row) we project that when $g_V = g' \approx 0.36$, di-lepton searches will have a reach of 6--7\,TeV, di-boson searches 3--4\,TeV, and $t \bar t$ searches roughly 2.5--3.5\,TeV. Likewise for the HE-LHC (centre row) we expect a mass reach of around 12--14\,TeV for di-leptons, 6--9\,TeV for di-bosons, and 4--7\,TeV for $t \bar t$. The FCC-hh (bottom row) has an impressive reach in the respective channels of approximately 40\,TeV, 15--20\,TeV, and 6--18\,TeV.

%-------------------------------------------------
\subsection{Model D: New Non-abelian Gauge Symmetry}
\label{subsec:4.2}
%-------------------------------------------------

We now consider an explicit model with the gauge group $G = SU(3)_C \times SU(2)_L\times SU(2)_R\times U(1)_X$ where $X=(B-L)/2$ \cite{Schmaltz:2010p2610} and the corresponding gauge couplings are $g_s, \ g_L, \ g_R$ and $g_X$, respectively. We do not assume left-right symmetry, so that $g_{L}$ is not necessarily equal to $g_{R}$.  Using the notation $(SU(2)_{L}, SU(2)_{R})_{X}$, the SM fermions transform as $Q\sim \(\mathbf{2},\mathbf{1}\)_{1/6}$, $Q_{R}=\(U, D\)^{T} \sim\(\mathbf{1},\mathbf{2}\)_{1/6}$, $L \sim \(\mathbf{2},\mathbf{1}\)_{-1/2}$ and $L_{R}=\(N, E\)^{T} \sim \(\mathbf{1},\mathbf{2}\)_{-1/2}$, where we have introduced three generations of right-handed neutrinos, $N$. We consider the minimal scalar sector compatible with the breaking pattern $SU(2)_{R}\times U(1)_{X}\to U(1)_{Y}$, electroweak symmetry breaking and with dimension-four Yukawa interactions. This consists of two scalar multiplets: a bidoublet $\Phi$ transforming as  $(\mathbf 2, \mathbf 2)_{0}$,
\begin{align} 
    \Phi = \begin{pmatrix}
        \phi_1^0 & \phi_2^+ \\
        \phi_1^- & \phi_2^0 
    \end{pmatrix} \,,
\end{align}
where the superscripts indicate the electric charge, and a doublet $H_{R}$ transforming as $(\mathbf 1, \mathbf 2)_{1/2}$.\footnote{Note that minimal left-right symmetric models often contain two Higgs triplets instead of a doublet as this allows for a Majorana mass term for the right-handed neutrinos and small neutrino masses \cite{Barenboim:2001vu}.} The SM Higgs is identified with the $+1/2$ hypercharge component of the bidoublet after $SU(2)_{R} \times U(1)_X$ breaking, $H = (\phi_2^+, \phi_2^0)^T$. At the renormalizable level the Lagrangian contains the terms
\begin{equation}\label{lagwc}
\bry{lll}
\dst \mathcal L&\supset&\dst -\frac{1}{4} W^a_{L\mu\nu} W_L^{a\mu\nu}  -\frac{1}{4} W^a_{R\mu\nu} W_R^{a\mu\nu}-\frac{1}{4} X_{\mu\nu} X^{\mu\nu}
+\sum_{\Psi=Q,Q_{R},L,L_{R}}i\overline{\Psi} \slashed{D} \Psi \vspace{2mm} \\
&&\dst + \,(D_\mu H_{R})^\dagger D^\mu H_{R}+\tr\left[\(D_\mu \Phi\)^\dagger D^\mu \Phi\right]-  V(H_{R},\Phi) \,,
\ery
\end{equation}
where
\be
\bry{l}
\dst D_{\mu} \Psi=\demub \Psi-ig_{L,R}W_{L,R\,\mu}^{a}\f{\sigma^{a}}{2}\Psi-i g_{X}\f{B-L}{2} X_{\mu}\Psi\,,\vspace{1mm}\\
\dst D_{\mu} H_{R}=\demub H_{R}-ig_{L,R}W_{L,R\,\mu}^{a}\f{\sigma^{a}}{2}H_{R}-i \f{g_X}{2} X_{\mu}H_{R}\,,\vspace{1mm}\\
\dst D_{\mu}\Phi=\demub\Phi-ig_{L}W_{L\,\mu}^{a}\f{\sigma^{a}}{2} \Phi+ig_{R}\Phi W_{R\,\mu}^{a}\f{\sigma^{a}}{2}\,.
\ery
\ee
We assume that the scalar potential, $V(H_{R},\Phi)$, is such that the two scalar multiplets acquire the vevs\footnote{The vev of $\Phi$ could in general be of the form $\langle\Phi \rangle = \(\bry{cc} k & 0 \\ 0 & k' e^{i\alpha_{1}}\ery\)$. However, for simplicity, we consider $\alpha_{1}=0$ and $2k^{2}=2k^{\prime\,2}=v^{2}\approx (246\text{\,GeV})^{2}$.}
\be
\langle H_{R}\rangle = \f{1}{\sqrt{2}}\(\bry{c} 0 \\ v_{R}\ery\)\,,\qquad 
\langle \Phi \rangle = \f{1}{\sqrt{2}}\(\bry{cc} v & 0 \\ 0 & v\ery\)\,,
\ee
where $\langle H_{R}\rangle$ is responsible for the spontaneous breaking of $SU(2)_R\times U(1)_X\to U(1)_Y$ while $\langle \Phi \rangle$ breaks the electroweak symmetry. Note that the hypercharge in this model is given by $Y = T^3_R + X$ with $T^3_R = \sigma^3/2 = \text{diag}(1/2,-1/2)$. The mass terms of the heavy gauge bosons after $SU(2)_R \times U(1)_X $ breaking are given by (in the unitary gauge for the heavy fields)
\be\label{eq:massesmodelC}
\mathcal L_{\text{mass}}\supset\f{v_{R}^{2}}{4}\left[g_{R}^{2}W^+_{R\,\mu}W^{-\,\mu}_{R}+\f{1}{2}\(g_{R}W^3_{R\,\mu}-g_{X}X_{\mu}\)^{2}\right].
\ee
After $SU(2)_R \times U(1)_X \to U(1)_Y$ breaking but before the electroweak $SU(2)_L \times U(1)_Y$ breaking, \cref{eq:massesmodelC} allows us to identify the degrees of freedom transforming as the neutral and charged heavy vector singlets,
\be\label{eq:fieldredefmodelC}
\bry{l}
\dst g_{N}V^{0}_{\mu}=g_{R}W^3_{R\,\mu}-g_{X}X_{\mu}\,,\vspace{1mm}\\
\dst g_{C}V^{+}_{\mu}=g_{R}W^+_{R\,\mu}\,.
\ery
\ee
where we identify $g_{C}\equiv g_{R}$ and $g_{N}\equiv \sqrt{g_{X}^{2}+g_{R}^{2}}$.  The heavy vector masses are 
\be
m_{V^{+}}=\f{g_{C} v_{R}}{2}\,,\qquad m_{V^{0}}= \f{g_{N} v_{R}}{2}\,.
\ee
The remaining neutral combination
\be
g_{N}B_{\mu}=g_{X}W^3_{R\,\mu}+g_{R}X_{\mu} \,,
\ee
with the identification $1/g^{\prime\,2}\equiv 1/g_{X}^{2}+1/g_{R}^{2}$, remains massless and can be identified with the $U(1)_{Y}$ gauge field. 
We can then re-express the Lagrangian in terms of the fields $B_{\mu},V^{0}_{\mu},V^{+}_{\mu}$ using the field redefinitions
\be\label{eq:fieldredefmodelC2}
\bry{l}
\dst g_{N}X_{\mu}=g_{R}B_{\mu}-g_{X}V^{0}_{\mu}\,,\vspace{1mm}\\
\dst g_{N}W^3_{R\,\mu}=g_{X}B_{\mu}+g_{R}V^{0}_{\mu}\,,\vspace{1mm}\\
\dst g_{R}W^{+}_{R\,\mu}=g_{C}V^{+}_{\mu}\,.
\ery
\ee
Under these field redefinitions the field strength tensors become
\begin{equation}
\bry{ll}
\dst W_{R\,\mu\nu}^{\pm} &=\partial_{[\mu}V^{\pm}_{\nu ]} \mp i\f{g_{R}}{g_{N}} \(g_{X}B_{[\mu}+g_{R}V_{[\mu}^{0}\)V_{\nu]}^{\pm}\,, \\
\dst g_{N}W_{R\,\mu\nu}^{3} & =g_{X}B_{\mu\nu}+g_{R}\(\partial_{[\mu}V^{0}_{\nu ]}-ig_{N}V_{[\mu}^{+}V_{\nu]}^{-}\)\,,
\vspace{2mm}\\
\dst g_{N}X_{\mu\nu} &= g_{R}B_{\mu\nu}-g_{X}\partial_{[\mu}V^{0}_{\nu ]}\,.
\ery
\end{equation}
Identifying
\begin{equation}
g=g_{L} \,, \qquad
g_{V}=g_{R}\,, \qquad
\frac{1}{g'^{2}}=\frac{1}{g_{X}^{2}}+\frac{1}{g_{R}^2}\,,
\end{equation}
where $g$ is the usual $SU(2)_L$ coupling and $g'$ is the usual $U(1)_Y$ coupling, and defining $k_{V}=\sqrt{1-g^{\prime\,2}/g_{V}^{2}}$, we can match this model onto our simplified model Lagrangian in \cref{sml} as summarised in \cref{Table:matching}.

We can now study the collider phenomenology of this model.  Note that in addition to the charged and neutral vectors, this model contains a second fairly light Higgs boson which is generally phenomenologically excluded.  However, it can decouple for zero $CP$ phases and a fine-tuned choice of couplings in the scalar sector \cite{Barenboim:2001vu}.  In this section, we only consider the phenomenology of the heavy vectors.

First, we can immediately use the results from \cref{sec:data-bounds} to find LHC limits on Model D. \Cref{Fig:BoundsCharged1} shows current limits on $V^+$ at $m_{V^+}=2$, 3 and 5\,TeV, with lines representing the matching relations of Model D, for different values of $g_V$. 
At $m_{V^+} \lesssim 3\,\text{TeV}$, Model D is excluded by $t \bar b$ searches for all couplings $g_V > g'$, but at $m_{V^+} = 5\,\text{TeV}$ searches targeting the charged vector cannot place any constraints.  We see that if a new $t\bar b$ search presented their results as a sequence of panels of exclusion contours for different values of $m_{V^+}$, this could easily be used to place limits on a range of models, as we have done with Model D here. 

Since Model D contains both a charged and a neutral heavy vector, we can also look at limits from searches for a neutral singlet, \cref{fig:BoundsNeutral4x3}, where di-lepton and di-boson searches provide the strongest bounds. In the top-right panel of \cref{fig:BoundsNeutral4x3}, we see that $|g_V| = 0.61$ is not quite excluded by di-lepton searches.  For this parameter point $m_{V^0}=5$\,TeV corresponds to $m_{V^+} \approx k_V m_{V^0} \approx 4.1\,\text{TeV}$, so it is hard to tell from these plots which search is most constraining on this model in this region. Since the limits are only weakly dependent on $\lambda_n$, we can however see that despite the large parameter space, for the parameter points shown in the right-hand panels of \cref{fig:BoundsNeutral4x3} Model D is almost excluded by the di-lepton search. 

%-------------------------------------------------------------
\begin{figure}
    \centering
    \includegraphics[width=0.49\textwidth]{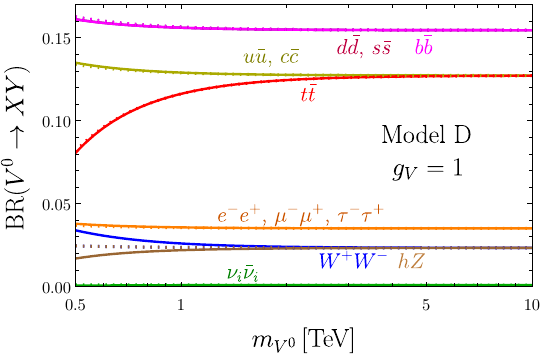}
    \includegraphics[width=0.49\textwidth]{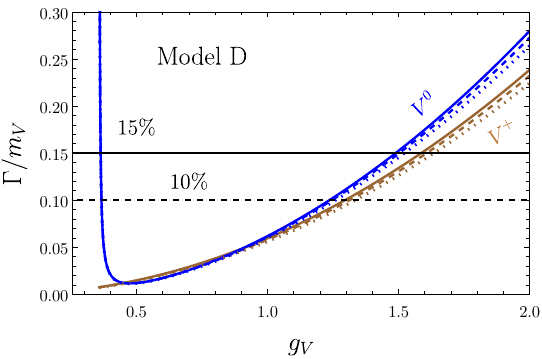}
    \caption{\small Model D branching ratios for the two body decays of the neutral vector $V^{0}$ (left) for a coupling $g_V=1$, and the width-to-mass ratio as a function of $g_V$ (right) for both the charged (brown) and neutral (blue) vectors. In the left panel, the solid lines show our full numerical expressions for the BRs, while the dotted lines correspond to the approximate BRs. In the right panel, the solid lines show $\Gamma/m_V$ for $m_V=5$\,TeV, while the dotted and dashed lines correspond to 500 and 750 GeV, respectively.}
    \label{fig:BRs-WtoM-ModelD}
\end{figure}
%-------------------------------------------------------------

We now compare the simplified model to a more detailed analysis of the explicit model.  In \cref{fig:BRs-WtoM-ModelD} (left) we show the branching ratios of the neutral heavy vectors (left) in Model D.  We see that for $g_V=1$ the branching ratios are similar to the case of universal couplings, \cref{fig:BRs-universal}.  In Model D the branching ratio to down-type quarks is slightly higher than that to up-type quarks, as their hypercharges mean that $c_d^{\text{eff}}$ is larger than $c_u^{\text{eff}}$ (see \cref{Table:matching}).  Also, the branching ratios to bosons (the $W^+W^-$ and $Zh$ channels) are slightly enhanced while the $\nu\bar{\nu}$ channel is suppressed. We do not show the branching ratios for the charged vector since in Model D, even after mixing with the SM gauge bosons, the charged vector only couples to $u \bar d, c \bar s$ and $t \bar b$ ($c_H^+ = 0$ in this model).  Since it couples to these pairs with equal strength, the branching ratio to each pair is just 1/3 (ignoring top quark mass effects).

In \cref{fig:BRs-WtoM-ModelD} (right) we show the width-to-mass ratio as a function of $g_V$ for the neutral and charged resonances in Model D. For the charged vector, the narrow width approximation applies in this model for $g_V \lesssim 1.6$, while for the neutral vector it is valid in the range $0.4 \lesssim g_V \lesssim 1.5$. Compared to the case of universal couplings, the charged heavy vector is slightly narrower, while the neutral vector is narrower at large $g_V$ and significantly broader for $g_V \lesssim 0.4$, where the width quickly becomes very large as $k_V \to 0$. 

To demonstrate the power of the simplified model, we derive the exclusion limits in the explicit model. We again use the \texttt{NNPDF} parton distribution functions to construct the parton luminosities, compute the HVS production cross-sections and branching ratios, and finally compare $\sigma \times \text{BR}$ to the experimental limits in Model D's parameter space.

%-------------------------------------------------------------
\begin{figure}
    \centering
    \includegraphics[width=0.49\textwidth]{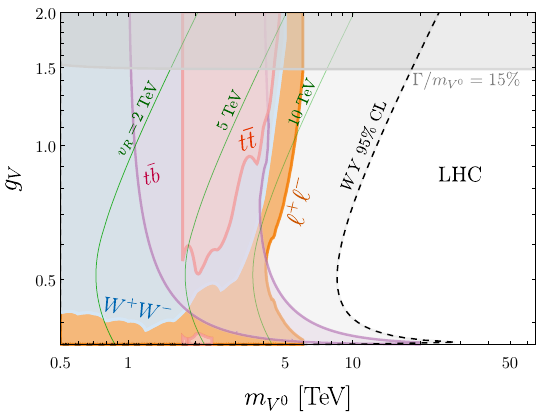}
    \includegraphics[width=0.49\textwidth]{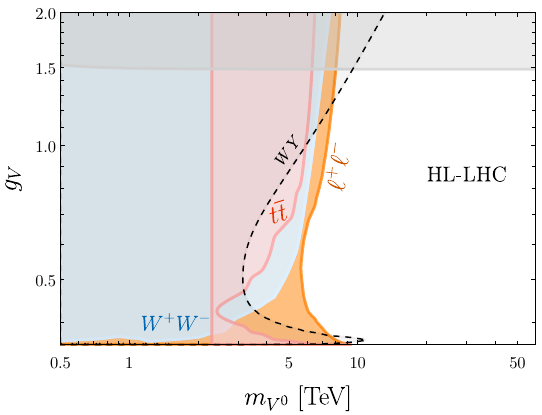}
    \includegraphics[width=0.49\textwidth]{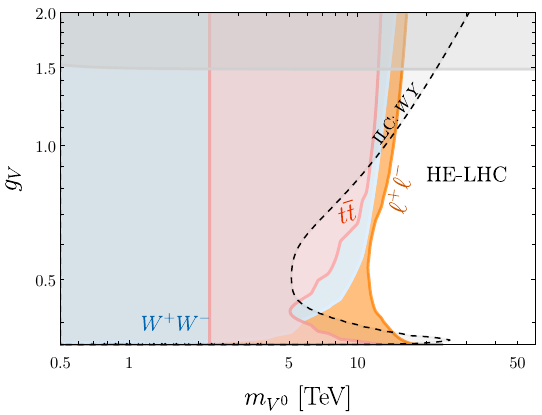}
    \includegraphics[width=0.49\textwidth]{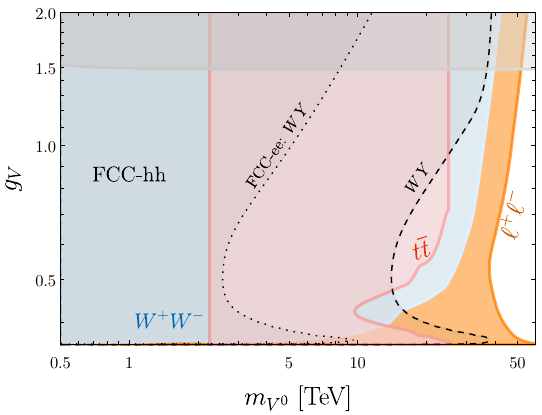}
    \caption{\small Experimental limits in the $(m_{V^0},g_V)$ plane for Model D. In the upper left panel the coloured regions represent the combined di-lepton search~\cite{ATLAS:2019erb} (orange), the semileptonic di-boson search \cite{ATLAS:2020fry} (blue), the $t \bar t$ search \cite{ATLAS:2020lks} (red) and the $t \bar b$ search \cite{CMS:2021mux} (purple). The green contours correspond to fixing the vev of $H_R$, $v_R$, to the values of 2, 5, and 10\,TeV. The remaining panels show the extrapolations of the neutral searches to the HL-LHC (upper right), the HE-LHC (bottom left) and the FCC-hh (bottom right). The dashed black contours depict current and projected EWPT constraints on $W$ and $Y$ at $95\%$ CL. The grey region indicates where $\Gamma/m_{V^0} \geq 15\%$.}
    \label{fig:non-abelianExtension}
\end{figure}
%-------------------------------------------------------------

\Cref{fig:non-abelianExtension} shows the exclusion regions of the most stringent di-lepton (orange), di-boson (blue) and $t \bar t$ (red) searches on Model D in the $(m_{V^0},g_V)$ parameter space. Since the charged and neutral vector masses are related, we also show the exclusion region of the $t \bar b$ search (purple).  The upper left panel shows current LHC constraints, and the remaining panels show our projections to the HL-LHC, HE-LHC and FCC-hh. At the LHC, we see that, depending on $g_V$, masses up to 4--6\,TeV are excluded by the di-lepton search, while masses up to 4--5\,TeV are reached by the di-boson and $t \bar t$ searches. The heavy vector singlet coupling to gauge bosons goes to zero for $g_V \to g' \sim 0.36$, as $k_V \to 0$ in this limit, and the di-boson searches lose sensitivity. In the same limit, the fermion couplings become very large, leading to sensitive constraints in the di-lepton channel. Because of this behaviour, $\sigma\times$BR in the di-fermion channel grows large both for $1 \lesssim g_V$ and $g_V\rightarrow g'$, and has a minimum around $g_V\simeq 0.53$ (for a fixed mass). 
At large $g_V$, where $m_{V^+} \approx m_{V^0}$, the $t \bar b$ search reaches $m_{V^0} \approx 4\,\text{TeV}$, but as $g_V \to g'$ this search becomes very sensitive, since in this regime $m_{V^+} \approx k_V m_{V^0} \ll m_{V^0}$.  For a fixed $m_{V^0}$, the charged vector becomes very light so gives strong constraints on the model. 
The green contours show three different benchmark values for $v_R$ in this space.  We see that $v_R \lesssim 10$\,TeV is almost entirely excluded.  
The dashed black contours show EWPT constraints at $95\%$ confidence level for $W\,Y$. Again we have no constraints for $S\,T\,U$, since the charged lepton Yukawa interaction is invariant under $U(1)_{(B-L)/2}$, i.e., $c_L^0 - c_E^0 - c_H^0 =0$ \cite{Strumia:2022qkt}.

We can now compare the limits obtained using the simplified model framework to those shown in \cref{fig:non-abelianExtension}.  This is complicated by the fact that there are relevant limits on both the charged and neutral vector, and that the ratio of the masses depends on $g_V$. 
Even though it is difficult to directly compare the different presentations, the same picture emerges.  Searches for $t \bar b$ exclude Model D for all $g_V$ for $m_{V^+} \lesssim 3\,\text{TeV}$ (recalling that $m_{V^+} = k_V m_{V^0}$), while the di-lepton searches are close to excluding $m_{V^0} = 5\,\text{TeV}$ with $0.43 \lesssim g_V \lesssim 0.61$.

Finally, we discuss the remaining panels in \cref{fig:non-abelianExtension}, which show the expected sensitivities of the 14\,TeV HL-LHC (upper right), the 27\,TeV HE-LHC (lower left) and the 100\,TeV FCC-hh (lower right).  We project that while higher luminosities will not dramatically increase the mass reach, higher centre-of-mass energies of 27 and 100\,TeV will probe masses around 16\,TeV and 50\,TeV, respectively, and surpass the indirect constraints. Note that the current EWPT constraints at the LHC are stronger in some or most regions of parameter space than the future projections. This is unrealistic and reflects the fact that the future projections are out-of-date and should be updated.

%-------------------------------------------------
\subsection{Model E: Minimal Composite Higgs Model}
\label{subsec:4.3}
%-------------------------------------------------

In composite Higgs models the Higgs boson is naturally light as it is a pseudo-Nambu Goldstone boson, which appears due to the spontaneous breaking of a global symmetry.  In the case of a minimal composite Higgs model, which we study here, there is a global $SO(5)$ symmetry which is broken to $SO(4) \simeq SU(2)_{L}\times SU(2)_{R}$ \cite{Contino:2011np,Contino:2013un}.  The $SU(2)_{L}$ is gauged and is the SM $SU(2)_{L}$ symmetry, while the $SU(2)_R$ will be broken to yield the SM gauge group.  In addition to the SM particles, these models predict a new heavy and strongly coupled sector, which is governed by this global symmetry.  
The underlying symmetry structure of the strong dynamics is described by the Callan-Coleman-Wess-Zumino (CCWZ) formalism \cite{Coleman:1969sm,Callan:1969sn}, and we here follow the discussion in Appendix A of Ref.~\cite{DeSimone:2012ul}.  Ultimately, the lightest particles in the new sector may be vector resonances, which can be described by the simplified model presented in this paper.

Following the $SO(5)$ breaking, there is a right-handed heavy vector triplet $\rho_\mu$ which transforms in the $(\mathbf{1},\mathbf{3})$ (irreducible) representation of the custodial $SU(2)_{L}\times SU(2)_{R}$ symmetry 
\begin{equation}\label{rhotrans}
{\rho}_\mu\equiv {\rho}_\mu^{a_{R}} t_{R}^{a_{R}}\;\rightarrow\;h_4{\rho}_\mu h_4^T-ih_4\partial_\mu h_4^T~~~{\textrm{for}}~~~a_{R}=1,2,3\,, 
\end{equation}
where $t_{R}^{a_{R}}$ are the generators of the vector representation of the $SU(2)_{R}$ subgroup of $SO(4)$, and $h_4$ is a non-linear $SO(4)$ transformation whose construction is described in Appendix A of Ref.~\cite{DeSimone:2012ul}. When the $SU(2)_R$ is broken, $\rho_\mu$ will lead to charged and neutral heavy vector singlets.  The vector $\rho_\mu$ is the right-handed counterpart of the left-handed heavy vector triplet $\rho_{L}$, which transforms in the $(\mathbf{3},\mathbf{1})$ representation. Note that unless some parity symmetry which exchanges the two $SU(2)$ symmetries is present, the masses of the left- and right-handed triplet are independent parameters. This means that one of the two could be much lighter than the other, so it makes sense to study these two representations in isolation. For a comprehensive study of the two representations together in the context of the minimal composite Higgs model see Ref.~\cite{Brooijmans:2014eja}, and for a detailed discussion of the phenomenology of $\rho_{L}$ and its matching onto a simplified Lagrangian, see \cite{Pappadopulo:2014tg}.

We then consider the following Lagrangian
\begin{equation}
\label{rholag}
    \mathcal L_{\rho}=
    -\frac{1}{4\hat g'^{2}}(B_{\mu\nu})^{2}
    -\frac{1}{4 \hat g^{2}}(W^{a_{L}}_{\mu\nu})^{2}
    -\frac{1}{4 g_\rho^{2}}({\rho}_{\mu\nu}^{a_{R}})^{2}+\frac{f^{2}}{4} d_\mu^{i}d^{\mu i}
    +\frac{m_{\rho}^{2}}{2g_\rho^{2}}\left({\rho}_\mu^{a_{R}}-e_\mu^{a_{R}}\right)^{2}\,,
\end{equation}
where $B_{\mu\nu}$ is the field strength of the gauge boson associated with the unbroken $U(1)_Y$ subgroup after $SU(2)_R$ breaking, $W_{\mu\nu}^{a_L}$ is the field strength of the gauge boson associated with $SU(2)_L$, and the field strength of the heavy vector is given by $\dst {\rho}_{\mu\nu}^{a}= \partial_\mu {\rho}_{\nu}^{a}- \partial_\nu {\rho}_{\mu}^{a}-\epsilon^{abc}{\rho}_{\mu}^{b}{\rho}_{\nu}^{c}$. The full expressions for the covariant variables $d$ and $e$ for $SO(5)/SO(4)$ can be derived in the CCWZ formalism and are given in Appendix A of Ref.~\cite{DeSimone:2012ul} and in \cite{Brooijmans:2014eja}. Here we only need approximate formulae in the large $f$ limit,
\begin{align}
\label{dmudmuANDemu}
    d_\mu^i d^{\mu\, i} &= 
    \frac{4}{f^{2}}|D_\mu H|^{2} 
    + O\left(\frac{1}{f^4} \right) \, , 
    \\
    \rho^{a_{R}}_\mu- e_\mu^{a_{R}} &=
    \rho_{\mu}^{a_{R}}
    + B_\mu \delta^{a_{R},3}
    -\frac{i}{f^{2}}J_{\mu}^{a_{R}} 
    + O\left(\frac{1}{f^4} \right)\,,
\end{align}
with
\be
J_{\mu}^{1}=-\f{J_{\mu}^{+}+J_{\mu}^{-}}{2\sqrt{2}}\,,\qquad J_{\mu}^{2}=-i\f{J_{\mu}^{+}-J_{\mu}^{-}}{2\sqrt{2}}\,,\qquad J_{\mu}^{3}=J_{\mu}^{0}\,,
\ee
where
\begin{equation}
J_{\mu}^{+}=\f{1}{\sqrt{2}} H^{c\,\dagger} {\overset{{}_{\leftrightarrow}}{D}}_{\mu} H\,,\qquad J_{\mu}^{-}=\f{1}{\sqrt{2}} H^{\dagger} {\overset{{}_{\leftrightarrow}}{D}}_{\mu} H^{c}\,,\qquad J_{\mu}^{0}=\f{1}{2} H^{\dagger} {\overset{{}_{\leftrightarrow}}{D}}_{\mu} H \,.
\end{equation}
We can then define a right-handed triplet $V_{\mu}^{a_{R}}$, which does not transform under the SM gauge group, as 
\begin{equation}
V_\mu^{a_{R}}\equiv\rho_{\mu}^{a_{R}}+B_\mu \delta^{a_{R},3}.
\end{equation}
In the charge eigenstate basis this field redefinition is
\begin{equation}
V_\mu^{\pm}\equiv\rho_{\mu}^{\pm}\,,\qquad V_\mu^{0}\equiv\rho_{\mu}^{3}+B_\mu.
\end{equation}
After this field redefinition, the field strength becomes
\begin{equation}
\rho_{\mu\nu}^{\pm}=\partial_{[\mu}V^{\pm}_{\nu ]}\pm i\(V_{[\mu}^{0}- B_{[\mu}\)V_{\nu]}^{\pm} \,,\qquad \rho_{\mu\nu}^{3}=\partial_{[\mu}V^{0}_{\nu ]}+iV_{[\mu}^{+}V_{\nu]}^{-}-B_{\mu\nu} \,.
\end{equation}
Using the large $f$ expansion, as used in \cref{dmudmuANDemu}, we can now match $\mathcal L_\rho$ with the ``tilde'' basis of \cref{AppA}, leading to
\begin{equation}
g_{V}=g_\rho,~~~\frac{1}{g'^{2}}=\frac{1}{\hat g'^{2}}+\frac{1}{g_\rho^{2}}~~~{\textrm{and}}~~~g=\hat g \,.
\end{equation}
After canonically normalizing the kinetic terms of the neutral and charged heavy vectors we obtain
\begin{equation}
\label{mathchingNSM}
\bry{l}
\dst \tilde m_{V^{0}}=\tilde m_{V^{+}} =m_\rho, \quad  \tilde c_{V B}^{0}=\tilde c_{VV B}^{+}=-\tilde c_{VVV}^{0}=-\tilde c_{VVV}^{+}=1, \vspace{2mm}\\
\dst \tilde c_{H}^{0}=2c_{H}^{+}=-\frac{m_\rho^{2}}{g_\rho^{2} f^{2}}\equiv - a_\rho^{2},\quad\tilde c_{VV HH}^{0}=c_{VV HH}^{+}=\tilde c_{\Psi}^{0}=\tilde c_{q}^{+}=0,
\ery
\end{equation}
where $a_\rho$ is an $O(1)$ free parameter as defined in Ref.~\cite{Contino:2011np}.
We can now use \eq{cnontilde} to convert these parameters into the basis without any kinetic mixing terms. The results are shown in \cref{Table:matching}.

We see that the matching relations in this strongly coupled model differ from those for the weakly coupled Model D discussed in \cref{subsec:4.2}. The main difference lies in the couplings of the charged heavy vector, which couples to the Higgs current but not directly to the fermionic current.  The fermion couplings of the neutral vector are now all proportional to the hypercharge, and note that while all couplings of the neutral vector are proportional to $1/\kappa_V$ (or $1/\kappa_V^2$), if the free parameter $a_\rho \sim 1$, then $c_H^0 \sim -k_V$ and $c_{VVHH}^0 \sim -g'^2/2g_V^2$.

Note that in order to perform the matching we ignored both higher-dimensional operators coming from subleading corrections to \cref{dmudmuANDemu} and higher derivative terms which could be added to the Lagrangian in \eq{rholag}. For a detailed discussion of higher-dimensional operators see, e.g., \cite{Contino:2011np,Pappadopulo:2014tg}.

%-------------------------------------------------------------
\begin{figure}
    \centering
    \includegraphics[width=0.49\textwidth]{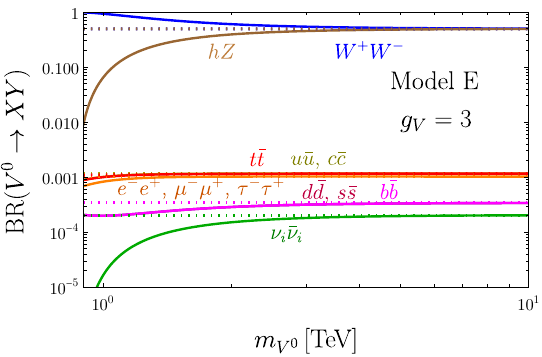}
    \includegraphics[width=0.49\textwidth]{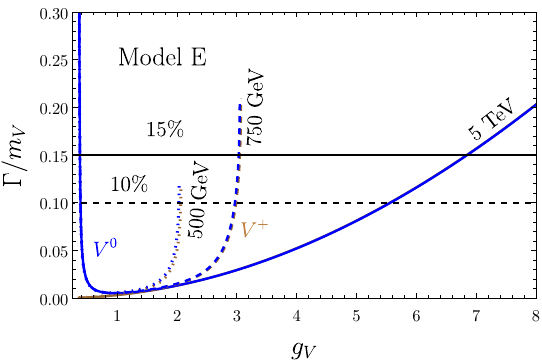}
    \caption{\small Model E branching ratios for the two body decays of the neutral vector $V^{0}$ (left) for a coupling of $g_V=3$ and $a_\rho = 1$, and the width-to-mass ratio as a function of $g_V$ (right) for both the charged (brown) and neutral (blue) vectors. In the left panel, the solid lines show our full numerical expressions for the branching ratios, and the dotted lines show the approximate  branching ratios. In the right panel, the solid lines show $\Gamma/m_V$ for $m_V=5$\,TeV, while the dotted and dashed lines correspond to 500 and 750 GeV, respectively. The lower-mass curves terminate around $g_V=2$ and $g_V=3$ where the model becomes theoretically excluded.}
    \label{fig:BRs-WtoM-ModelE}
\end{figure}
%-------------------------------------------------------------

We now turn to the phenomenology of this model and make use of the results from \cref{sec:data-bounds} to find the current LHC limits. From \cref{Table:matching} we see that the ratios $\lambda_{d,e,n}$ are independent of the resonance mass and $g_V$. They are fixed to $\lambda_d = 0.54$, $\lambda_e = 1.63$, and $\lambda_n = \lambda_e/\sqrt{5}$. Choosing these values on the grid of plots in \cref{fig:BoundsNeutral4x3} allows us to represent Model E as a line in the simplified model parameter space, reflecting the fact that $g_V$ is still a free parameter after fixing $\lambda_e$ and $\lambda_d$. We also set $a_\rho = 1$.  The most stringent constraint on Model E comes from di-lepton searches which limit $g_V c_u^\text{eff} < 0.35$ for $m_{V^0} = 5$\,TeV indicating that $g_V > 0.61$.

We now compare the simplified model to a more detailed analysis of the explicit model. For small values of $g_V$ the phenomenology is similar to Model D, but in \cref{fig:BRs-WtoM-ModelE} we show the branching ratios of the neutral vector with  $g_V = 3$ (we expect $g_V \gtrsim 1$ in strongly coupled models) and $a_\rho = 1$ (left) and the total widths of both vectors (right) in this model.  We see that in this case the neutral vector predominantly decays to di-bosons, and all other decays are at the per mille level or lower.  This is in contrast to the other models we have looked at, where the leading branching ratios have been to quarks.  This suppresses the production cross-section of the neutral singlet, and boosts the relative importance of the di-boson search channels.  In  \cref{fig:BRs-WtoM-ModelE} (right) we see that for multi-TeV vectors, the narrow width approximation holds for $g_V \lesssim 7$, but breaks down at lower values of $g_V$ for lighter resonances. We do not show the branching ratios for the charged vector. Its coupling to quarks is zero, $c_q^+ =0$, which implies a negligible DY production cross-section. Production via vector boson fusion can be relevant but is left for future work.

%-------------------------------------------------------------
\begin{figure}
    \centering
    \includegraphics[width=0.49\textwidth]{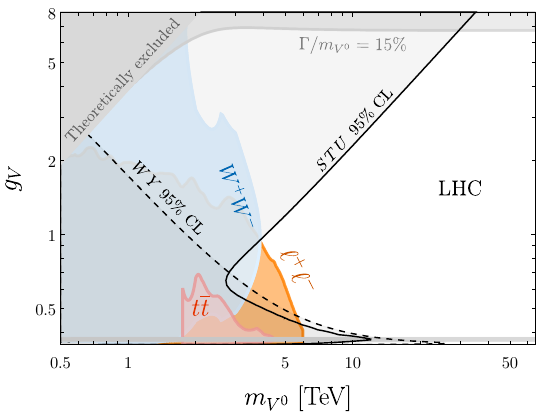}
    \includegraphics[width=0.49\textwidth]{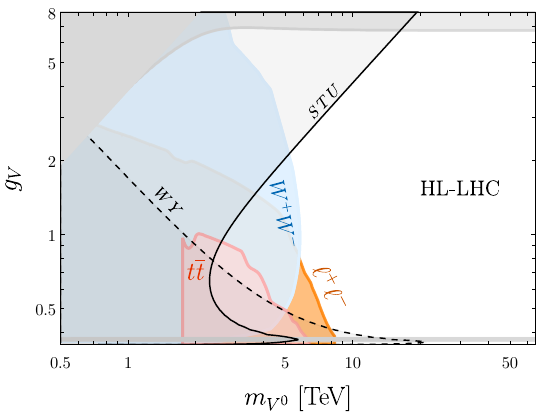}
    \includegraphics[width=0.49\textwidth]{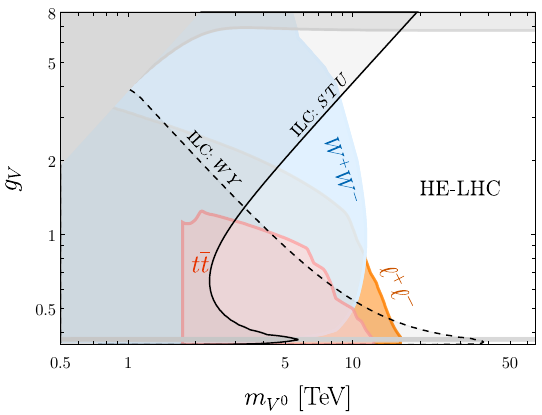}
    \includegraphics[width=0.49\textwidth]{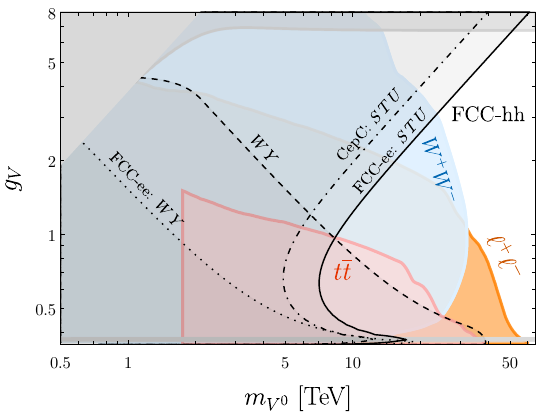}
    \caption{\small Experimental limits in the $(m_{V^0},g_V)$ plane for Model E. In the upper left panel we show the combined di-lepton search~\cite{ATLAS:2019erb} (orange), the semileptonic di-boson search \cite{ATLAS:2020fry} (blue), and the $t \bar t$ search \cite{ATLAS:2020lks} (red). The remaining panels show the extrapolations of the searches to the HL-LHC (upper right), the HE-LHC (bottom left) and the FCC-hh (bottom right). The black contours depict current and projected EWPT constraints on $S$, $T$, and $U$ (solid), and on $W$ and $Y$ (dashed), at $95\%$ CL.  The solid grey region in the upper-left is theoretically excluded and the grey contour shows the boundary where $\Gamma/m_{V^0}=15\%$.}
    \label{fig:stronglyCoupledExtension}
\end{figure}
%-------------------------------------------------------------

Again we go through the process of setting limits on this particular model. 
 \Cref{fig:stronglyCoupledExtension} shows the excluded parameter space in the $(m_{V^0},g_V)$ plane for the same searches as above (di-lepton, di-boson, and $t \bar t$). The upper left panel corresponds to the current LHC limits. In the strongly coupled region, $1 \lesssim g_V$, the strongest direct search is the $W^+W^-$ channel which reaches up to around 4\,TeV.  This gets weaker as $g_V$ increases from 1, since when $a_\rho = 1$ the cross-section times branching ratio is proportional to $1/g_V^2$.  The narrow width approximation starts to fail as $g_V$ reaches around 7 (and at lower values for $m_{V^0} \lesssim 2$\,TeV).  For these large values of $g_V$, perturbativity also starts to break down (not shown).  Beyond the direct collider limits, we see that the indirect EWPTs provide very strong constraints on this model at large $g_V$, extending beyond 10\,TeV.  However, since this constraint can be modified by further heavy particle content in a complete model, which we expect in a composite model, it is not as robust as the direct constraints, which would only change with additional light particles. For completeness we also show the limits at $g_V \lesssim 1$.  Of the direct searches, the di-lepton search is the most constraining, as it was for Model D.  We see that in this regime EWPTs also provide reasonably strong indirect constraints via $W\,Y$, although di-lepton constraints are still stronger for $g_V\gtrsim0.5$.  Note that this plot also confirms the conclusions we drew from \cref{fig:BoundsNeutral4x3}. At 5\,TeV the di-lepton search sets the most stringent constraint, and direct searches allow $g_V > 0.61$.

We also show our extrapolations of the direct limits to the HL-LHC, HE-LHC and FCC-hh in the remaining panels of \cref{fig:stronglyCoupledExtension}, along with projected EWPT constraints at various future colliders. We see that future direct searches will be able to probe the region that is unconstrained by EWPTs, where the di-boson and di-lepton limits reach 5--9\,TeV, 10--17\,TeV and 30--50\,TeV, respectively. For large couplings, $g_V\gtrsim 3$, these colliders could only exclude narrow $V^0$ resonances below 3\,TeV, 6\,TeV and 14\,TeV, respectively, so EWPTs still provide a useful complement to direct searches. 
The current EWPT constraints at the LHC are again stronger in some or most regions of parameter space than the future projections, reflecting the fact that the future projections are out-of-date and should be updated.

%-------------------------------------------------
\subsection{Summary}
%-------------------------------------------------

Overall, we see that while a model specific analysis can give more information on any particular model, experimental results presented in the parameter space of the simplified model can give useful information on a wide range of models and allow for a quick comparison of any specific model to the data. The simplified model approach works particularly well for the charged singlet. Since the charged parameter space is smaller (there are only two coupling parameters) it is easier and very useful to present searches for the charged vector on a simplified model parameter space.  

The large number of free parameters makes this approach more challenging for the neutral singlet.
However, with the effective parameters we defined and noting the weak dependence on $\lambda_n$ of the leading di-lepton and di-boson constraints, the neutral vector simplified model can also be useful. Experimental results could be presented in a set of plots for, e.g., $\lambda_e = 0.5, 1, 1.5, 2$ with the limits labeled by the resonance mass and $\lambda_d$. While this may not always give an accurate comparison of an explicit model to the data, it can give a quick and rough idea of the viable parameter space in a certain model. If a parameter point of particular interest in an explicit model lies close to the boundary of an experimental limit, a detailed recasting of the experimental analysis would be required. 

%=================================================
\section{Relation to the Heavy Vector Triplet Model}
\label{sec:hvt}
%=================================================

In this section we explore the relation between this model of heavy vector singlets and the HVT model, discussed in Ref.~\cite{Pappadopulo:2014tg}. In particular, we discuss two matching procedures to relate the parameter space of the singlet and the triplet. Although the parameter space of the HVS is significantly larger than that of the HVT, we show that in certain cases experimental limits derived for the HVT can give (perhaps rough) limits on the HVS. In what follows, HVT parameters from Ref.~\cite{Pappadopulo:2014tg} will be labeled with a `T' superscript and HVS parameters, where appropriate, with an `S' superscript.

%-------------------------------------------------
\subsection{Exact Matching at the Lagrangian Level}
\label{Exact HVT matching}
%-------------------------------------------------

The most direct way to relate the two models is to compare the parameters at the level of the Lagrangian. For limits depending on Drell-Yan production, exact matching cannot be done for the charged vectors since the two models can only be related if the fermion couplings are zero,
\be\label{eq:Charged fermion matching}
    c_F^\text{T} = c_q^+ = 0 \,.
\ee
This is because the vectors couple to either right- or left-handed currents (that is, currents of the form $\overline{u_R} \gamma^\mu d_R$ or $\overline{f_L} \gamma^\mu \tau^a f_L$) in the singlet and triplet Lagrangian, respectively. 
Although non-zero couplings to the Higgs current can be related,
\be
\label{eq: Exact matching charged}
    (c_H^\text{T})^2 = (2c_H^+)^2 \,,
\ee
this term alone leads to mixing suppressed Drell-Yan production which is not large enough to provide relevant limits.

%-------------------------------------------------------------
\begin{figure}
\begin{center}
    \includegraphics[width=0.5\textwidth]{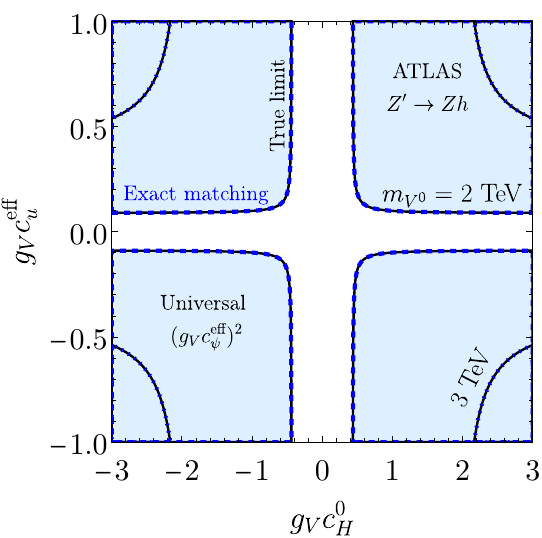}
    \caption{\small Di-boson limits initially obtained for the HVT (taken from figure 5 of Ref.~\cite{ATLAS:2017ptz}) presented in the $(g_Vc_H^0,g_Vc_u^\text{eff})$ plane of the HVS parameter space for $m_{V^0} = 2$\,TeV (blue, dashed) and 3\,TeV (blue, dotted) using the exact mapping following \cref{Exact HVT matching}.  The solid black lines show the exact limit obtained by computing $\sigma \times \text{BR}$ in this model.
    }
    \label{fig: HVT Matching 1}
\end{center}
\end{figure}
%-------------------------------------------------------------

For the neutral vector, the fermion couplings can be non-zero. Since the HVS parameter space is substantially larger than that of the HVT, we make use of the effective parameter space defined in \cref{sec: Ch3 neutral vector}, and relate these effective couplings to the HVT parameters. Matching the Lagrangians leads to the following relations,
\be \label{eq: Exact matching HVT/HVS}
\bry{lll}
(g^2c_q^\text{T}/g_V)^2 &= (g_Vc_u^\text{eff})^2 = (g_Vc_d^\text{eff})^2\,, \vspace{1mm} \\
(g^2c_{q3}^\text{T}/g_V)^2 &= (g_Vc_{u3}^\text{eff})^2 = (g_Vc_{d3}^\text{eff})^2\,, \vspace{1mm} \\
(g^2c_l^\text{T}/g_V)^2 &= (g_Vc_e^\text{eff})^2 = (g_Vc_L^0)^2\,, \vspace{1mm} \\
(g_Vc_H^\text{T})^2 &= (g_Vc_H^0)^2 \,,
\ery
\ee
which lead to the requirements,
\be \label{eq: HVS matching restrictions}
(c_U^0)^2 = (c_D^0)^2\,, \qquad
(c_E^0)^2 = 0\,.
\ee
If these conditions are satisfied, the translation of experimental bounds from the HVT to the HVS parameter space is exact. 
In \Cref{fig: HVT Matching 1} we take the di-boson limits obtained in \cite{ATLAS:2017ptz}, which were presented in the $(c_H^T, c_F^T)$ parameter space, and in the dashed and dotted blue lines we directly translate them onto the effective HVS parameter space. Ref.~\cite{ATLAS:2017ptz} assumes a universal HVT coupling to fermions $c_F^\text{T}$, so we show the limits for universal effective couplings $c_\psi^\text{eff}$ in the HVS parameter space.
We also show the exact exclusion after computing the limit on $\sigma \times \text{BR}$ in solid black, and we see that they agree exactly with the limits taken from the HVT exclusions.

One limitation of this mapping is that the above conditions need to be satisfied in any simplified or explicit model. One explicit singlet model that satisfies these requirements is the $U(1)_B$ model (see \cref{Table:matching}). In that case, experimental limits on the HVT parameters can be directly mapped onto the parameters of the $U(1)_B$ model. 
However, since there are only a few models for which \cref{eq: HVS matching restrictions} holds, we now discuss a more general numerical matching.

%-------------------------------------------------
\subsection{Approximate Matching at the Level of $\sigma \times \text{BR}$}
\label{subsec: Numerical matching}
%-------------------------------------------------

A more general method to map the HVS model onto the HVT model is to compare the production cross-section times branching ratio on a channel-by-channel basis. If we set $(\sigma \times \text{BR})^\text{T} = (\sigma \times \text{BR})^\text{S}$ separately for di-jet, di-lepton, di-boson, and heavy di-quark final states, the couplings of the HVT may be expressed in terms of the HVS parameters and vice versa.\footnote{Motivated by explicit models, the fermion couplings of the HVT contain the factor $g^2/g_V$, as opposed to the factor $g_V$ in the HVS. Of course, one can absorb $g_V$ into the couplings such that $g^2c_F^\text{T}/g_V = g_F^\text{T}$ and $g_V c_\Psi^\text{S} = g_\Psi^\text{S}$, and this is already done for many resonance searches.} 

%-------------------------------------------------------------
\begin{table}
    \fontsize{11}{11}\selectfont
    \centering
    \renewcommand{\arraystretch}{4}
    \begin{tabular}{c|c}
    \toprule \hline 
    \parbox{2cm}{\centering $V$ Decay Channel} & Matching relations \\
    \hline
    $jj$ & $\dst \(\frac{g^2c_q^\text{T}}{g_V}\)^4 = g_V^4 \dfrac{\Gamma_\text{tot}^\text{T}}{\Gamma_\text{tot}^\text{S}} (c_q^+)^4$ \\
    $t \bar b$ & $\dst \(\frac{g^4 c_q^\text{T} c_{q3}^\text{T}}{g_V^2}\)^2 = g_V^4 \dfrac{\Gamma_\text{tot}^\text{T}}{\Gamma_\text{tot}^\text{S}} (c_q^+)^2(c_{q3}^+)^2$ \\
    
    $W_L^+Z_L/W_L^+h$ & $\dst (g^2 c_q^\text{T}c_H^\text{T})^2 = g_V^4 \dfrac{\Gamma_\text{tot}^\text{T}}{\Gamma_\text{tot}^\text{S}} 4(c_q^+c_H^+)^2$ \\

    \hline 
    
    $jj$ & $ \dst\(\frac{g^2c_q^\text{T}}{g_V}\)^4 = g_V^4 \dfrac{1}{2} \dfrac{\Gamma_\text{tot}^\text{T}}{\Gamma_\text{tot}^\text{S}} \dfrac{(c_u^\text{eff})^2 \(\frac{dL_{u\bar{u}}}{d\hat{s}}\) + (c_d^\text{eff})^2 \(\frac{dL_{d\bar{d}}}{d\hat{s}}\)}{\frac{dL_{u\bar{u}}}{d\hat{s}} + \frac{dL_{d\bar{d}}}{d\hat{s}}} \( (c_u^\text{eff})^2 + (c_d^\text{eff})^2 \)$ \\
    
    $b\bar{b}$ & $\dst \(\frac{g^4 c_q^\text{T}c_{q3}^\text{T}}{g_V^2}\)^2 = g_V^4 \dfrac{\Gamma_\text{tot}^\text{T}}{\Gamma_\text{tot}^\text{S}} \dfrac{(c_u^\text{eff})^2 \(\frac{dL_{u\bar{u}}}{d\hat{s}}\) + (c_d^\text{eff})^2 \(\frac{dL_{d\bar{d}}}{d\hat{s}}\)}{\frac{dL_{u\bar{u}}}{d\hat{s}} + \frac{dL_{d\bar{d}}}{d\hat{s}}} (c_{d3}^\text{eff})^2$ \\
    
    $t\bar{t}$ & $\dst \(\frac{g^4 c_q^\text{T} c_{q3}^\text{T}}{g_V^2}\)^2 = g_V^4 \dfrac{\Gamma_\text{tot}^\text{T}}{\Gamma_\text{tot}^\text{S}} \dfrac{(c_u^\text{eff})^2 \(\frac{dL_{u\bar{u}}}{d\hat{s}}\) + (c_d^\text{eff})^2 \(\frac{dL_{d\bar{d}}}{d\hat{s}}\)}{\frac{dL_{u\bar{u}}}{d\hat{s}} + \frac{dL_{d\bar{d}}}{d\hat{s}}} (c_{u3}^\text{eff})^2$ \\
    
    $l^+l^-$ & $\dst \(\frac{g^4 c_q^\text{T}c_l^\text{T}}{g_V^2}\)^2 = g_V^4 \dfrac{\Gamma_\text{tot}^\text{T}}{\Gamma_\text{tot}^\text{S}} \dfrac{(c_u^\text{eff})^2 \(\frac{dL_{u\bar{u}}}{d\hat{s}}\) + (c_d^\text{eff})^2 \(\frac{dL_{d\bar{d}}}{d\hat{s}}\)}{\frac{dL_{u\bar{u}}}{d\hat{s}} + \frac{dL_{d\bar{d}}}{d\hat{s}}} (c_e^\text{eff})^2$ \\
    
    $W_L^+W_L^-/Z_Lh$ & $ \dst\(g^2 c_q^\text{T} c_H^\text{T} \)^2 = g_V^4 \dfrac{\Gamma_\text{tot}^\text{T}}{\Gamma_\text{tot}^\text{S}} \dfrac{(c_u^\text{eff})^2 \(\frac{dL_{u\bar{u}}}{d\hat{s}}\) + (c_d^\text{eff})^2 \(\frac{dL_{d\bar{d}}}{d\hat{s}}\)}{\frac{dL_{u\bar{u}}}{d\hat{s}} + \frac{dL_{d\bar{d}}}{d\hat{s}}} (c_H^0)^2$ \\

    \hline \bottomrule
    \end{tabular}
    \caption{\small Relations for all relevant search channels that lead to $(\sigma \times \text{BR})^\text{T} = (\sigma \times \text{BR})^\text{S}$, where the superscripts T and S refer to the heavy vector triplet and singlet models, respectively. 
 Note that in the top half of the table the widths refer to the charged vector, while in the bottom half they refer to the neutral vector.}
    \label{tab: Matchings}
\end{table}
%-------------------------------------------------------------

\Cref{tab: Matchings} shows the relations that lead to equal cross-section times branching ratio for the charged and neutral vectors for each experimentally relevant final state. Since the charged singlet and triplet sectors have the same number of free parameters, the charged relations in the top part of the table are fairly simple. However, the relations for the neutral vectors are more involved. Because the neutral singlet couples to both left- and right-handed fermion currents, its couplings to $u\bar u$ and $d \bar d$ in the production cross-section can be different. This implies a different production cross-section for the neutral singlet and triplet, so we need to retain the parton luminosities in these matching relations. Note that for the $t \bar t$ channel, we neglect the effects of the top mass.

%-------------------------------------------------------------
\begin{figure}
\begin{center}
    \includegraphics[width=0.49\textwidth]{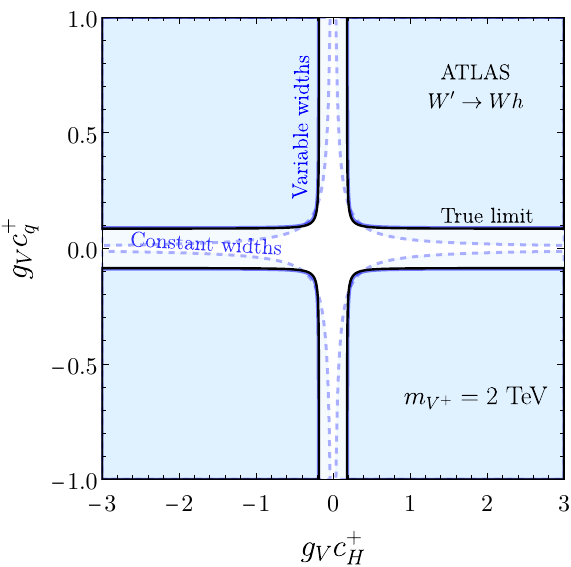}
    \includegraphics[width=0.49\textwidth]{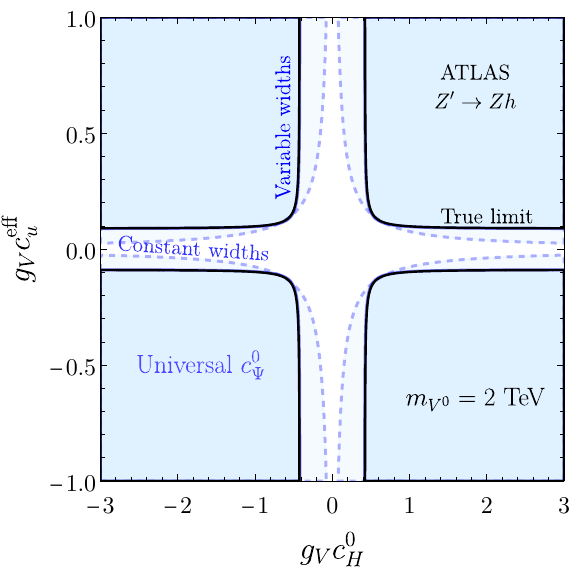}
    \caption{\small Di-boson limits initially obtained for the HVT  (taken from figure 5 of Ref.~\cite{ATLAS:2017ptz}) presented in the $(g_Vc_H^+,g_Vc_q^+)$ (left) and $(g_Vc_H^0,g_Vc_u^\text{eff})$ (right) planes of the HVS parameter space for a mass of 2\,TeV. Limits from the relations in \cref{tab: Matchings} with total widths that are functions of the simplified model parameters are shown in solid blue lines, the dashed blue lines correspond to the relations in \cref{tab: Matchings} when the total widths are fixed and the black lines depict the exact limit obtained from $\sigma \times \text{BR}$.}
    \label{fig: HVT Matching 2}
\end{center}
\end{figure}
%-------------------------------------------------------------

Experimental limits on a heavy vector triplet are presented in both the $(c_H^\text{T},c_F^\text{T})$ and $(c_q^\text{T},c_l^\text{T})$ planes. The relations in \cref{tab: Matchings} allow us to translate these experimental bounds directly into the singlet parameter space. An example of this for a charged (left) and neutral (right) di-boson search \cite{ATLAS:2017ptz} is shown in \cref{fig: HVT Matching 2}. The mapping using the relations in \cref{tab: Matchings} is shown with solid blue lines. They agree exactly with the black lines, which show the exact limit obtained from $\sigma \times \text{BR}$. While the agreement here is very good, note that discrepancies can arise in principle when the HVT is taken as a signal template in the experimental analyses, as this encodes HVT properties in the acceptances. However, since the HVS is also a colourless spin-1 resonance, just like the HVT, we expect this effect to be small.   

A further subtlety arises from the treatment of the total widths.  In the solid blue lines we compute the total widths as a function of the simplified model parameters for the triplet and the singlet. On the other hand, if we treat the widths in \cref{tab: Matchings} as fixed experimental quantities, $\Gamma_\text{tot}^\text{T}/\Gamma_\text{tot}^\text{S} = 1$, we obtain the dashed blue lines. 
We see that the correct treatment of the total widths yields exclusion limits that are much closer to the exact result. 
If a quick and rough limit is required, treating the total widths as constants gives a reasonable approximation to the exact limit, and it is particularly easy to obtain.  For a more accurate result, the tools we provide can be used to compute the widths.

In this section we have seen that limits on a heavy vector triplet can easily be used to set limits on heavy vector singlets.  However, if limits are set using a combination of channels, such as $Wh$ and $Zh$, then this procedure does not work.  It is therefore useful if limits are presented using these channels individually, possibly as well as in combination.

%=================================================
\section{Conclusions}
\label{sec:conclusions}
%=================================================

In this work we describe a model-independent approach to study colourless spin-one particles that transform as singlets under $SU(2)_L$, and have zero and unit hypercharge. 
We start from a dimension-4 simplified model Lagrangian that couples a charged and neutral HVS to the SM in the most general way. We compute the mixing angles and physical masses after electroweak symmetry breaking. We take the physical heavy vector masses, the $Z$-boson mass, the Fermi constant and the fine structure constant as input parameters and discuss an inversion procedure to express all physical quantities in terms of these inputs.
The narrow width approximation is assumed throughout this work and we discuss its limitations within the context of a heavy resonance search. We present a semi-analytic expression for the production cross-sections of the heavy vectors
and provide analytic expressions for the partial decay widths neglecting mixing effects from electroweak symmetry breaking, as well as accompanying \texttt{FeynRules} and \texttt{MadGraph} UFO model files that implement the widths numerically including all mixing effects. We emphasise that the effects of electroweak symmetry breaking on the branching ratios can be sizeable even for masses over a TeV.
We also describe the impact of the HVS on the electroweak oblique parameters.

We present the current LHC limits on the simplified model with universal couplings to the SM and find that di-lepton searches are the most constraining, followed closely by di-boson final states. The experimental limits are also presented in the full simplified model parameter space. This works particularly well for the two free parameters of the charged heavy singlet (in addition to the mass). The six-dimensional parameter space of the neutral singlet can be reduced to four dimensions without sacrificing much accuracy and experimental bounds can be presented in a row or a grid of two-dimensional plots. We have also implemented an extrapolation procedure based on the number of background events for projecting LHC limits to the HL-LHC, HE-LHC, SPPC and FCC-hh. We project that the HL-LHC will be able to probe 1--2\,TeV higher than current limits, while the 100\,TeV FCC-hh will be able to reach masses of 40--50\,TeV.

To demonstrate the power of the simplified model, we match three benchmark models onto our simplified model: an abelian and non-abelian weakly coupled extension of the SM gauge group, and a strongly coupled composite Higgs model. The explicit models correspond to points or lines in the simplified model parameter space where it is particularly easy to see whether they are experimentally excluded or allowed. In addition, we also translate current and future experimental bounds into the mass vs coupling plane of the explicit models. 
Depending on the coupling strengths, we see that weakly coupled gauge extensions are generally excluded for masses up to 5--6\,TeV. While the limit is similar in a composite Higgs model for small couplings, the exclusion limit drops down to 2\,TeV for stronger couplings.

Finally, we study the relation between the HVS and HVT simplified models.
By mapping the two models either at the level of the Lagrangian or at the level of the production cross-section times branching ratio, we find that experimental limits on the heavy vector triplet can easily constrain certain singlet parameter combinations, without the implementation of the entire singlet model.

%=================================================
\section*{Acknowledgements}
%=================================================

This research was partially supported by the University of Massachusetts Amherst, by the University of Melbourne and by the Australian Government through the Australian Research Council Centre of Excellence for Dark Matter Particle Physics (CDM, CE200100008). MJB, TM and AT would like to thank the IPPP at Durham University for kind hospitality during part of this work.

\appendix

%=================================================
\section{The Tilde Basis for the Neutral Singlet}
\label{AppA}
%=================================================

We mentioned in \cref{sec:explicit-models} that a kinetic mixing between the neutral singlet and the SM $B_\mu$ field may be present, and can be absorbed by a field redefinition. Here we discuss the relation between the parameters in the two bases where the kinetic mixing is present or not. We distinguish the parameters in the basis with kinetic mixing with a tilde and refer to the corresponding basis as the ``tilde basis''.

We define the Lagrangian in the ``tilde basis'', $\tilde{\mathcal{L}}$, as the Lagrangian in $\mathcal{L}_{\text{ph}}$ given in \cref{sml} with all couplings replaced by ``tilde''
\begin{equation}
c\to \tilde c, \quad m_V\to \tilde m_V \, ,
\end{equation}
and with the addition of the kinetic mixing term
\begin{equation}
\tilde{\mathcal{L}} \supset \tilde c_{VB}^{0} \frac{g'}{2g_{V} }\partial_{[\mu}V^0_{\nu ]}B^{\mu\nu}.
\end{equation}
Using the field redefinitions
\be\label{fieldred}
\left\{
\bry{l}
B_\mu\rightarrow B_\mu + \alpha V_\mu^0\\
V_\mu^0\rightarrow \beta V_\mu^0
\ery
\right.\,,
\ee
with
\begin{equation}
\dst \alpha = \frac{g'~\tilde{c}_{VB}^{0}}{\sqrt{g_{V} ^{2}  -\tilde{c}_{VB}^{0\,2}g^{\prime\,2}}}~~~{\textrm{and}}~~~
\dst \beta=\frac{g_{V} }{\sqrt{g_{V}^{2}  -\tilde{c}_{VB}^{0\,2}g^{\prime\, 2}}},
\end{equation}
we get the following relations between the parameters in the two bases 
\be \label{cnontilde}
\bry{lll}
\dst m_{V^{0}}^{2}=\frac{g_{V} ^{2}}{g_{V} ^{2}  -\tilde{c}_{VB}^{0\,2} g^{\prime 2}} \tilde m_{V^{0}}^{2}\,,\vspace{2mm}\\
\dst m_{V^{+}}^{2}= \tilde m_{V^{+}}^{2} \,,\vspace{2mm}\\
\dst c_{VV B}^{+}= \tilde{c}_{VVB}^{+} \,,\vspace{2mm}\\
\dst c_{VVV}^{0}= \f{g_{V}}{\sqrt{g_{V} ^{2}  -\tilde{c}_{VB}^{0\,2} g^{\prime \, 2}}} \(\tilde{c}_{VVV}^{0} + \f{g'^{2}}{g_{V}^{2}} \tilde{c}_{VB}^{0} \tilde{c}_{VV B}^{+}  \) \,,\vspace{2mm}\\
\dst c_{VVV}^{+}= \frac{g_{V} }{\sqrt{g_{V} ^{2}  -\tilde{c}_{VB}^{0\,2}g^{\prime \, 2}}} \( \tilde{c}_{VVV}^{+} + \f{g^{\prime \, 2}}{g_{V}^{2}} \tilde{c}_{VB}^{0}\) \,,\vspace{2mm}\\
c_H^{0} = \dst \frac{g_{V} }{\sqrt{g_{V} ^{2}  -\tilde{c}_{VB}^{0\,2}g^{\prime \, 2}}} \( \tilde c_{H}^{0} + \f{g^{\prime 2}}{g_{V}^{2}}  \tilde{c}_{VB}^{0}\)\,,\vspace{2mm}\\
\dst c_{H}^{+}= \tilde c_{H}^{+}\,,\vspace{2mm}\\
\dst c_{VVHH}^{0} = \frac{g_{V} ^{2}}{g_{V} ^{2}  -\tilde{c}_{VB}^{0\,2} g^{\prime 2}} \(\tilde c_{VVHH}^{0} + \f{g^{\prime 2}}{2g_{V} ^{2}} \tilde{c}_{VB}^{0} \tilde{c}_{H}^{0}  + \f{g^{\prime 4}}{4 g_{V} ^{4}}  \tilde{c}_{VB}^{0\,2}\) \,,\vspace{2mm}\\
\dst c_{VVHH}^{+} = \tilde c_{VVHH}^{+}\,,\vspace{2mm}\\
\dst c_{\Psi}^{0}= \frac{g_{V} }{\sqrt{g_{V} ^{2}  -\tilde{c}_{VB}^{0\,2}g^{\prime \, 2}}} \(\tilde{c}_{\Psi}^{0} +  \f{g^{\prime 2}}{g_{V} ^{2}} \tilde{c}_{VB}^{0} 2 Y_\Psi \)\,,\vspace{2mm}\\
\dst c_{q}^{+}= \tilde{c}_{q}^{+}\,.
\ery
\ee
where $Y_\Psi$ stands collectively for the appropriate hypercharge. 

If the neutral and charged singlets belong to a single $SU(2)_{R}$ triplet, as in Model D and E above, there are various relations between the charged and neutral tilde coefficients depending on the details of the model (which imply relations between the non-tilde coefficients,  seen in \cref{Table:matching} for Models D and E).

%=================================================
\section{Input Parameters for \texttt{FeynRules} and \texttt{MadGraph} Models}
\label{Appendix:InputParameters}
%=================================================

We provide a \texttt{FeynRules} \cite{Alloul:2013vc,Christensen:2008py} and corresponding \texttt{MadGraph} \cite{Alwall:2011fk} UFO model with the simplified model Lagrangian in \cref{sml} implemented in the mass eigenstate basis and in the unitary gauge. The model files are registered in the HEPMDB model database \cite{hepmdb} with the unique identifier hepmdb:0724.0349 and are available to download \href{https://hepmdb.soton.ac.uk/hepmdb:0724.0349}{here}. Here we explain the input parameters for these model files.

We take the physical heavy vector masses $m_{V^{0}}$, $m_{V^{+}}$, the $Z$-boson mass $m_Z$ and the Fermi constant, $G_F^{\ell\ell}$, and the fine structure constant as input parameters. We use these inputs to iteratively rewrite 
$\hat m_{V^{+}}$ in terms of $m_{V^{+}}$,
$\hat v$ in terms of $m_Z$,
$\hat m_{V^{0}}$ in terms of $m_{V^{0}}$ and 
the SM coupling $g$ in terms of the Fermi constant $G_F^{\ell\ell}$.
In this parameter inversion we work to a precision of order $\mathcal{O}(g^2 g_V^2\hat{v}^4/\hat{m}_{V^{0,+}}^4)$. We can then express all quantities, such as partial widths and cross-sections, in terms of these five physical input parameters. 

As a result of this parameter inversion, the model files impose the constraints
\begin{align}
    |c_H^+| &< \frac{m_{V^+}^{2}}{g g_V \hat{v}^2} \,, \\
    |c_H^0| &< \frac{g^2 \sqrt{m_{V^0}^{2} - m_Z^{2}}}{2 g_V m_Z \sqrt{g^2 - 4 \pi \alpha_{\text{EM}}}} \,.
\end{align}
where $\alpha_\text{EM}=e^2/4\pi$ is the fine structure constant.\footnote{Numerically, we use $\alpha_\text{EM}\left(m_Z\right)\approx 1/127.9$ and neglect the small effects from renormalization group running.} Depending on the input parameters these constraints roughly correspond to $|c_H^+| < 25 \, (228)$ and $|c_H^0| < 4 \, (12)$ for heavy vector masses around $1 (3)\,$TeV and $g_V=1$. From the expression for $G_F^{\ell\ell}$, \cref{eq:Gf}, we can infer that 
\begin{align}
    |c_H^+| &< \frac{\hat{m}_{V^+}}{g_V \hat{v}} \,,
\end{align}
which implies $|c_H^+| \lesssim 4 \, (12)$ for a mass around $1 \, (3)\,$TeV. The actual limit is even somewhat below this value since a $|c_H^+|$ close to its upper bound leads to deviations in $G_F^{\ell\ell}$ and $g$ that are so large that they render the model inconsistent. \texttt{MadGraph} will give an error for non-physical values.

Furthermore we find that
\begin{equation}
   g_V^2 < \frac{8 \, G_F  m_{V^+}^{2} m_{V^0}^{2} }{4 \sqrt{2} c_H^{+ \,2} m_{V^0}^{2}  - \sqrt{2} c_H^{0 \,2} m_{V^+}^{2} } \qquad \text{if} \qquad |c_H^0| < 2 |c_H^+| \frac{m_{V^0} }{m_{V^+}} \,, \\
\end{equation}
corresponding to $g_V < 9 \, (4.5)$ for $c_H^+ = c_H^0 = 0.5 \, (1)$ and masses around $1\,$TeV. At $3\,$TeV this changes to $g_V < 13 \, (14)$. 

Besides the physical inputs listed above, the \texttt{FeynRules} and \texttt{MadGraph} model files also take the overall coupling $g_{V}$ and all charged and neutral $c_{i}$ parameters as inputs for the heavy singlets. 
The Higgs mass is also an input parameter fixed to a default value of $125.25\,$GeV and the Higgs coupling parameters $a,b,c,d_{3},d_{4}$ are implemented following the notation of Ref.~\cite{Contino:2010vc}. The default values $a=b=c=d_{3}=d_{4}=1$ render the Higgs sector SM like. The model files take full account of the mixing between the heavy vector singlets and SM gauge bosons after electroweak symmetry breaking.

%=================================================
\section{Extrapolation Procedure}
\label{Appendix:extrapolation}
%=================================================

In this appendix we describe in more detail the extrapolation procedure used in \cref{sec:future-projections} to estimate the sensitivity of future experiments. We start from the experimental exclusion limit at a given resonance mass, $m_0$, determined by an existing LHC analysis. This exclusion is based on a certain number of background events. We now define an equivalent mass, $m$, at a future collider, which describes the resonance mass where the number of background events at the future collider is equal to the number of background events at $m_0$ in the LHC analysis. That is, we find the equivalent mass $m$ that satisfies
\begin{equation}
    \label{backgrounds-app}
    B(s_0,L_0,m_0) = B(s,L,m),
\end{equation}
where $B$ is the number of background events, $\sqrt{s_0}$ and $L_0$ are the LHC centre-of-mass energy and integrated luminosity, and $\sqrt{s}$ and $L$ correspond to that of the new collider.  The number of background events for a certain process can be written as
\begin{equation}
    \label{backgrounds parton lumi-app}
    B(m;s,L) \propto L \cdot \sum_{\{i,j\}} \int d\hat{s} \dfrac{1}{\hat{s}}\dfrac{d\mathcal{L}_{ij}}{d\hat{s}}(\sqrt{\hat{s}};\sqrt{s})[\hat{s}\hat{\sigma}_{ij}(\hat{s})],
\end{equation}
where the integral is performed in a small window of partonic invariant mass squared $\hat{s} \in [m^2 - \Delta\hat{s}/2, m^2 + \Delta\hat{s}/2]$ (of fixed relative width such that $\Delta \hat{s}/\hat{s} \ll 1$) centred around the resonance mass. The sum is performed over all partonic interactions that contribute to the background. Here $d\mathcal{L}_{ij}/d\hat{s}$ are the parton luminosities for initial partons $i,j$ with centre-of-mass energy $\sqrt{\hat{s}}$, and $\hat{\sigma}_{ij}$ is the parton-level cross-section for the partonic interactions contributing to background processes. At energies far above the SM masses, all cross-sections $\hat{\sigma}_{ij}$ are proportional to $1/\hat{s}$, so we can write at tree-level,
\begin{equation}
    \label{c_ij constants-app}
   c_{ij} = \lim_{\hat s \to \infty} [\hat{s}\hat{\sigma}_{ij}(\hat{s})] ,
\end{equation}
where the $c_{ij}$ are process-dependent constants. Since we assume the background to be computed in a narrow window $\Delta \hat s \ll m^2$, the parton luminosities can be taken to be constant over the integration region and the expected background can be written as
\begin{equation}
    B(m;s,L) \propto L\cdot \dfrac{\Delta\hat{s}}{m^2} \cdot \sum_{\{i,j\}} c_{ij} \dfrac{d\mathcal{L}_{ij}}{d\hat{s}}(m;\sqrt{s})\,.
\end{equation}
Substituting this into \cref{backgrounds-app} and assuming fixed relative widths $\Delta \hat s/m^2 = \Delta \hat s_0/m_0^2$ gives
\begin{equation}
    \label{backgrounds sum c_ij-app}
    \sum_{\{i,j\}} c_{ij} \dfrac{d\mathcal{L}_{ij}}{d\hat{s}}(m;\sqrt{s}) = \dfrac{L_0}{L}\sum_{\{i,j\}} c_{ij} \dfrac{d\mathcal{L}_{ij}}{d\hat{s}}(m_0;\sqrt{s_0})\,.
\end{equation}
Since the number of background events will be the same for a heavy vector with mass $m$ at a future collider as for a heavy vector with mass $m_0$ at the LHC, we could obtain the expected limit on the cross-section times branching ratio at the equivalent mass $m$ by rescaling with the integrated luminosities,
\begin{equation}
    \text{limit}
    [\sigma \times \text{BR}](m;s,L) = \dfrac{L_0}{L}
    \text{limit}[\sigma \times \text{BR}]_0(m_0;s_0,L_0)
    \,.
\end{equation}
This holds as long as the signal acceptances and efficiencies of the two experiments are similar, which we will assume.  However, there is a subtlety at low masses. For a collider with higher centre-of-mass energy, $s>s_0$, and larger luminosity, $L>L_0$, than the LHC, the equivalent mass $m$ is  greater than the original $m_0$, so an increase in the integrated luminosity could mean that there will be no predicted limits at the lowest masses considered in the original analysis.
This is dealt with in  Ref.~\cite{Thamm:2015zwa} by defining an intermediate luminosity $L'$ which smoothly varies over the range $L'\leq L$. This leads to a set of extrapolated limits for each $L'$ and we can choose the strongest limit at each mass point $m$.  While this procedure produces limits down to the lowest mass considered in the original analysis, it is conservative at low masses since the low mass limit is obtained with a data set, $L'$, smaller than the one corresponding to the expected luminosity, $L$. An improvement to this procedure provides a more realistic projection at lower masses \cite{Buttazzo:2015bka}. Instead of rescaling the limit on $[\sigma \times \text{BR}](m_0;s_0,L_0)$ by $L_0/L'$, it is rescaled by $L_0/\sqrt{LL'}$, such that 
\begin{equation}
    [\sigma \times \text{BR}](m;s,L,L') = \dfrac{L_0}{\sqrt{L L'}}
    \text{limit}[\sigma \times \text{BR}]_0(m_0;s_0,L_0)
    \,.
\end{equation}
The limit is then given by the minimum of $[\sigma \times \text{BR}](m;s,L,L')$ at each equivalent mass point $m$ over the range of $L'$,
\begin{equation}
    \text{limit}[\sigma \times \text{BR}](m;s,L) = \min_{L' \leq L}[\sigma \times \text{BR}](m;s,L,L')
    \,.
\end{equation}

\bibliographystyle{JHEP}
\bibliography{refs}

\providecommand{\href}[2]{#2}\begingroup\raggedright\begin{thebibliography}{100}

\bibitem{Barger:1980ix}
V.~D. Barger, W.-Y. Keung, and E.~Ma, {\it {A Gauge Model With Light $W$ and
  $Z$ Bosons}},  {\em Phys. Rev. D} {\bf 22} (1980) 727.

\bibitem{Hewett:1989dr}
J.~L. Hewett and T.~G. Rizzo, {\it {Low-energy phenomenology of
  superstring-inspired $E_{6}$ models}},  {\em Phys. Rept.} {\bf 183} (1989)
  193--381. [\href{http://inspirehep.net/record/268529}{Inspire}].

\bibitem{Cvetic:1995zs}
M.~Cvetic and S.~Godfrey, {\em {Discovery and identification of extra gauge
  bosons}}, pp.~383--415.
\newblock 3, 1995.
\newblock \href{http://arxiv.org/abs/hep-ph/9504216}{{\tt hep-ph/9504216}}.

\bibitem{Rizzo:2006wq}
T.~G. Rizzo, {\it {$Z'$ Phenomenology and the LHC}},
  \href{http://arxiv.org/abs/hep-ph/0610104}{{\tt hep-ph/0610104}}.
  [\href{http://inspirehep.net/record/728548}{Inspire}].

\bibitem{Agashe:2007hh}
K.~Agashe, H.~Davoudiasl, S.~Gopalakrishna, T.~Han, G.-Y. Huang, G.~Perez,
  Z.-G. Si, and A.~Soni, {\it {LHC Signals for Warped Electroweak Neutral Gauge
  Bosons}},  {\em Phys. Rev.} {\bf D 76} (2007) 115015,
  [\href{http://arxiv.org/abs/0709.0007}{{\tt arXiv:0709.0007}}].
  [\href{http://inspirehep.net/record/759584}{Inspire}].

\bibitem{Langacker:2008yv}
P.~Langacker, {\it The {P}hysics of {H}eavy ${Z}^\prime$ {G}auge {B}osons},
  {\em Rev. Mod. Phys.} {\bf 81} (2009) 119--1228,
  [\href{http://arxiv.org/abs/0801.1345}{{\tt arXiv:0801.1345}}].
  [\href{http://inspirebeta.net/record/777086}{Inspire}].

\bibitem{Salvioni:2010p2769}
E.~Salvioni, A.~Strumia, G.~Villadoro, and F.~Zwirner, {\it Non-universal
  minimal {Z}' models: present bounds and early {LHC} reach},  {\em JHEP} {\bf
  03} (2010) 010, [\href{http://arxiv.org/abs/0911.1450}{{\tt
  arXiv:0911.1450}}]. [\href{http://inspirebeta.net/record/836375}{Inspire}].

\bibitem{Accomando:2013ve}
E.~Accomando, D.~Becciolini, A.~S. Belyaev, S.~Moretti, and C.~H.
  Shepherd-Themistocleous, {\it {$Z'$ at the LHC: Interference and Finite Width
  Effects in Drell-Yan}},  {\em JHEP} {\bf 10} (2013) 153,
  [\href{http://arxiv.org/abs/1304.6700}{{\tt arXiv:1304.6700}}].
  [\href{http://inspirehep.net/record/1229792}{Inspire}].

\bibitem{Agashe:2009bj}
K.~Agashe, S.~Gopalakrishna, T.~Han, G.-Y. Huang, and A.~Soni, {\it {LHC
  signals for warped electroweak charged gauge bosons}},  {\em Phys. Rev.} {\bf
  D 80} (2009) 075007, [\href{http://arxiv.org/abs/0810.1497}{{\tt
  arXiv:0810.1497}}]. [\href{http://inspirehep.net/record/798910}{Inspire}].

\bibitem{Schmaltz:2010p2610}
M.~Schmaltz and C.~Spethmann, {\it Two {S}imple ${W}^\prime$ {M}odels for the
  {E}arly {LHC}},  {\em JHEP} {\bf 07} (2011) 046,
  [\href{http://arxiv.org/abs/1011.5918}{{\tt arXiv:1011.5918}}].
  [\href{http://inspirehep.net/record/878831}{Inspire}].

\bibitem{Grojean:2011vu}
C.~Grojean, E.~Salvioni, and R.~Torre, {\it A weakly constrained ${W}'$ at the
  early {LHC}},  {\em JHEP} {\bf 07} (2011) 002,
  [\href{http://arxiv.org/abs/1103.2761}{{\tt arXiv:1103.2761}}].
  [\href{http://inspirehep.net/record/892770}{Inspire}].

\bibitem{Langacker:1989xa}
P.~Langacker and S.~U. Sankar, {\it {Bounds on the Mass of W(R) and the
  W(L)-W(R) Mixing Angle xi in General SU(2)-L x SU(2)-R x U(1) Models}},  {\em
  Phys. Rev. D} {\bf 40} (1989) 1569--1585.

\bibitem{Frank:2010p2250}
M.~Frank, A.~Hayreter, and I.~Turan, {\it Production and {D}ecays of ${W}_{R}$
  bosons at the {LHC}},  {\em Phys. Rev.} {\bf D 83} (2011) 035001,
  [\href{http://arxiv.org/abs/1010.5809}{{\tt arXiv:1010.5809}}].
  [\href{http://inspirebeta.net/record/874800}{Inspire}].

\bibitem{Accomando:2011up}
E.~Accomando, D.~Becciolini, S.~de~Curtis, D.~Dominici, L.~Fedeli, and C.~H.
  Shepherd-Themistocleous, {\it {Interference effects in heavy $W'$-boson
  searches at the LHC}},  {\em Phys. Rev.} {\bf D 85} (2011) 115017,
  [\href{http://arxiv.org/abs/1110.0713}{{\tt arXiv:1110.0713}}].
  [\href{http://inspirehep.net/record/930438}{Inspire}].

\bibitem{Accomando:2011gt}
E.~Accomando, D.~Becciolini, S.~D. Curtis, D.~Dominici, and L.~Fedeli, {\it
  {$W'$ production at the LHC in the 4-site Higgsless model}},  {\em Phys.
  Rev.} {\bf D 84} (2011) 115014, [\href{http://arxiv.org/abs/1107.4087}{{\tt
  arXiv:1107.4087}}]. [\href{http://inspirehep.net/record/919245}{Inspire}].

\bibitem{Dobrescu:2021vak}
B.~A. Dobrescu and F.~Yu, {\it {Dijet and electroweak limits on a Z' boson
  coupled to quarks}},  {\em Phys. Rev. D} {\bf 109} (2024), no.~3 035004,
  [\href{http://arxiv.org/abs/2112.05392}{{\tt arXiv:2112.05392}}].

\bibitem{Chanowitz:1993fc}
M.~S. Chanowitz and W.~Kilgore, {\it {Complementarity of Resonant and
  Nonresonant Strong $WW$ Scattering at the LHC}},  {\em Phys. Lett.} {\bf B
  322} (1993) 147--153, [\href{http://arxiv.org/abs/hep-ph/9311336}{{\tt
  hep-ph/9311336}}]. [\href{http://inspirehep.net/record/360435}{Inspire}].

\bibitem{Barbieri:2008cc}
R.~Barbieri, G.~Isidori, V.~S. Rychkov, and E.~Trincherini, {\it {Heavy Vectors
  in Higgs-less models}},  {\em Phys. Rev. D} {\bf 78} (2008) 036012,
  [\href{http://arxiv.org/abs/0806.1624}{{\tt arXiv:0806.1624}}].

\bibitem{Barbieri:2009p33}
R.~Barbieri, A.~E. {C{\'a}rcamo Hern{\'a}ndez}, G.~Corcella, R.~Torre, and
  E.~Trincherini, {\it Composite vectors at the large hadron collider},  {\em
  JHEP} {\bf 03} (2010) 068, [\href{http://arxiv.org/abs/0911.1942}{{\tt
  arXiv:0911.1942}}]. [\href{http://inspirehep.net/record/836568}{Inspire}].

\bibitem{Agashe:2009dg}
K.~Agashe and R.~Contino, {\it {Composite Higgs-Mediated FCNC}},  {\em Phys.
  Rev.} {\bf D 80} (2009) 075016, [\href{http://arxiv.org/abs/0906.1542}{{\tt
  arXiv:0906.1542}}]. [\href{http://inspirehep.net/record/822501}{Inspire}].

\bibitem{Agashe:2009ve}
K.~Agashe, A.~Azatov, T.~Han, Y.~Li, Z.-G. Si, and L.~Zhu, {\it {LHC Signals
  for Coset Electroweak Gauge Bosons in Warped/Composite PGB Higgs Models}},
  {\em Phys. Rev.} {\bf D 81} (2010) 096002,
  [\href{http://arxiv.org/abs/0911.0059}{{\tt arXiv:0911.0059}}].
  [\href{http://inspirehep.net/record/835687}{Inspire}].

\bibitem{Cata:2009iy}
O.~Cata, G.~Isidori, and J.~F. Kamenik, {\it {Drell-Yan production of Heavy
  Vectors in Higgsless models}},  {\em Nucl. Phys. B} {\bf 822} (2009)
  230--244, [\href{http://arxiv.org/abs/0905.0490}{{\tt arXiv:0905.0490}}].

\bibitem{Barbieri:2010mn}
R.~Barbieri, S.~Rychkov, and R.~Torre, {\it {Signals of composite
  electroweak-neutral Dark Matter: LHC/Direct Detection interplay}},  {\em
  Phys. Lett. B} {\bf 688} (2010) 212--215,
  [\href{http://arxiv.org/abs/1001.3149}{{\tt arXiv:1001.3149}}].

\bibitem{CarcamoHernandez:2010wpm}
A.~E. Carcamo~Hernandez, {\it {Top quark effects in composite vector pair
  production at the LHC}},  {\em Eur. Phys. J. C} {\bf 72} (2012) 2154,
  [\href{http://arxiv.org/abs/1008.1039}{{\tt arXiv:1008.1039}}].

\bibitem{CarcamoHernandez:2010qxf}
A.~E. Carcamo~Hernandez and R.~Torre, {\it {A 'Composite' scalar-vector system
  at the LHC}},  {\em Nucl. Phys. B} {\bf 841} (2010) 188--204,
  [\href{http://arxiv.org/abs/1005.3809}{{\tt arXiv:1005.3809}}].

\bibitem{Falkowski:2011ua}
A.~Falkowski, C.~Grojean, A.~Kaminska, S.~Pokorski, and A.~Weiler, {\it {If no
  Higgs then what?}},  {\em JHEP} {\bf 11} (2011) 028,
  [\href{http://arxiv.org/abs/1108.1183}{{\tt arXiv:1108.1183}}].
  [\href{http://inspirehep.net/record/922186}{Inspire}].

\bibitem{Contino:2011np}
R.~Contino, D.~Marzocca, D.~Pappadopulo, and R.~Rattazzi, {\it On the effect of
  resonances in composite {H}iggs phenomenology},  {\em JHEP} {\bf 10} (2011)
  081, [\href{http://arxiv.org/abs/1109.1570}{{\tt arXiv:1109.1570}}].
  [\href{http://inspirehep.net/record/926810}{Inspire}].

\bibitem{Chanowitz:2011ew}
M.~S. Chanowitz, {\it {A heavy little Higgs and a light $Z'$ under the radar}},
   {\em Phys. Rev.} {\bf D 84} (2011) 035014,
  [\href{http://arxiv.org/abs/1102.3672}{{\tt arXiv:1102.3672}}].
  [\href{http://inspirehep.net/record/889847}{Inspire}].

\bibitem{Bellazzini:2012tv}
B.~Bellazzini, C.~Csaki, J.~Hubisz, J.~Serra, and J.~Terning, {\it {Composite
  Higgs Sketch}},  {\em JHEP} {\bf 11} (2012) 003,
  [\href{http://arxiv.org/abs/1205.4032}{{\tt arXiv:1205.4032}}].
  [\href{http://inspirehep.net/record/1115304}{Inspire}].

\bibitem{Accomando:2012us}
E.~Accomando, L.~Fedeli, S.~Moretti, S.~D. Curtis, and D.~Dominici, {\it
  {Charged di-boson production at the LHC in a 4-site model with a composite
  Higgs boson}},  {\em Phys.Rev.} {\bf D86} (2012) 115006,
  [\href{http://arxiv.org/abs/1208.0268}{{\tt arXiv:1208.0268}}].
  [\href{http://inspirehep.net/record/1124599}{Inspire}].

\bibitem{Low:2015uha}
M.~Low, A.~Tesi, and L.-T. Wang, {\it {Composite spin-1 resonances at the
  LHC}},  {\em Phys. Rev. D} {\bf 92} (2015), no.~8 085019,
  [\href{http://arxiv.org/abs/1507.07557}{{\tt arXiv:1507.07557}}].

\bibitem{Accomando:2016mvz}
E.~Accomando, D.~Barducci, S.~De~Curtis, J.~Fiaschi, S.~Moretti, and C.~H.
  Shepherd-Themistocleous, {\it {Drell-Yan production of multi Z$^{'}$-bosons
  at the LHC within Non-Universal ED and 4D Composite Higgs Models}},  {\em
  JHEP} {\bf 07} (2016) 068, [\href{http://arxiv.org/abs/1602.05438}{{\tt
  arXiv:1602.05438}}].

\bibitem{Liu:2018hum}
D.~Liu, L.-T. Wang, and K.-P. Xie, {\it {Prospects of searching for composite
  resonances at the LHC and beyond}},  {\em JHEP} {\bf 01} (2019) 157,
  [\href{http://arxiv.org/abs/1810.08954}{{\tt arXiv:1810.08954}}].

\bibitem{Liu:2019bua}
D.~Liu, L.-T. Wang, and K.-P. Xie, {\it {Broad composite resonances and their
  signals at the LHC}},  {\em Phys. Rev. D} {\bf 100} (2019), no.~7 075021,
  [\href{http://arxiv.org/abs/1901.01674}{{\tt arXiv:1901.01674}}].

\bibitem{DeCurtis:2021fdm}
S.~De~Curtis and D.~Dominici, {\it {Spin-1 resonances}},  {\em Eur. Phys. J.
  ST} {\bf 231} (2022), no.~7 1299--1308,
  [\href{http://arxiv.org/abs/2110.01907}{{\tt arXiv:2110.01907}}].

\bibitem{Greco:2014aza}
D.~Greco and D.~Liu, {\it {Hunting composite vector resonances at the LHC:
  naturalness facing data}},  {\em JHEP} {\bf 12} (2014) 126,
  [\href{http://arxiv.org/abs/1410.2883}{{\tt arXiv:1410.2883}}].

\bibitem{Liu:2023jta}
D.~Liu, L.-T. Wang, and K.-P. Xie, {\it {Composite resonances at a 10 TeV muon
  collider}},  \href{http://arxiv.org/abs/2312.09117}{{\tt arXiv:2312.09117}}.

\bibitem{Baker:2022zxv}
M.~J. Baker, T.~Martonhelyi, A.~Thamm, and R.~Torre, {\it {The role of vector
  boson fusion in the production of heavy vector triplets at the LHC and
  HL-LHC}},  {\em JHEP} {\bf 11} (2022) 066,
  [\href{http://arxiv.org/abs/2207.05091}{{\tt arXiv:2207.05091}}].

\bibitem{ATLAS:2019fgd}
{\bf ATLAS} Collaboration, G.~Aad et~al., {\it {Search for new resonances in
  mass distributions of jet pairs using 139 fb$^{-1}$ of $pp$ collisions at
  $\sqrt{s}=13$ TeV with the ATLAS detector}},  {\em JHEP} {\bf 03} (2020) 145,
  [\href{http://arxiv.org/abs/1910.08447}{{\tt arXiv:1910.08447}}].

\bibitem{ATLAS:2017eqx}
{\bf ATLAS} Collaboration, M.~Aaboud et~al., {\it {Search for new phenomena in
  dijet events using 37 fb$^{-1}$ of $pp$ collision data collected at
  $\sqrt{s}=$13 TeV with the ATLAS detector}},  {\em Phys. Rev. D} {\bf 96}
  (2017), no.~5 052004, [\href{http://arxiv.org/abs/1703.09127}{{\tt
  arXiv:1703.09127}}].

\bibitem{ATLAS:2018uca}
{\bf ATLAS} Collaboration, M.~Aaboud et~al., {\it {Search for $W' \rightarrow
  tb$ decays in the hadronic final state using pp collisions at $\sqrt{s}=13$
  TeV with the ATLAS detector}},  {\em Phys. Lett. B} {\bf 781} (2018)
  327--348, [\href{http://arxiv.org/abs/1801.07893}{{\tt arXiv:1801.07893}}].

\bibitem{CMS:2021mux}
{\bf CMS} Collaboration, A.~M. Sirunyan et~al., {\it {Search for W' bosons
  decaying to a top and a bottom quark at $\sqrt{s}=13$ TeV in the hadronic
  final state}},  {\em Phys. Lett. B} {\bf 820} (2021) 136535,
  [\href{http://arxiv.org/abs/2104.04831}{{\tt arXiv:2104.04831}}].

\bibitem{CMS:2017zod}
{\bf CMS} Collaboration, A.~M. Sirunyan et~al., {\it {Search for heavy
  resonances decaying to a top quark and a bottom quark in the lepton+jets
  final state in proton\textendash{}proton collisions at 13 TeV}},  {\em Phys.
  Lett. B} {\bf 777} (2018) 39--63,
  [\href{http://arxiv.org/abs/1708.08539}{{\tt arXiv:1708.08539}}].

\bibitem{ATLAS:2018iui}
{\bf ATLAS} Collaboration, M.~Aaboud et~al., {\it {Search for resonant $WZ$
  production in the fully leptonic final state in proton-proton collisions at
  $\sqrt{s} = 13$ TeV with the ATLAS detector}},  {\em Phys. Lett. B} {\bf 787}
  (2018) 68--88, [\href{http://arxiv.org/abs/1806.01532}{{\tt
  arXiv:1806.01532}}].

\bibitem{ATLAS:2019nat}
{\bf ATLAS} Collaboration, G.~Aad et~al., {\it {Search for diboson resonances
  in hadronic final states in 139 fb$^{-1}$ of $pp$ collisions at $\sqrt{s} =
  13$ TeV with the ATLAS detector}},  {\em JHEP} {\bf 09} (2019) 091,
  [\href{http://arxiv.org/abs/1906.08589}{{\tt arXiv:1906.08589}}]. [Erratum:
  JHEP 06, 042 (2020)].

\bibitem{ATLAS:2016hal}
{\bf ATLAS} Collaboration, M.~Aaboud et~al., {\it {Searches for heavy diboson
  resonances in $pp$ collisions at $\sqrt{s}=13$ TeV with the ATLAS detector}},
   {\em JHEP} {\bf 09} (2016) 173, [\href{http://arxiv.org/abs/1606.04833}{{\tt
  arXiv:1606.04833}}].

\bibitem{CMS:2022pjv}
{\bf CMS} Collaboration, {\it {Search for new heavy resonances decaying to WW,
  WZ, ZZ, WH, or ZH boson pairs in the all-jets final state in proton-proton
  collisions at $\sqrt{s}$ = 13 TeV}},
  \href{http://arxiv.org/abs/2210.00043}{{\tt arXiv:2210.00043}}.

\bibitem{CMS:2017fgc}
{\bf CMS} Collaboration, A.~M. Sirunyan et~al., {\it {Search for massive
  resonances decaying into $WW$, $WZ$, $ZZ$, $qW$, and $qZ$ with dijet final
  states at $\sqrt{s}=13\text{ }\text{ }\mathrm{TeV}$}},  {\em Phys. Rev. D}
  {\bf 97} (2018), no.~7 072006, [\href{http://arxiv.org/abs/1708.05379}{{\tt
  arXiv:1708.05379}}].

\bibitem{CMS:2016rqm}
{\bf CMS} Collaboration, A.~M. Sirunyan et~al., {\it {Search for massive
  resonances decaying into WW, WZ or ZZ bosons in proton-proton collisions at
  $\sqrt{s} = $ 13 TeV}},  {\em JHEP} {\bf 03} (2017) 162,
  [\href{http://arxiv.org/abs/1612.09159}{{\tt arXiv:1612.09159}}].

\bibitem{ATLAS:2020fry}
{\bf ATLAS} Collaboration, G.~Aad et~al., {\it {Search for heavy diboson
  resonances in semileptonic final states in pp collisions at $\sqrt{s}=13$ TeV
  with the ATLAS detector}},  {\em Eur. Phys. J. C} {\bf 80} (2020), no.~12
  1165, [\href{http://arxiv.org/abs/2004.14636}{{\tt arXiv:2004.14636}}].

\bibitem{ATLAS:2017jag}
{\bf ATLAS} Collaboration, M.~Aaboud et~al., {\it {Search for $WW/WZ$ resonance
  production in $\ell \nu qq$ final states in $pp$ collisions at $\sqrt{s} =$
  13 TeV with the ATLAS detector}},  {\em JHEP} {\bf 03} (2018) 042,
  [\href{http://arxiv.org/abs/1710.07235}{{\tt arXiv:1710.07235}}].

\bibitem{CMS:2021klu}
{\bf CMS} Collaboration, A.~Tumasyan et~al., {\it {Search for heavy resonances
  decaying to WW, WZ, or WH boson pairs in the lepton plus merged jet final
  state in proton-proton collisions at $\sqrt{s}$ = 13 TeV}},  {\em Phys. Rev.
  D} {\bf 105} (2022), no.~3 032008,
  [\href{http://arxiv.org/abs/2109.06055}{{\tt arXiv:2109.06055}}].

\bibitem{CMS:2018dff}
{\bf CMS} Collaboration, A.~M. Sirunyan et~al., {\it {Search for a heavy
  resonance decaying to a pair of vector bosons in the lepton plus merged jet
  final state at $ \sqrt{s}=13 $ TeV}},  {\em JHEP} {\bf 05} (2018) 088,
  [\href{http://arxiv.org/abs/1802.09407}{{\tt arXiv:1802.09407}}].

\bibitem{ATLAS:2017otj}
{\bf ATLAS} Collaboration, M.~Aaboud et~al., {\it {Searches for heavy $ZZ$ and
  $ZW$ resonances in the $\ell\ell qq$ and $\nu\nu qq$ final states in $pp$
  collisions at $\sqrt{s}=13$ TeV with the ATLAS detector}},  {\em JHEP} {\bf
  03} (2018) 009, [\href{http://arxiv.org/abs/1708.09638}{{\tt
  arXiv:1708.09638}}].

\bibitem{CMS:2021xor}
{\bf CMS} Collaboration, A.~Tumasyan et~al., {\it {Search for heavy resonances
  decaying to ZZ or ZW and axion-like particles mediating nonresonant ZZ or ZH
  production at $\sqrt{s}$ = 13 TeV}},
  \href{http://arxiv.org/abs/2111.13669}{{\tt arXiv:2111.13669}}.

\bibitem{CMS:2018sdh}
{\bf CMS} Collaboration, A.~M. Sirunyan et~al., {\it {Search for a heavy
  resonance decaying into a Z boson and a Z or W boson in 2\ensuremath{\ell}2q
  final states at $ \sqrt{s}=13 $ TeV}},  {\em JHEP} {\bf 09} (2018) 101,
  [\href{http://arxiv.org/abs/1803.10093}{{\tt arXiv:1803.10093}}].

\bibitem{CMS:2021itu}
{\bf CMS} Collaboration, A.~Tumasyan et~al., {\it {Search for heavy resonances
  decaying to Z($\nu\bar{\nu}$)V(q$\bar{q}$') in proton-proton collisions at
  $\sqrt{s}$ = 13 TeV}},  {\em Phys. Rev. D} {\bf 106} (2022), no.~1 012004,
  [\href{http://arxiv.org/abs/2109.08268}{{\tt arXiv:2109.08268}}].

\bibitem{CMS:2018ygj}
{\bf CMS} Collaboration, A.~M. Sirunyan et~al., {\it {Search for a heavy
  resonance decaying into a Z boson and a vector boson in the $ \nu
  \overline{\nu}\mathrm{q}\overline{\mathrm{q}} $ final state}},  {\em JHEP}
  {\bf 07} (2018) 075, [\href{http://arxiv.org/abs/1803.03838}{{\tt
  arXiv:1803.03838}}].

\bibitem{ATLAS:2018sxj}
{\bf ATLAS} Collaboration, M.~Aaboud et~al., {\it {Search for heavy resonances
  decaying to a photon and a hadronically decaying $Z/W/H$ boson in $pp$
  collisions at $\sqrt{s}=13$ $\mathrm{TeV}$ with the ATLAS detector}},  {\em
  Phys. Rev. D} {\bf 98} (2018), no.~3 032015,
  [\href{http://arxiv.org/abs/1805.01908}{{\tt arXiv:1805.01908}}].

\bibitem{CMS:2021zxu}
{\bf CMS} Collaboration, A.~Tumasyan et~al., {\it {Search for W$\gamma$
  resonances in proton-proton collisions at $\sqrt{s} =$ 13 TeV using hadronic
  decays of Lorentz-boosted W bosons}},  {\em Phys. Lett. B} {\bf 826} (2022)
  136888, [\href{http://arxiv.org/abs/2106.10509}{{\tt arXiv:2106.10509}}].

\bibitem{ATLAS:2017xel}
{\bf ATLAS} Collaboration, M.~Aaboud et~al., {\it {Search for heavy resonances
  decaying into a $W$ or $Z$ boson and a Higgs boson in final states with
  leptons and $b$-jets in 36 fb$^{-1}$ of $\sqrt s = 13$ TeV $pp$ collisions
  with the ATLAS detector}},  {\em JHEP} {\bf 03} (2018) 174,
  [\href{http://arxiv.org/abs/1712.06518}{{\tt arXiv:1712.06518}}]. [Erratum:
  JHEP 11, 051 (2018)].

\bibitem{ATLAS:2017ptz}
{\bf ATLAS} Collaboration, M.~Aaboud et~al., {\it {Search for heavy resonances
  decaying to a $W$ or $Z$ boson and a Higgs boson in the
  $q\bar{q}^{(\prime)}b\bar{b}$ final state in $pp$ collisions at $\sqrt{s} =
  13$ TeV with the ATLAS detector}},  {\em Phys. Lett. B} {\bf 774} (2017)
  494--515, [\href{http://arxiv.org/abs/1707.06958}{{\tt arXiv:1707.06958}}].

\bibitem{ATLAS:2019erb}
{\bf ATLAS} Collaboration, G.~Aad et~al., {\it {Search for high-mass dilepton
  resonances using 139 fb$^{-1}$ of $pp$ collision data collected at
  $\sqrt{s}=$13 TeV with the ATLAS detector}},  {\em Phys. Lett. B} {\bf 796}
  (2019) 68--87, [\href{http://arxiv.org/abs/1903.06248}{{\tt
  arXiv:1903.06248}}].

\bibitem{ATLAS:2017fih}
{\bf ATLAS} Collaboration, M.~Aaboud et~al., {\it {Search for new high-mass
  phenomena in the dilepton final state using 36 fb$^{-1}$ of proton-proton
  collision data at $\sqrt{s}=13$ TeV with the ATLAS detector}},  {\em JHEP}
  {\bf 10} (2017) 182, [\href{http://arxiv.org/abs/1707.02424}{{\tt
  arXiv:1707.02424}}].

\bibitem{CMS:2021ctt}
{\bf CMS} Collaboration, A.~M. Sirunyan et~al., {\it {Search for resonant and
  nonresonant new phenomena in high-mass dilepton final states at $ \sqrt{s} $
  = 13 TeV}},  {\em JHEP} {\bf 07} (2021) 208,
  [\href{http://arxiv.org/abs/2103.02708}{{\tt arXiv:2103.02708}}].

\bibitem{CMS:2018ipm}
{\bf CMS} Collaboration, A.~M. Sirunyan et~al., {\it {Search for high-mass
  resonances in dilepton final states in proton-proton collisions at
  $\sqrt{s}=$ 13 TeV}},  {\em JHEP} {\bf 06} (2018) 120,
  [\href{http://arxiv.org/abs/1803.06292}{{\tt arXiv:1803.06292}}].

\bibitem{CMS:2016cfx}
{\bf CMS} Collaboration, V.~Khachatryan et~al., {\it {Search for narrow
  resonances in dilepton mass spectra in proton-proton collisions at $\sqrt{s}$
  = 13 TeV and combination with 8 TeV data}},  {\em Phys. Lett. B} {\bf 768}
  (2017) 57--80, [\href{http://arxiv.org/abs/1609.05391}{{\tt
  arXiv:1609.05391}}].

\bibitem{ATLAS:2017eiz}
{\bf ATLAS} Collaboration, M.~Aaboud et~al., {\it {Search for additional heavy
  neutral Higgs and gauge bosons in the ditau final state produced in 36
  fb$^{-1}$ of pp collisions at $ \sqrt{s}=13 $ TeV with the ATLAS detector}},
  {\em JHEP} {\bf 01} (2018) 055, [\href{http://arxiv.org/abs/1709.07242}{{\tt
  arXiv:1709.07242}}].

\bibitem{CMS:2016xbv}
{\bf CMS} Collaboration, V.~Khachatryan et~al., {\it {Search for heavy
  resonances decaying to tau lepton pairs in proton-proton collisions at $
  \sqrt{s}=13 $ TeV}},  {\em JHEP} {\bf 02} (2017) 048,
  [\href{http://arxiv.org/abs/1611.06594}{{\tt arXiv:1611.06594}}].

\bibitem{CMS:2022zoc}
{\bf CMS} Collaboration, A.~Tumasyan et~al., {\it {Search for narrow resonances
  in the b-tagged dijet mass spectrum in proton-proton collisions at
  s=13\,\,TeV}},  {\em Phys. Rev. D} {\bf 108} (2023), no.~1 012009,
  [\href{http://arxiv.org/abs/2205.01835}{{\tt arXiv:2205.01835}}].

\bibitem{ATLAS:2023taw}
{\bf ATLAS} Collaboration, G.~Aad et~al., {\it {Search for top-philic heavy
  resonances in pp collisions at $\sqrt{s}=13$~$\text {TeV}$ with the ATLAS
  detector}},  {\em Eur. Phys. J. C} {\bf 84} (2024), no.~2 157,
  [\href{http://arxiv.org/abs/2304.01678}{{\tt arXiv:2304.01678}}].

\bibitem{ATLAS:2020lks}
{\bf ATLAS} Collaboration, G.~Aad et~al., {\it {Search for $ t\overline{t} $
  resonances in fully hadronic final states in $pp$ collisions at $ \sqrt{s} $
  = 13 TeV with the ATLAS detector}},  {\em JHEP} {\bf 10} (2020) 061,
  [\href{http://arxiv.org/abs/2005.05138}{{\tt arXiv:2005.05138}}].

\bibitem{CMS:2017ucf}
{\bf CMS} Collaboration, A.~M. Sirunyan et~al., {\it {Search for $
  \mathrm{t}\overline{\mathrm{t}} $ resonances in highly boosted lepton+jets
  and fully hadronic final states in proton-proton collisions at $ \sqrt{s}=13
  $ TeV}},  {\em JHEP} {\bf 07} (2017) 001,
  [\href{http://arxiv.org/abs/1704.03366}{{\tt arXiv:1704.03366}}].

\bibitem{Choudhury:2011cg}
D.~Choudhury, R.~M. Godbole, and P.~Saha, {\it {Dijet resonances, widths and
  all that}},  {\em JHEP} {\bf 01} (2012) 155,
  [\href{http://arxiv.org/abs/1111.1054}{{\tt arXiv:1111.1054}}].

\bibitem{Pappadopulo:2014tg}
D.~Pappadopulo, A.~Thamm, R.~Torre, and A.~Wulzer, {\it {Heavy Vector Triplets:
  Bridging Theory and Data}},  {\em JHEP} {\bf 09} (2014) 060,
  [\href{http://arxiv.org/abs/1402.4431}{{\tt arXiv:1402.4431}}].

\bibitem{Sullivan:2002p2617}
Z.~Sullivan, {\it Fully differential ${W}^\prime$ production and decay at
  next-to-leading order in {QCD}},  {\em Phys. Rev.} {\bf D 66} (2002) 075011,
  [\href{http://arxiv.org/abs/hep-ph/0207290}{{\tt hep-ph/0207290}}].
  [\href{http://inspirebeta.net/record/591241}{Inspire}].

\bibitem{Rizzo:2007bk}
T.~G. Rizzo, {\it {The Determination of the Helicity of $W'$ Boson Couplings at
  the LHC}},  {\em JHEP} {\bf 05} (2007) 037,
  [\href{http://arxiv.org/abs/0704.0235}{{\tt arXiv:0704.0235}}].
  [\href{http://inspirehep.net/record/747947}{Inspire}].

\bibitem{Salvioni:2009mt}
E.~Salvioni, G.~Villadoro, and F.~Zwirner, {\it Minimal {Z}' models: present
  bounds and early {LHC} reach},  {\em JHEP} {\bf 11} (2009) 068,
  [\href{http://arxiv.org/abs/0909.1320}{{\tt arXiv:0909.1320}}].
  [\href{http://inspirebeta.net/record/830532}{Inspire}].

\bibitem{Alves:2011dz}
{\bf LHC New Physics Working Group} Collaboration, D.~Alves et~al., {\it
  {Simplified Models for LHC New Physics Searches}},  {\em J. Phys.} {\bf G 39}
  (2012) 105005, [\href{http://arxiv.org/abs/1105.2838}{{\tt
  arXiv:1105.2838}}]. [\href{http://inspirehep.net/record/900212}{Inspire}].

\bibitem{Bauer:2009p1295}
C.~W. Bauer, Z.~Ligeti, M.~Schmaltz, J.~Thaler, and D.~G.~E. Walker, {\it
  {S}upermodels for early {LHC}},  {\em Phys. Lett.} {\bf B 690} (2010)
  280--288, [\href{http://arxiv.org/abs/0909.5213}{{\tt arXiv:0909.5213}}].
  [\href{http://inspirebeta.net/record/832461}{Inspire}].

\bibitem{Accomando:2010p2249}
E.~Accomando, A.~S. Belyaev, L.~Fedeli, S.~F. King, and C.~H.
  Shepherd-Themistocleous, {\it {$Z'$ physics with early LHC data}},  {\em
  Phys. Rev.} {\bf D 83} (2012) 075012,
  [\href{http://arxiv.org/abs/1010.6058}{{\tt arXiv:1010.6058}}].
  [\href{http://inspirehep.net/record/874714}{Inspire}].

\bibitem{delAguila:2010mx}
F.~del Aguila, J.~de~Blas, and M.~Perez-Victoria, {\it {Electroweak Limits on
  General New Vector Bosons}},  {\em JHEP} {\bf 09} (2010) 033,
  [\href{http://arxiv.org/abs/1005.3998}{{\tt arXiv:1005.3998}}].

\bibitem{Han:2010p2696}
T.~Han, I.~Lewis, and Z.~Liu, {\it Colored {R}esonant {S}ignals at the {LHC}:
  {L}argest {R}ate and {S}implest {T}opology},  {\em JHEP} {\bf 12} (2010) 085,
  [\href{http://arxiv.org/abs/1010.4309}{{\tt arXiv:1010.4309}}].
  [\href{http://inspirebeta.net/record/873674}{Inspire}].

\bibitem{Barbieri:2011p2759}
R.~Barbieri and R.~Torre, {\it Signals of single particle production at the
  earliest {LHC}},  {\em Phys. Lett.} {\bf B 695} (2011) 259--263,
  [\href{http://arxiv.org/abs/1008.5302}{{\tt arXiv:1008.5302}}].
  [\href{http://inspirebeta.net/record/866732}{Inspire}].

\bibitem{Chiang:2011kq}
C.-W. Chiang, N.~D. Christensen, G.-J. Ding, and T.~Han, {\it {Discovery in
  Drell-Yan Processes at the LHC}},  {\em Phys. Rev.} {\bf D 85} (2012) 015023,
  [\href{http://arxiv.org/abs/1107.5830}{{\tt arXiv:1107.5830}}].
  [\href{http://inspirehep.net/record/921469}{Inspire}].

\bibitem{DeSimone:2012ul}
A.~{De Simone}, O.~Matsedonskyi, R.~Rattazzi, and A.~Wulzer, {\it {A First Top
  Partner's Hunter Guide}},  {\em JHEP} {\bf 04} (2013) 004,
  [\href{http://arxiv.org/abs/1211.5663}{{\tt arXiv:1211.5663}}].
  [\href{http://inspirehep.net/record/1203860}{Inspire}].

\bibitem{deBlas:2012tc}
J.~{de Blas}, J.~M. Lizana, and M.~Perez-Victoria, {\it {Combining searches of
  $Z'$ and $W'$ bosons}},  {\em JHEP} {\bf 01} (2013) 166,
  [\href{http://arxiv.org/abs/1211.2229}{{\tt arXiv:1211.2229}}].
  [\href{http://inspirehep.net/record/1201947}{Inspire}].

\bibitem{AguilarSaavedra:2013hg}
J.~A. Aguilar-Saavedra, R.~Benbrik, S.~Heinemeyer, and M.~Perez-Victoria, {\it
  {A handbook of vector-like quarks: mixing and single production}},  {\em
  Phys. Rev.} {\bf D 88} (2013) 094010,
  [\href{http://arxiv.org/abs/1306.0572}{{\tt arXiv:1306.0572}}].
  [\href{http://inspirehep.net/record/1236810}{Inspire}].

\bibitem{Buchkremer:2013uj}
M.~Buchkremer, G.~Cacciapaglia, A.~Deandrea, and L.~Panizzi, {\it {Model
  Independent Framework for Searches of Top Partners}},  {\em Nucl. Phys.} {\bf
  B 876} (2013) 376--417, [\href{http://arxiv.org/abs/1305.4172}{{\tt
  arXiv:1305.4172}}]. [\href{http://inspirehep.net/record/1233883}{Inspire}].

\bibitem{Lizana:2013vz}
J.~M. Lizana and M.~Perez-Victoria, {\it {Vector triplets at the LHC}},  {\em
  EPJ Web Conf.} {\bf 60} (2013) 17008,
  [\href{http://arxiv.org/abs/1307.2589}{{\tt arXiv:1307.2589}}].
  [\href{http://inspirehep.net/record/1242121}{Inspire}].

\bibitem{Chivukula:2017lyk}
R.~S. Chivukula, P.~Ittisamai, K.~Mohan, and E.~H. Simmons, {\it {Broadening
  the Reach of Simplified Limits on Resonances at the LHC}},  {\em Phys. Rev.
  D} {\bf 96} (2017), no.~5 055043,
  [\href{http://arxiv.org/abs/1707.01080}{{\tt arXiv:1707.01080}}].

\bibitem{Chivukula:2021foa}
R.~S. Chivukula, P.~Ittisamai, J.~Osborne, and E.~H. Simmons, {\it {Narrow
  Resonances Revisited -- Simplifying Multidimensional Constraints}},  {\em
  Phys. Rev. D} {\bf 103} (2021), no.~9 095008,
  [\href{http://arxiv.org/abs/2103.06283}{{\tt arXiv:2103.06283}}].

\bibitem{Saez:2018off}
B.~D. S\'aez, F.~Rojas-Abatte, and A.~R. Zerwekh, {\it {Dark Matter from a
  Vector Field in the Fundamental Representation of $SU(2)_L$}},  {\em Phys.
  Rev. D} {\bf 99} (2019), no.~7 075026,
  [\href{http://arxiv.org/abs/1810.06375}{{\tt arXiv:1810.06375}}].

\bibitem{Belyaev:2018xpf}
A.~Belyaev, G.~Cacciapaglia, J.~Mckay, D.~Marin, and A.~R. Zerwekh, {\it
  {Minimal Spin-one Isotriplet Dark Matter}},  {\em Phys. Rev. D} {\bf 99}
  (2019), no.~11 115003, [\href{http://arxiv.org/abs/1808.10464}{{\tt
  arXiv:1808.10464}}].

\bibitem{Dawson:2024ozw}
S.~Dawson, M.~Forslund, and M.~Schnubel, {\it {SMEFT Matching to $Z^\prime$
  Models at Dimension-8}},  \href{http://arxiv.org/abs/2404.01375}{{\tt
  arXiv:2404.01375}}.

\bibitem{Panico:2015jxa}
G.~Panico and A.~Wulzer, {\em {The Composite Nambu-Goldstone Higgs}}, vol.~913.
\newblock Springer, 2016.

\bibitem{MuLan:2010shf}
{\bf MuLan} Collaboration, D.~M. Webber et~al., {\it {Measurement of the
  Positive Muon Lifetime and Determination of the Fermi Constant to
  Part-per-Million Precision}},  {\em Phys. Rev. Lett.} {\bf 106} (2011)
  041803, [\href{http://arxiv.org/abs/1010.0991}{{\tt arXiv:1010.0991}}].

\bibitem{MuLan:2012sih}
{\bf MuLan} Collaboration, V.~Tishchenko et~al., {\it {Detailed Report of the
  MuLan Measurement of the Positive Muon Lifetime and Determination of the
  Fermi Constant}},  {\em Phys. Rev. D} {\bf 87} (2013), no.~5 052003,
  [\href{http://arxiv.org/abs/1211.0960}{{\tt arXiv:1211.0960}}].

\bibitem{Alloul:2013vc}
A.~Alloul, N.~D. Christensen, C.~Degrande, C.~Duhr, and B.~Fuks, {\it
  {FeynRules 2.0 - A complete toolbox for tree-level phenomenology}},  {\em
  Comput. Phys. Commun.} {\bf 185} (2014) 2250--2300,
  [\href{http://arxiv.org/abs/1310.1921}{{\tt arXiv:1310.1921}}].

\bibitem{Christensen:2008py}
N.~D. Christensen and C.~Duhr, {\it {FeynRules - Feynman rules made easy}},
  {\em Comput. Phys. Commun.} {\bf 180} (2009) 1614--1641,
  [\href{http://arxiv.org/abs/0806.4194}{{\tt arXiv:0806.4194}}].

\bibitem{Alwall:2011fk}
J.~Alwall, M.~Herquet, F.~Maltoni, O.~Mattelaer, and T.~Stelzer, {\it {MadGraph
  5: going beyond}},  {\em JHEP} {\bf 06} (2011) 128,
  [\href{http://arxiv.org/abs/1106.0522}{{\tt arXiv:1106.0522}}].
  [\href{http://inspirehep.net/record/912611}{Inspire}].

\bibitem{hepmdb}
M.~Bondarenko, A.~Belyaev, J.~Blandford, L.~Basso, E.~Boos, V.~Bunichev,
  et~al., {\it {High Energy Physics Model Database : Towards decoding of the
  underlying theory (within Les Houches 2011: Physics at TeV Colliders New
  Physics Working Group Report)}},  \href{http://arxiv.org/abs/1203.1488}{{\tt
  arXiv:1203.1488}}.

\bibitem{ParticleDataGroup:2024pth}
S.~N. et~al. (Particle Data~Group), {\it {The Review of Particle Physics
  (2024)}},  {\em to be published in Phys. Rev. D 110, 030001 (2024)}.

\bibitem{Hartland:2012ia}
N.~P. Hartland and E.~R. Nocera, {\it {A Mathematica interface to NNPDFs}},
  {\em Nucl. Phys. B Proc. Suppl.} {\bf 234} (2013) 54--57,
  [\href{http://arxiv.org/abs/1209.2585}{{\tt arXiv:1209.2585}}].

\bibitem{Chanowitz:1985hj}
M.~S. Chanowitz and M.~K. Gaillard, {\it {The TeV Physics of Strongly
  Interacting W's and Z's}},  {\em Nucl.Phys.} {\bf B261} (1985).
  [\href{http://inspirebeta.net/record/216026}{Inspire}].

\bibitem{Wulzer:2013mza}
A.~Wulzer, {\it {An Equivalent Gauge and the Equivalence Theorem}},  {\em Nucl.
  Phys. B} {\bf 885} (2014) 97--126,
  [\href{http://arxiv.org/abs/1309.6055}{{\tt arXiv:1309.6055}}].

\bibitem{Cacciapaglia:2006pk}
G.~Cacciapaglia, C.~Csaki, G.~Marandella, and A.~Strumia, {\it {The Minimal Set
  of Electroweak Precision Parameters}},  {\em Phys. Rev.} {\bf D 74} (2006)
  033011, [\href{http://arxiv.org/abs/hep-ph/0604111}{{\tt hep-ph/0604111}}].
  [\href{http://inspirehep.net/record/714334}{Inspire}].

\bibitem{CMS:2022krd}
{\bf CMS} Collaboration, A.~Tumasyan et~al., {\it {Search for new physics in
  the lepton plus missing transverse momentum final state in proton-proton
  collisions at $\sqrt{s} =$ 13 TeV}},  {\em JHEP} {\bf 07} (2022) 067,
  [\href{http://arxiv.org/abs/2202.06075}{{\tt arXiv:2202.06075}}].

\bibitem{Strumia:2022qkt}
A.~Strumia, {\it {Interpreting electroweak precision data including the W-mass
  CDF anomaly}},  {\em JHEP} {\bf 08} (2022) 248,
  [\href{http://arxiv.org/abs/2204.04191}{{\tt arXiv:2204.04191}}].

\bibitem{Farina:2016rws}
M.~Farina, G.~Panico, D.~Pappadopulo, J.~T. Ruderman, R.~Torre, and A.~Wulzer,
  {\it {Energy helps accuracy: electroweak precision tests at hadron
  colliders}},  {\em Phys. Lett. B} {\bf 772} (2017) 210--215,
  [\href{http://arxiv.org/abs/1609.08157}{{\tt arXiv:1609.08157}}].

\bibitem{Torre:2020aiz}
R.~Torre, L.~Ricci, and A.~Wulzer, {\it {On the W\&Y interpretation of
  high-energy Drell-Yan measurements}},  {\em JHEP} {\bf 02} (2021) 144,
  [\href{http://arxiv.org/abs/2008.12978}{{\tt arXiv:2008.12978}}].

\bibitem{ATLAS:2022jsi}
{\bf ATLAS} Collaboration, {\it {Combination of searches for heavy resonances
  using 139 fb$^{-1}$ of proton\textendash{}proton collision data at $\sqrt{s}$
  = 13 TeV with the ATLAS detector}}, .

\bibitem{FCC:2018bvk}
{\bf FCC} Collaboration, A.~Abada et~al., {\it {HE-LHC: The High-Energy Large
  Hadron Collider}: {Future Circular Collider Conceptual Design Report Volume
  4}},  {\em Eur. Phys. J. ST} {\bf 228} (2019), no.~5 1109--1382.

\bibitem{CidVidal:2018eel}
X.~Cid~Vidal et~al., {\it {Report from Working Group 3}: {Beyond the Standard
  Model physics at the HL-LHC and HE-LHC}},  {\em CERN Yellow Rep. Monogr.}
  {\bf 7} (2019) 585--865, [\href{http://arxiv.org/abs/1812.07831}{{\tt
  arXiv:1812.07831}}].

\bibitem{FCC:2018vvp}
{\bf FCC} Collaboration, A.~Abada et~al., {\it {FCC-hh: The Hadron Collider}:
  {Future Circular Collider Conceptual Design Report Volume 3}},  {\em Eur.
  Phys. J. ST} {\bf 228} (2019), no.~4 755--1107.

\bibitem{CEPC-SPPCStudyGroup:2015csa}
{\bf CEPC-SPPC} Collaboration, {\it {CEPC-SPPC Preliminary Conceptual Design
  Report. 1. Physics and Detector}}, .

\bibitem{Thamm:2015zwa}
A.~Thamm, R.~Torre, and A.~Wulzer, {\it {Future tests of Higgs compositeness:
  direct vs indirect}},  {\em JHEP} {\bf 07} (2015) 100,
  [\href{http://arxiv.org/abs/1502.01701}{{\tt arXiv:1502.01701}}].

\bibitem{ZurbanoFernandez:2020cco}
I.~Zurbano~Fernandez et~al., {\it {High-Luminosity Large Hadron Collider
  (HL-LHC): Technical design report}}, .

\bibitem{Helsens:2019bfw}
C.~Helsens, D.~Jamin, M.~L. Mangano, T.~G. Rizzo, and M.~Selvaggi, {\it {Heavy
  resonances at energy-frontier hadron colliders}},  {\em Eur. Phys. J. C} {\bf
  79} (2019) 569, [\href{http://arxiv.org/abs/1902.11217}{{\tt
  arXiv:1902.11217}}].

\bibitem{deBlas:2016ojx}
J.~de~Blas, M.~Ciuchini, E.~Franco, S.~Mishima, M.~Pierini, L.~Reina, and
  L.~Silvestrini, {\it {Electroweak precision observables and Higgs-boson
  signal strengths in the Standard Model and beyond: present and future}},
  {\em JHEP} {\bf 12} (2016) 135, [\href{http://arxiv.org/abs/1608.01509}{{\tt
  arXiv:1608.01509}}].

\bibitem{Barenboim:2001vu}
G.~Barenboim, M.~Gorbahn, U.~Nierste, and M.~Raidal, {\it {Higgs Sector of the
  Minimal Left-Right Symmetric Model}},  {\em Phys. Rev. D} {\bf 65} (2002)
  095003, [\href{http://arxiv.org/abs/hep-ph/0107121}{{\tt hep-ph/0107121}}].

\bibitem{Contino:2013un}
R.~Contino, C.~Grojean, D.~Pappadopulo, R.~Rattazzi, and A.~Thamm, {\it {Strong
  Higgs Interactions at a Linear Collider}},  {\em JHEP} {\bf 02} (2014) 006,
  [\href{http://arxiv.org/abs/1309.7038}{{\tt arXiv:1309.7038}}].
  [\href{http://inspirehep.net/record/1255676}{Inspire}].

\bibitem{Coleman:1969sm}
S.~R. Coleman, J.~Wess, and B.~Zumino, {\it {Structure of phenomenological
  Lagrangians. 1.}},  {\em Phys. Rev.} {\bf 177} (1969) 2239--2247.

\bibitem{Callan:1969sn}
C.~G. Callan, Jr., S.~R. Coleman, J.~Wess, and B.~Zumino, {\it {Structure of
  phenomenological Lagrangians. 2.}},  {\em Phys. Rev.} {\bf 177} (1969)
  2247--2250.

\bibitem{Brooijmans:2014eja}
G.~Brooijmans et~al., {\it {Les Houches 2013: Physics at TeV Colliders: New
  Physics Working Group Report}},  \href{http://arxiv.org/abs/1405.1617}{{\tt
  arXiv:1405.1617}}.

\bibitem{Contino:2010vc}
R.~Contino, C.~Grojean, M.~Moretti, F.~Piccinini, and R.~Rattazzi, {\it
  {S}trong {D}ouble {H}iggs {P}roduction at the {LHC}},  {\em JHEP} {\bf 05}
  (2010) 089, [\href{http://arxiv.org/abs/1002.1011}{{\tt arXiv:1002.1011}}].
  [\href{http://inspirebeta.net/record/845195}{Inspire}].

\bibitem{Buttazzo:2015bka}
D.~Buttazzo, F.~Sala, and A.~Tesi, {\it {Singlet-like Higgs bosons at present
  and future colliders}},  {\em JHEP} {\bf 11} (2015) 158,
  [\href{http://arxiv.org/abs/1505.05488}{{\tt arXiv:1505.05488}}].

\end{thebibliography}\endgroup

\end{document}